\documentclass{pasj00}
\draft
\color{black}

\begin{document}
\SetRunningHead{U. Morita et al.}
{Chandra and XMM-Newton Observations of HCG~62} 
\Received{2005/mm/dd}
\Accepted{2005/mm/dd}

\title{Chandra and XMM-Newton Observations of \\
a Group of Galaxies HCG~62}

 \author{%
   Umeyo \textsc{Morita},\altaffilmark{1}
   Yoshitaka \textsc{Ishisaki},\altaffilmark{1}
   Noriko Y. \textsc{Yamasaki},\altaffilmark{2} 
   Naomi \textsc{Ota},\altaffilmark{3} \\
   Naomi \textsc{Kawano},\altaffilmark{4} 
   Yasushi \textsc{Fukazawa},\altaffilmark{4}
   and
   Takaya \textsc{Ohashi}\altaffilmark{1}
   }
 \altaffiltext{1}{Department of Physics, Tokyo Metropolitan University, 
   1-1 Minami-Osawa, Hachioji, Tokyo 192-0397}
 \email{umeyo@phys.metro-u.ac.jp}
 \altaffiltext{2}{Institute of Space and Astronautical Science (ISAS/JAXA),\\
   3-1-1 Yoshinodai, Sagamihara, Kanagawa 229-8510}
 \altaffiltext{3}{Cosmic Radiation Laboratory, 
   RIKEN, 2-1 Hirosawa, Wako, Saitama 351-0198}
 \altaffiltext{4}{Department of Physical Science, School of Science, 
   Hiroshima University, \\
   1-3-1 Kagamiyama, Higashi-Hiroshima, Hiroshima 739-8526}

\maketitle

\begin{abstract}
We present results from Chandra and XMM-Newton observations of the
bright group of galaxies HCG~62\@. There are two cavities at about
$30''$ northeast and $20''$ southwest of the central galaxy in the
Chandra image. The energy spectrum shows no significant change in the
cavity compared with that in the surrounding region. The radial X-ray
profile is described by a sum of 3-$\beta$ components with core radii
about 2, 10, and 160 kpc, respectively. We studied radial
distributions of temperature and metal abundance with joint spectral
fit for the Chandra and XMM-Newton data, and two temperatures were
required in the inner $r< 2'$ (35~kpc) region.
The sharp drop of temperature at $r\sim 5'$ implies the gravitational
mass density even lower than the gas density, suggesting the gas may
not be in hydrostatic equilibrium. Fe and Si abundances are
1--2 solar at the center and drop to about 0.1 solar at $r \sim 10'$.
O abundance is less than 0.5 solar and shows a flatter profile.
Observed metal distribution supports the view that iron and silicon
are produced by type Ia supernova in the central galaxy,
while galactic winds by type II supernova have caused
wide distribution of oxygen.
The supporting mechanism of the cavity is discussed.  Pressure for the sum
of electrons and magnetic field is too low to displace the hot group gas,
and the required pressure due to high energy protons are nearly
700 times higher than the electron pressure.
This leaves the origin of the cavities a puzzle,
and we discuss other possible origins of the cavities.
\end{abstract}

\section{Introduction}

Groups of galaxies hold significantly less amount of hot gas compared
with rich clusters on the average, and the apparent deficiency of
baryons in these low-mass systems is a problem in explaining the
baryon budget in rich systems in terms of hierarchical merging
scenario \citep{Voit2005}. The observations of compact groups of
galaxies are important to search for the hidden form of baryons and
their release mechanism, which would be strongly connected with
dynamics of gas and galaxies.  In particular, some groups dominated by
bright central galaxies have shown gas features which are strongly
affected by the activity of the central galaxies.  High sensitivity
X-ray observations of groups of galaxies are powerful method to look
into the role of central galaxies in terms of metal distribution and
gas morphologies.

The striking gas features most likely caused by the central galaxies
are the X-ray cavities. The cavities are circular regions showing a
significant depression of X-ray surface brightness.
Nearly 20 cavities have been recognized in clusters and groups with high
resolution images taken by ROSAT and Chandra
(\cite{Birzan2004}, hereafter B04;
\cite{Dunn2004}; \cite{Dunn2005}, hereafter D05).
They are located typically at 10--30 kpc from the central galaxies,
and strong correlation with radio lobes are seen in about 10 systems.
Remarkable cases are seen in the Perseus
(e.g., \cite{Boehringer1993,Fabian2000}) and
Hydra A (e.g.\ \cite{McNamara2000}) clusters,
both showing strong correlation with the 1.4~GHz radio lobes.
Several giant elliptical galaxies with radio robes,
e.g., M84 \citep{Finoguenov2001}, NGC~4636 \citep{Ohto2003},
are also known as nesting X-ray cavities.
The remaining half of the cavities, on the other hand, are not associated
with radio lobes, and they are designated as ghost cavities.
The one in A~2597 \citep{McNamara2001} or
the outer depressions in Perseus \citep{Fabian2000} are examples.
Cavities are thought to be produced by jets or buoyant bubbles
which are directly connected
with the activity of central radio galaxies. 

A subsonic displacement of the gas would create a low density,
rising bubble keeping the pressure balance with the surrounding ICM\@.
It appears to be in general supposed for cavities
that non-thermal pressure originated in relativistic particles
and/or magnetic fields in the radio lobe is probably large enough
to balance with the surrounding ICM gas pressure \citep{Fabian2002}.
This pseudo-pressure balance is justified by the fact
that there are no evidence for shock-heated gas around the
radio lobes in almost all of the X-ray cavities observed so far,
except for MKW~3s \citep{Mazzotta2002}.
This general scenario has been modeled theoretically,
and has at least qualitatively reproduced the morphology of cavities
(e.g., \cite{Churazov2001}).
Energy density of
relativistic electrons inferred from the synchrotron radio emission is
almost always smaller than that required to offset the hot gas by
orders of magnitude, and it is discussed that energy density of
protons are higher than those due to electrons by
factors of 100--1000 (D05). However, there is no
direct evidence indicating that such a high energy density is really
carried by protons. This situation is the severest in the case of
ghost cavities. In this view, it is important to examine ghost
cavities in groups of galaxies where the gas is relatively cool and
non-thermal effect can be recognized somewhat easily.

Detailed studies on the metal distribution in clusters and groups have
been carried out using ASCA, BeppoSAX, Chandra and
XMM-Newton. Distribution of iron and silicon indicate strong central
concentration in clusters and groups characterized by bright central
galaxies, and the excess iron mass is found to correlate with the
luminosity of the cD galaxy \citep{DeGrandi2004}. This indicates
that iron and silicon (main products from type Ia supernova; SN~Ia)
trace the enhanced star-formation activity in bright galaxies.
On the other hand, distribution of oxygen
(i.e.\ type II supernova product; SN~II) is not well understood.
\citet{Matsushita2003} showed that oxygen
distribution around M~87 is flatter than those of iron and silicon,
with the level about half as much as the others. Such low oxygen
abundances are also derived in the centers of other clusters and
groups (e.g.\ \cite{Buote2003b,Xue2004}).  For the study of oxygen
distribution, low temperature systems such as groups of galaxies are
suitable targets.

In this paper, we report the results from Chandra and XMM-Newton
observations of HCG~62, which is one of the nearest Hickson compact
galaxy groups \citep{Hickson1989}.
The whole group consists of 63 galaxies \citep{Mulchaey2003}
within a radius of 50$'$ (900~kpc),
but the central region is dominated by 4 galaxies.
HCG~62 is the brightest group of galaxies in the X-ray band, and
the extended X-ray emission was first discovered by
\citet{Ponman_Bertram1993} from
the ROSAT PSPC observation.  Based on the ASCA observation,
\citet{Fukazawa2001} detected excess hard X-ray emission, and
\citet{Finoguenov_Ponman1999} report strong central concentration
of iron. Using the high resolution image of Chandra,
\authorcite{Vrtilek2001}~(\yearcite{Vrtilek2001},\yearcite{Vrtilek2002})
detected two ghost cavities, 
which is the first report of cavities in groups of galaxies.

This paper is organized as follows: In \S\,2 we describe the
Chandra and XMM-Newton observations and the data reduction. 
In \S\,3 we give the image of HCG~62 of 
both Chandra and XMM-Newton, 
in \S\,4 we describe the X-ray cavity structure using Chandra image.
In \S\,5 we present our results on the temperature profiles and 
the abundances profiles of Fe and $\alpha$-elements (Si, Mg, and O). 
In \S\,6--8 we give discussions of the obtained results,
and finally we summarize our conclusions in \S\,9.
Throughout this paper we adopt 
$\Omega_{\rm \Lambda}=1-\Omega_{\rm M}=0.73$ and
$h_{70}\equiv H_0/(70~{\rm km~s^{-1}\,Mpc^{-1}})=1$; 
$1\arcmin$ corresponds to 17.8~kpc at $z=0.0145$.  
The quoted errors indicate the 90\% confidence range, unless otherwise stated. 
We use the solar abundance ratio of \citet{Anders_Grevesse_1989}.

\section{Observation and Data Reduction}

\subsection{Chandra Observation}
\label{subsec:Chandra obs}

HCG~62 was observed on 25 January 2000 with
the Advanced CCD imaging Spectrometer (ACIS)
I2, I3, S2, S3, and S4 chips operated
at the CCD temperature of $-110{\rm ^{\circ}C}$ 
with a frame readout time of 3.24~s.
We used the data of only the ACIS-S3 chip ($8.4'\times 8.4'$)
covering the central part of HCG~62 in this paper.
The pointing coordinates were
(\timeform{12h59m05.70s}, \timeform{-09D12'20.00''}) (J2000)
and the total exposure was 49.15~ks.
According to the ``Chandra Aspect'' web page,
http://cxc.harvard.edu/cal/ASPECT/,
the astrometry offset of RA $= -0.03''$, Dec $=  0.22''$
has been corrected for in the data, providing a radius of
0.6 arcsec as the absolute position accuracy in 90\% confidence.
The CCD temperature of ACIS was reduced to $-120{\rm ^{\circ}C}$
soon after the observation of HCG~62 in January 2000
due to the increase of charge transfer inefficiency (CTI) caused 
by the radiation damage in orbit (\cite{Grant2005}).
The data were telemetered in the Faint mode,
and events with the ASCA grades of 0, 2, 3, 4, and 6 were used.
Bad pixels, bad columns, and the columns next to bad columns
and to the chip node boundaries are excluded.
In order to remove periods of anomalous background levels,
we further filtered the events using the 0.3--10 keV band
light curve of the whole ACIS-S3 chip in 200~s bin,
and discarded periods which exceeded by $3\sigma$ above the
mean quiescent rate of 4.5 c/s/chip.
The net exposure time after the screening was 48,013~s. 
The data reduction was performed using CIAO version 3.1
with CALDB version 2.29,
and the spectral fitting was done by XSPEC version 11.3.0t.

All of the X-ray spectra were extracted using the
pulse-height invariant (PI) values,
which were recomputed using the latest gain file 
{\tt acisD1999-09-16gainN0005.fits},
appropriate for the CCD temperature of $-110{\rm ^{\circ}C}$.
In the spectral fitting, we initially generated
the response matrix file (RMF) of the ACIS-S3
using the CIAO ``mkrmf'' task with the input
FEF (FITS Embedded Function) file of
\verb|acisD1999-09-16fef_phaN0002.fits|,
which was chosen by default.
However, we found that the Si line of the IGM emission was
significantly broader than the response ($\sigma=42^{+9}_{-8}$~eV),
which caused the fit statistics not acceptable.
In order to examine whether this broad Si line is
a target specific issue or a common calibration problem,
we further checked the neutral Si line, $\rm K_{\alpha 1}=1739.98$~eV
and $\rm K_{\alpha 2}=1739.38$~eV, which is originated in the instrumental
background of the CCD, using the blank-sky data
obtained at $-110{\rm ^{\circ}C}$ (figure~\ref{fig:bg spec}~(a)).
The fit result gave the Gaussian $\sigma=29\pm 4$~eV 
($\rm FWHM=68\pm 9$~eV) and the line center energy of
$1753\pm 3$~eV, which is significantly broad, too. 
We splitted the blank-sky spectrum into two,
dividing the integration region into half,
i.e.\ upper rows and lower rows of the CCD,
although the results were similar.
This result indicates that the charge transfer inefficiency (CTI)
correction is working well.

We therefore concluded that the broad Si line was resulted because
the generated RMF had too sharp Gaussian core.
In fact, the derived Si line width for the blank-sky is consistent
with a calibration document, ACIS Memo \#182 by \citet{LaMarr2000}.
The FWHM of the Gaussian core for the RMF denoted in the FEF file
is G1\_FWHM = 73~eV for ACIS-S3 (CCD\_ID=7) at the energy of 1.8~keV,
while it should be around 125~eV according to the memo.
Our blank-sky data indicated $\sqrt{73^2+68^2} = 100$~eV,
which is slightly better than the value in the LaMarr's memo,
possibly due to the updated gain file.
We therefore created a new FEF file to adopt broader Gaussian core
over the whole energy range, modifying the G1\_FWHM column into
$\rm (16 \times ENERGY + G1\_FWHM)$\@.
Using the new FEF file, we generated the RMF files
with the ``mkrmf'' task for the following analysis.

Regarding the ``blank-sky'' background,
we adopted a series of observations compiled by
\cite{Markevitch_http} (http://cxc.harvard.edu/contrib/maxim/acisbg/),
when the CCD temperature was $-110{\rm ^{\circ}C}$.
The ``blank-sky'' background events were also screened to
remove background flares in the same way as described above.
The additional component due to the Galactic soft background are
considered in \S\,\ref{subsec:soft bg}.

\subsection{XMM-Newton observation}\label{subsec:xmm obs}

The XMM-Newton observation of HCG~62 was carried out
on 15 January 2003, assigned for 12.6~ks. 
The EPIC cameras were operated in full-frame mode for MOS,
and in extended-full-frame mode for pn.
The medium filters were used for both cameras.
XMM-Newton covers the wider field of view of $r\lesssim 15'$
than Chandra, while the angular resolution of $15''$
half-power diameter is broader.
Data reduction was performed using SAS version 6.0,
and the spectral fitting was done by XSPEC version 11.3.0t.
We selected events with pattern 0--12 for MOS 
and 0--4 for pn, and flag = 0 for both.
Bad pixels and bad columns were excluded.
We calculated the count rate distribution with 100~s intervals
over the 0.3--10~keV range using the whole chip of each sensor,
and rejected periods by requiring all the count rates
to be within $\pm 2\sigma$ around the mean.  
We iterated the process until the number of rejected data
in a step reached less than 5\% of the $2\sigma$ compared with 
the previous value.  
Because we did not find large time variation nor flares, 
the net exposure time after the screening was
12.6, 12.5, and 9.2 ks for MOS1, MOS2, and pn, respectively.

To correct for the vignetting effect,
the SAS ``evigweight'' task was applied for each event file.
All the X-ray spectra were extracted
using the vignetting-weighted events.
The response files for spectral fittings were generated
in the standard way with the SAS ``rmfgen'' and ``arfgen'' tasks
at on-axes of the X-ray mirrors.
The MOS1 and MOS2 spectra are summed up, and the energy range of
1.4367--1.5367~keV, where the background Al-K$_\alpha$ (1.4867~keV)
is strong as seen in figure~\ref{fig:bg spec}~(b) and (c),
is ignored in the spectral fitting for both MOS and pn.

The background event dataset created by \citet{Read2003} 
was adopted as the ``blank-sky'' background
for the XMM-Newton observation.
We applied the same selection criteria as the source events
to the ``blank-sky'' background.
The ``blank-sky'' background was scaled to 0.94 for MOS1,
1.00 for MOS2, and 0.84 for pn, respectively.
These factors were determined by the count rate ratio
within $r < 14'$ from the central galaxy HCG~62a
in 10--12~keV (MOS) or 12--14~keV (pn) band,
where the instrumental background is dominant for each sensor.
\citet{Katayama2004} and \citet{Nevalainen2005} have
studied the background data of XMM-Newton in detail,
and it is reported that the
the 90\% confidence background uncertainty is $\pm 5$\%
in 4--7~keV and $\pm 20$\% in 0.8--1~keV\@.
It is confirmed that our results does not change significantly
within errors when this level of uncertainty is considered.

\subsection{Background estimation}\label{subsec:soft bg}

\begin{figure*}[tbg]
\begin{center}
    \FigureFile(0.33\textwidth,0.33\textwidth){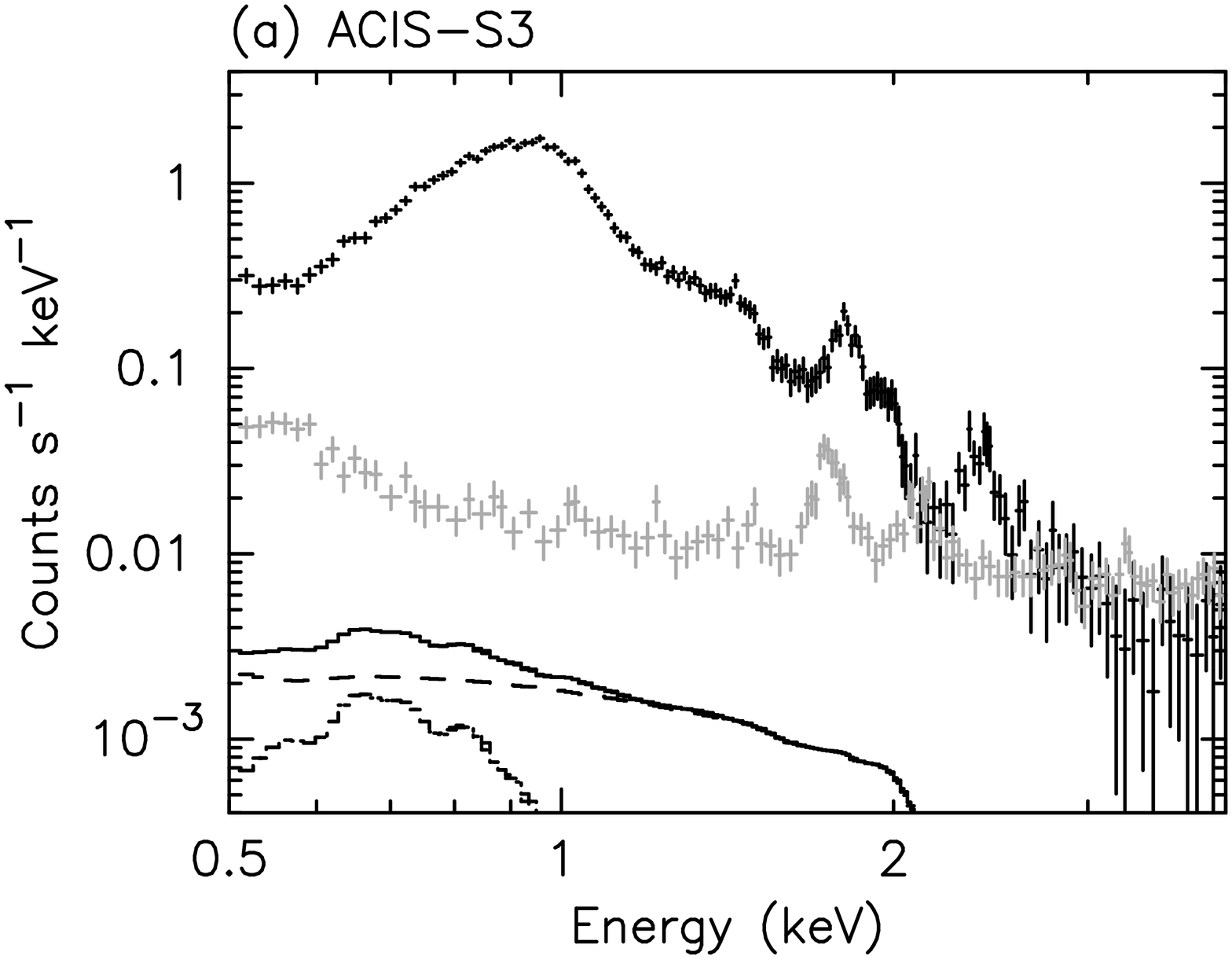}\hfill
    \FigureFile(0.33\textwidth,0.33\textwidth){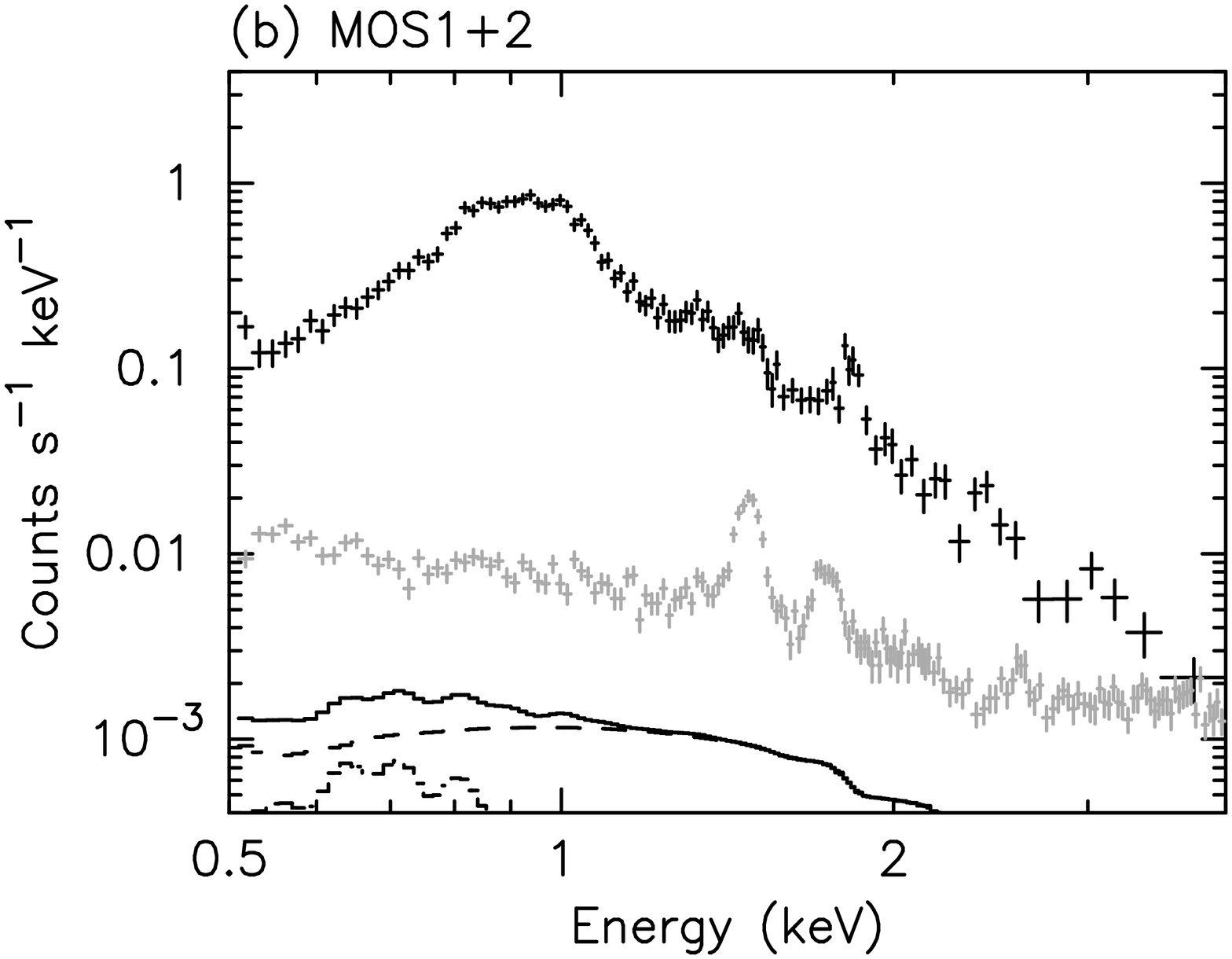}\hfill
    \FigureFile(0.33\textwidth,0.33\textwidth){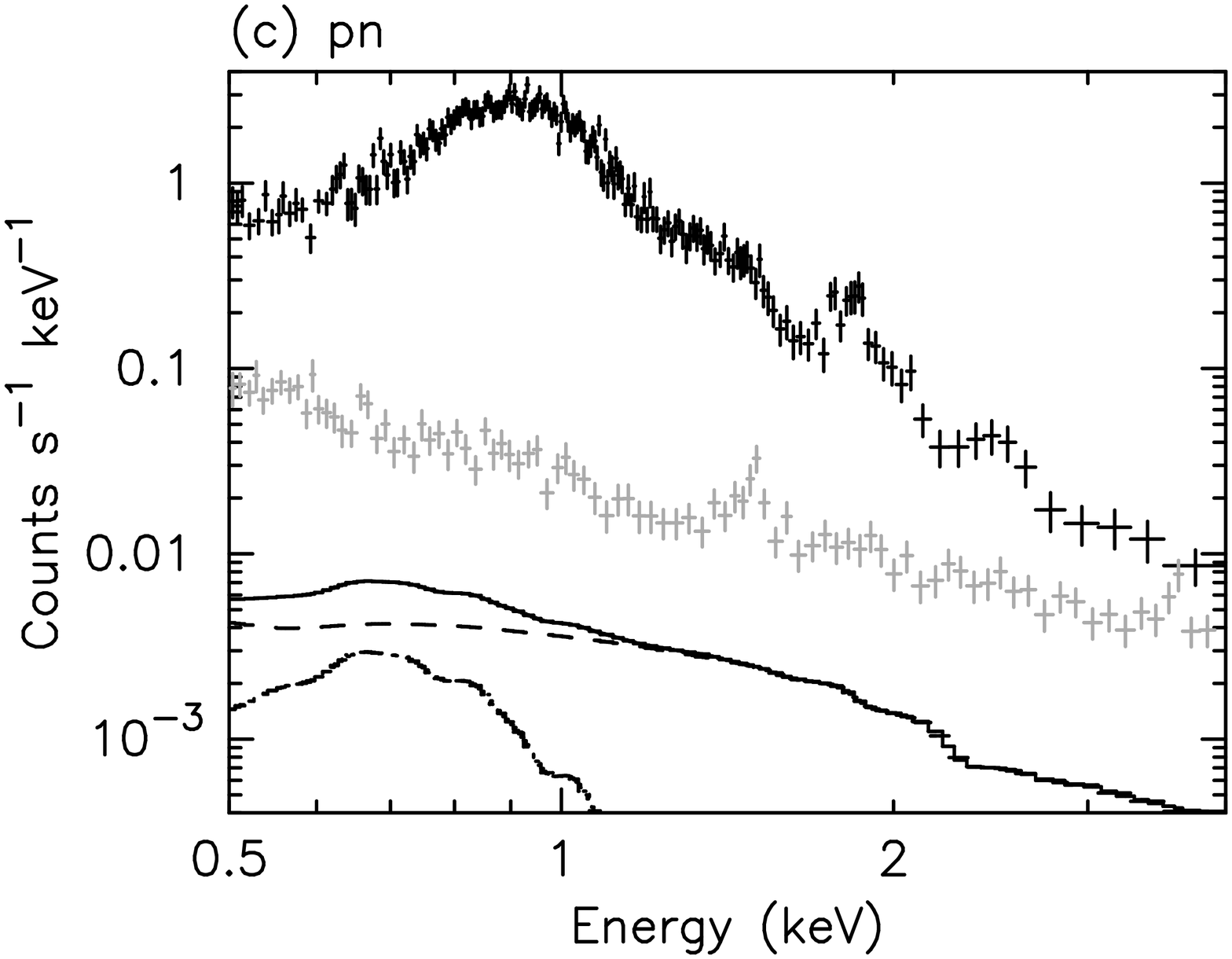}
\end{center}
\caption{
X-ray spectra of (a) Chandra ACIS-S3, (b) XMM-Newton MOS1+2,
and (c) pn, in 0.5--4~keV within $r < 2'$ from the central galaxy HCG~62a, 
subtracted by the ``blank-sky'' background. 
The ``blank-sky'' spectra are indicated by gray lines.
The additional background components of
the 0.3~keV Mekal and the $\Gamma=1.5$ power-law models
are shown by dashed black lines, and their sum is
indicated by solid black lines.
}\label{fig:bg spec}
\end{figure*}

HCG~62 is located near the edge of the North Polar Spur (NPS), which
is a large soft X-ray Galactic structure.  The ROSAT All-Sky Survey
(RASS) image at 3/4 keV band in this region indicates a soft X-ray
excess around the position of HCG~62, although significant fraction
of the excess is probably caused by HCG~62 itself.
This emission looks more extended than the Chandra and XMM-Newton
fields of view in the previous observations with ROSAT and ASCA\@.
Therefore, we cannot use the events in outer region as the background,
and we need to estimate the influence of the soft X-ray background
to our HCG~62 data using the ``blank-sky'' background.

There are four fields near HCG~62 already observed with Chandra,
Q1246$-$0542, NGC~4697, NGC~4594, and NGC~4782.
We found that the intensities below 2~keV in source-free regions in
these fields are higher than the level of the ``blank-sky'' background,
while the hard band intensities show no excess.
By fitting the energy spectra of the soft excess component
with the XSPEC Mekal model, we obtained the temperature
to be $\sim 0.3$~keV, consistent with the previous result
for the Galactic soft emission from NPS \citep{Inoue1980}.
The surface brightness in the four nearby fields differ significantly,
ranging in (5--$10) \times 10^{-10}$ photons~cm$^{-2}$~s$^{-1}$~arcsec$^{-2}$
in 0.5--1~keV band, and we need to estimate the soft X-ray background
component based on the data of the HCG~62 field.

We examined the radial profile at $r < 15'$ of the XMM-Newton data
in 0.5--1~keV band by fitting with a sum of a double $\beta$ model
and a constant intensity representing the soft X-ray background.
In fitting the profile, parameters for the double $\beta$ component
were fixed at the ROSAT values \citep{Zabludoff2000},
and only the level of the constant component was varied as a free parameter.
The resultant surface brightness is $2\times 10^{-11}$
photons~cm$^{-2}$~s$^{-1}$~arcsec$^{-2}$ in 0.5--1~keV band.
In the following analysis, the excess soft component
is assumed to be a thermal emission with the temperature of 0.3~keV
and abundance of 1~solar, with the normalization fixed
at the level of the radial profile fit above.
We do not apply interstellar absorption for this component, 
since its origin is supposed to be a nearby region in our galaxy.
When the normalization of the 0.3~keV component is varied
in the spectral fit, it agrees with the level derived from
the radial profile, within the 90\% confidence limit
for both Chandra and XMM-Newton.
In fact, the estimated additional soft component is much fainter than 
those in the four nearby fields,
even less than the ``blank-sky'' background component.
The ``blank-sky'' background also contains a certain fraction
of the Galactic soft emission, and the HCG~62 field appears
to require only the same level of soft X-ray background.
We have also confirmed that our result did not change significantly
within errors even doubling the intensity of the estimated soft
background component.

As for the additional hard X-ray background component,
we take into account the excess emission detected
with ASCA \citep{Fukazawa2001}.  In the 2--10 keV range,
the observed flux of the $\Gamma=1.5$ power-law component
with ASCA is $1.0\times 10^{-12}$ ergs~cm$^{-2}$~s$^{-1}$
at the ring-like area with inner and outer radii to be $5'$ and $15'$.
We assumed the same power-law spectrum in our analysis, 
with the spatial distribution uniform over the whole HCG~62 field for 
both Chandra and XMM-Newton.

In the following analysis, the soft and the hard components were
added as additional spectral models with fixed parameters.
We however confirmed that the results did not vary significantly
at the 90\% confidence level, even if we did not apply these backgrounds.
In figure~\ref{fig:bg spec}, the soft and the hard X-ray background
components as well as the ``blank sky'' background are compared
with the IGM spectrum for each instrument.

\section{X-ray Image}\label{sec:image}

\subsection{Chandra image}\label{subsec:image-acis}

\begin{figure*}[tbg]
\begin{center}
    \FigureFile(0.5\textwidth,0.5\textwidth){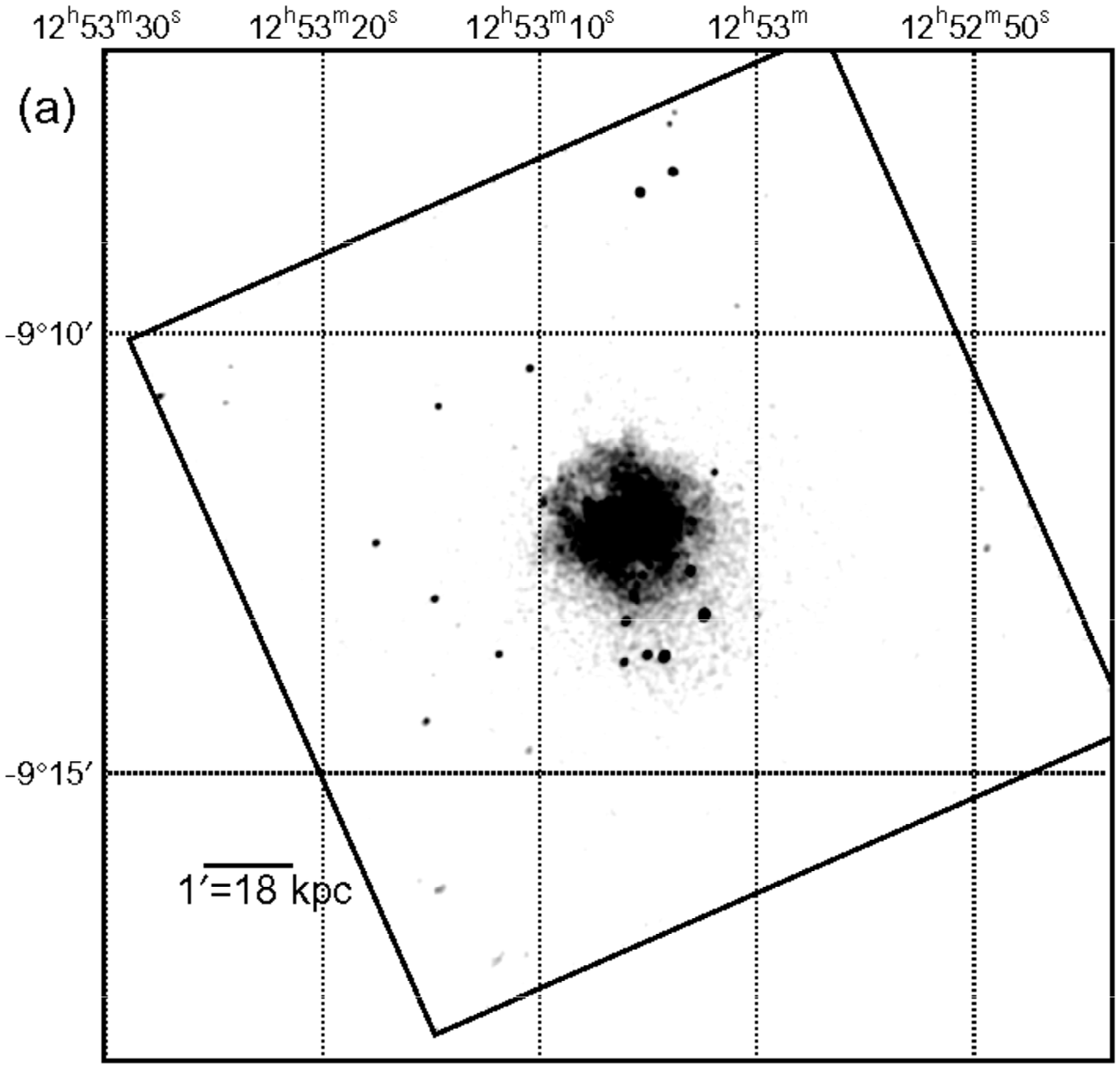}\hfill
    \FigureFile(0.5\textwidth,0.5\textwidth){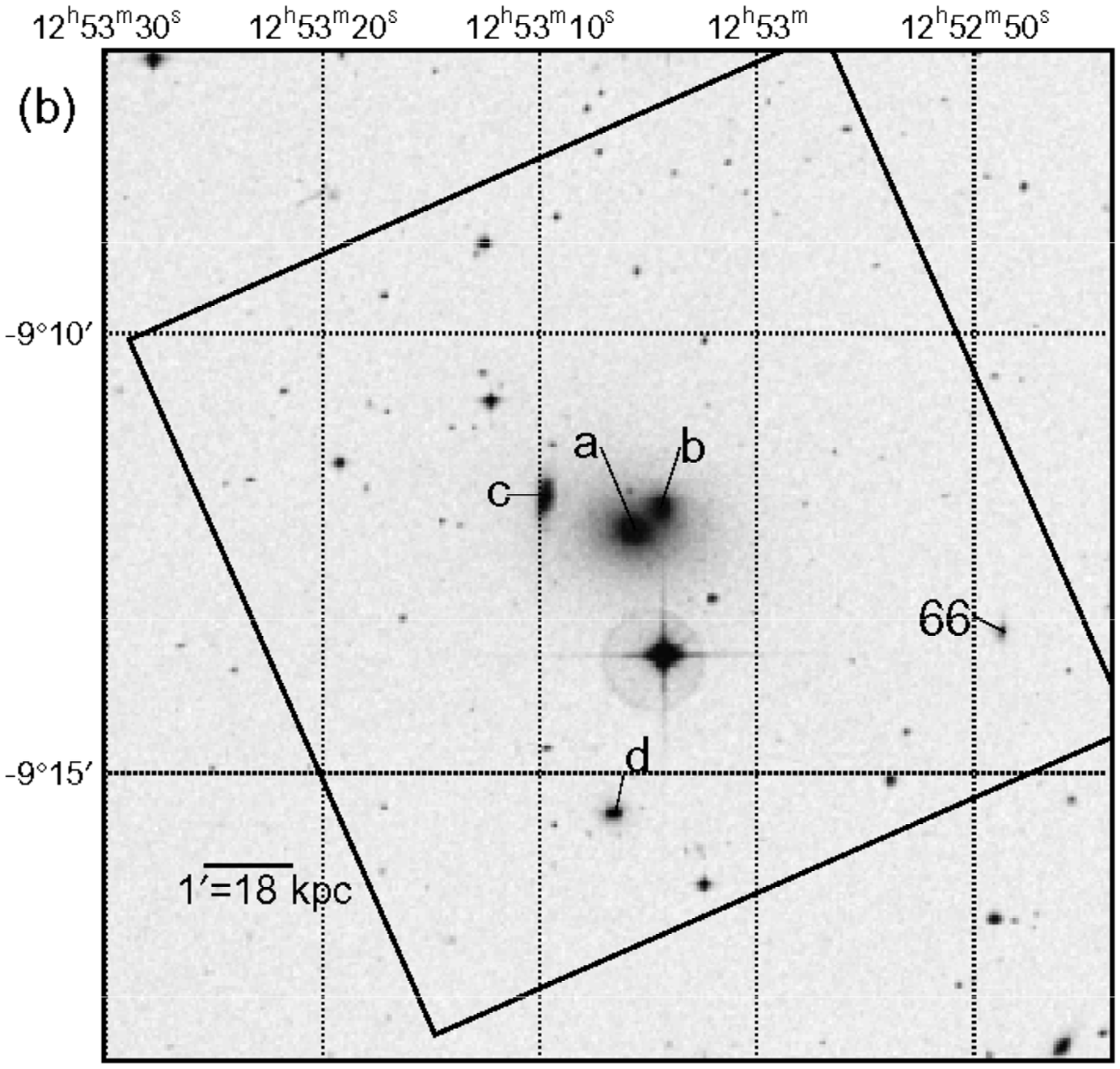}
\end{center}
\vspace*{-4ex}
\caption{
(a) Gaussian smoothed Chandra ACIS-S3 X-ray image 
in the 0.5--4 keV band. 
The smoothing scale is $\sigma=1.5''$  and the image is
corrected for exposure and background. 
(b) STScI Digitized Sky Survey (DSS) image of HCG 62.
The ACIS-S3 field is indicated by a square, and 
the member galaxies are denoted by a, b, c, d, and 66.
Coordinates of images are J2000.
}\label{fig:image-acis}
\end{figure*}

\begin{table*}
\begin{center}
\caption{Optical properties of HCG~62 and member galaxies.}
\label{tab1}
\begin{tabular}{lllcccc}\hline\hline
\makebox[5.8cm][l]{Object} & \multicolumn{2}{c}{Optical coords. (J2000)
\makebox[0cm][l]{\footnotemark[*]}} & 
$z$ \makebox[0cm][l]{\footnotemark[*]} &
Diameter \makebox[0cm][l]{\footnotemark[$\dagger$]} & 
$B$ \makebox[0cm][l]{\footnotemark[$\dagger$]}  &
Type \makebox[0cm][l]{\footnotemark[$\ddagger$]} \\ 
& \multicolumn{1}{c}{R.A.} & \multicolumn{1}{c}{Dec.} &  &  (arcsec) & (mag) & \\
\hline
HCG~62    $\dotfill$ & \timeform{12h53m06.1s} & \timeform{-09D12'16.3''} & 
0.0145 & --- & --- & Group   \\
HCG~62a   $\dotfill$ & \timeform{12h53m05.6s} & \timeform{-09D12'13''} & 
0.0143 & 59.9 & 13.79 & \makebox[1.5em][r]{E3} / \makebox[1.5em][l]{S0} \\
HCG~62b   $\dotfill$ & \timeform{12h53m04.4s} & \timeform{-09D11'59''} & 
0.0119 & 42.9 & 14.21 & \makebox[1.5em][r]{S0} / \makebox[1.5em][l]{S0} \\
HCG~62c   $\dotfill$ & \timeform{12h53m09.7s} & \timeform{-09D11'51''} & 
0.0148 & 40.2 & 15.00 & \makebox[1.5em][r]{S0} / \makebox[1.5em][l]{E} \\
HCG~62d   $\dotfill$ & \timeform{12h53m06.6s} & \timeform{-09D15'26''} & 
0.0136 & 19.7 & 16.30 & \makebox[1.5em][r]{E2} / \makebox[1.5em][l]{---} \\
HCG~62-66 $^\S$ $\dotfill$ & \timeform{12h52m48.7s} & \timeform{-09D13'22''} & 
0.0161 & --- & --- & \makebox[1.5em][r]{---} / \makebox[1.5em][l]{---} \\
\hline\\[-1ex]
\multicolumn{7}{@{}l@{}}{\hbox to 0pt{\parbox{180mm}{\footnotesize
\par\noindent\footnotemark[*]
Optical coordinates and redshift $z$ by
\citet{Mulchaey2003} for HCG~62 and by
\citet{Zabludoff2000} for member galaxies.
\par\noindent\footnotemark[$\dagger$]
$B$ band effective diameter, $\scriptstyle D_B=\sqrt{A_B/\pi}$, and magnitude
within $\mu_B=24.5$ mag~arcsec$^{-2}$ isophote by \citet{Hickson1989}.
\par\noindent\footnotemark[$\ddagger$]
Hubble morphological type classification
by \citet{Hickson1989} or \citet{Shimada2000}.
\par\noindent\footnotemark[$\S$]
Because HCG~62-66 is not catalogued by \citet{Hickson1989},
Diameter, $B$, and Type are left blank.
}\hss}}
\end{tabular}
\end{center}
\end{table*}

\begin{table*}
\begin{center}
\caption{X-ray properties of detected galaxies by Chandra ACIS-S3.}
\label{tab2}
\begin{tabular}{lllcrc}\hline\hline
\makebox[5.8cm][l]{Object} &
\multicolumn{2}{c}{X-ray position (J2000) \makebox[0cm][l]{\footnotemark[*]}} &
\hspace*{-3mm}
Pos.\ diff. \makebox[0cm][l]{\footnotemark[$\dagger$]} &
Obs.\ count \makebox[0cm][l]{\footnotemark[$\ddagger$]} &
Extended \\
& \multicolumn{1}{c}{R.A.} & \multicolumn{1}{c}{Dec.} &
(arcsec) & \multicolumn{1}{c}{(cts)} & Y/N \\
\hline
HCG~62a   $\dotfill$ & \timeform{12h53m05.63} & \timeform{-09D12'13.7''} & 0.8
& $4158\pm \makebox[1em][r]{65}$\hspace*{2mm} & Y \\
HCG~62b   $\dotfill$ & \timeform{12h53m04.43} & \timeform{-09D11'59.4''} & 0.6
& $1344\pm \makebox[1em][r]{37}$\hspace*{2mm} & Y \\
HCG~62c   $\dotfill$ & \timeform{12h53m09.76s} & \timeform{-09D11'55.7''} & 4.8
& $ 331\pm \makebox[1em][r]{19}$\hspace*{2mm} & Y ? \\
HCG~62d \makebox[0cm][l]{\footnotemark[$\S$]} $\dotfill$ & 
\multicolumn{1}{c}{---} & \multicolumn{1}{c}{---} & ---
& $  36\pm \makebox[1em][r]{ 7}$\hspace*{2mm} & Y ? \\
HCG~62-66 $\dotfill$ & \timeform{12h52m48.57s} & \timeform{-09D13'25.7''} & 4.2
& $  30\pm \makebox[1em][r]{ 7}$\hspace*{2mm} & Y ? \\
\hline\\[-1ex]
\multicolumn{6}{@{}l@{}}{\hbox to 0pt{\parbox{180mm}{\footnotesize
\par\noindent\footnotemark[*]
Detected position by the CIAO ``wavdetect'' task,
with 90\% confidence position accuracy of 0.6$''$ radius.
\par\noindent\footnotemark[$\dagger$]
Positional difference between X-ray and optical.
\par\noindent\footnotemark[$\ddagger$]
Observed count within a radius of 10$''$ in
0.5--4~keV including IGM emission but subtracted by the ``blank-sky''.
\par\noindent\footnotemark[$\S$]
HCG~62d was not detected by the ``wavdetect'' task,
so that the optical position was utilized.
}}}
\end{tabular}
\end{center}
\end{table*}

The Chandra image of HCG~62 taken with ACIS-S3 in the 0.5--4 keV
energy band is shown in figure~\ref{fig:image-acis}~(a).
We corrected for background and exposure, and applied Gaussian smoothing.  
The positional dependence of the telescope and the detector responses 
were corrected with an exposure map.  
The IGM emission is clearly observed, and it is extended around the 
central galaxy HCG~62a.  
The brightest region has a radius of about $1'$ from HCG~62a.

We searched for discrete X-ray sources in the ACIS-S3 field using the
CIAO ``wavdetect'' task. Choosing a significance parameter of
$10^{-6}$ for images in the energy bands 0.5--10~keV, 0.5--2.0~keV,
and 2.0--10.0~keV, we detected 50 sources in total,
including HCG~62a, 62b, 62c, and 62-66.
We have detected all the point sources previously identified
in the SEXSI catalog \citep{Harrison2003} using the same Chandra data.
Optical properties of the member galaxies within
the Chandra ACIS-S3 field of view are summarized in
table~\ref{tab1}, and the optical image is shown in
figure~\ref{fig:image-acis}~(b).
The X-ray detected positions and the ACIS-S3 count
are summarized in table~\ref{tab2}.
The X-ray position of HCG~62a and HCG~62b well agree with the optical
coordinate given in \citet{Zabludoff2000} within $< 1''$.
However, HCG~62c and HCG~62-66 indicate different positions
by about $5''$ and $4''$. 
This is due probably to the extent of the galaxies
and/or to the gas stripping,
because the probability of miss-identification of background or foreground
object within 5$''$ radius is calculated to be only 1.5\%.

The X-ray emission of HCG~62a is dominated by the
extended gas, for which we will describe the properties later.
The HCG~62b galaxy was also found to be significantly more extended
than the point spread function of Chandra.
We have fitted the radial brightness profile, $S(r)$, of HCG~62b
with a single $\beta$ model,
\begin{eqnarray}
S(r)=S_0 \left[ 1 + (r/R_{\rm c})^2 \right]^{-3\beta+1/2},
\end{eqnarray}
and obtained a core radius, $R_{\rm c} = 1.5^{+1.1}_{-0.6}$~arcsec,
$\beta = 0.5^{+0.2}_{-0.1}$,
and the normalization, $S_0 = 0.8^{+0.5}_{-0.3}\times 10^{-6}$
photons~cm$^{-2}$~s$^{-1}$~arcsec$^{-2}$.
There is also an indication for the HCG~62c galaxy
to be extended in the Chandra image
in the direction of the optical major axis (north -- south),
although it is not clear whether it is due to the extended IGM emission
because of the poor statistics of HCG~62c.
The HCG~62d galaxy was not detected by the ``wavdetect'' task,
however, the ``blank-sky'' subtracted count exhibited a clear excess
when integrated within 10$''$ around the optical position
as shown in table~\ref{tab2}. It is supposed that the extended
feature of HCG~62d may have hampered the detection by ``wavdetect''.

In the following analysis, the detected sources were excluded
except for HCG~62a and HCG~62d.
Since HCG~62a and HCG~62b are closely located, we need to separate
the HCG~62b component in order to examine the IGM structure.
It is notable that the redshift of HCG~62b is smaller than
the group redshift of $z=0.0145$ by 0.0026 ($\sim 11$~Mpc),
hence HCG~62b is probably not interacting with the IGM around the group core.
We produced two radial profiles in 0.5--4~keV band centered
on HCG~62a and HCG~62b, and fitted them with a double $\beta$
(HCG~62a) and a single $\beta$ (HCG~62b) models, respectively.
With these fits, we evaluated the relative intensity of HCG~62b compared 
with the IGM emission as a function of the distance from HCG~62b.
The contamination from HCG~62b becomes less than 10\% of the IGM intensity
when one goes to outside of a radius of $10''$.  
Thus, we extracted a circle with $10''$ radius around HCG~62b and 
carried out the analysis for the IGM emission.

The central HCG~62 image is shown in figure~\ref{fig:image-acis-xmm}~(a).
This image indicates surface brightness depressions in the northeast
and southwest directions at 20$''$--30$''$ from HCG~62a, the so-called
cavities as reported by
\authorcite{Vrtilek2001} (\yearcite{Vrtilek2001},\yearcite{Vrtilek2002}).
We look into the properties of these cavities in \S\,\ref{sec:cavity}.

\subsection{XMM-Newton image}

\begin{figure*}[tbg]
\begin{minipage}[t]{0.5\textwidth}
    \FigureFile(\textwidth,\textwidth){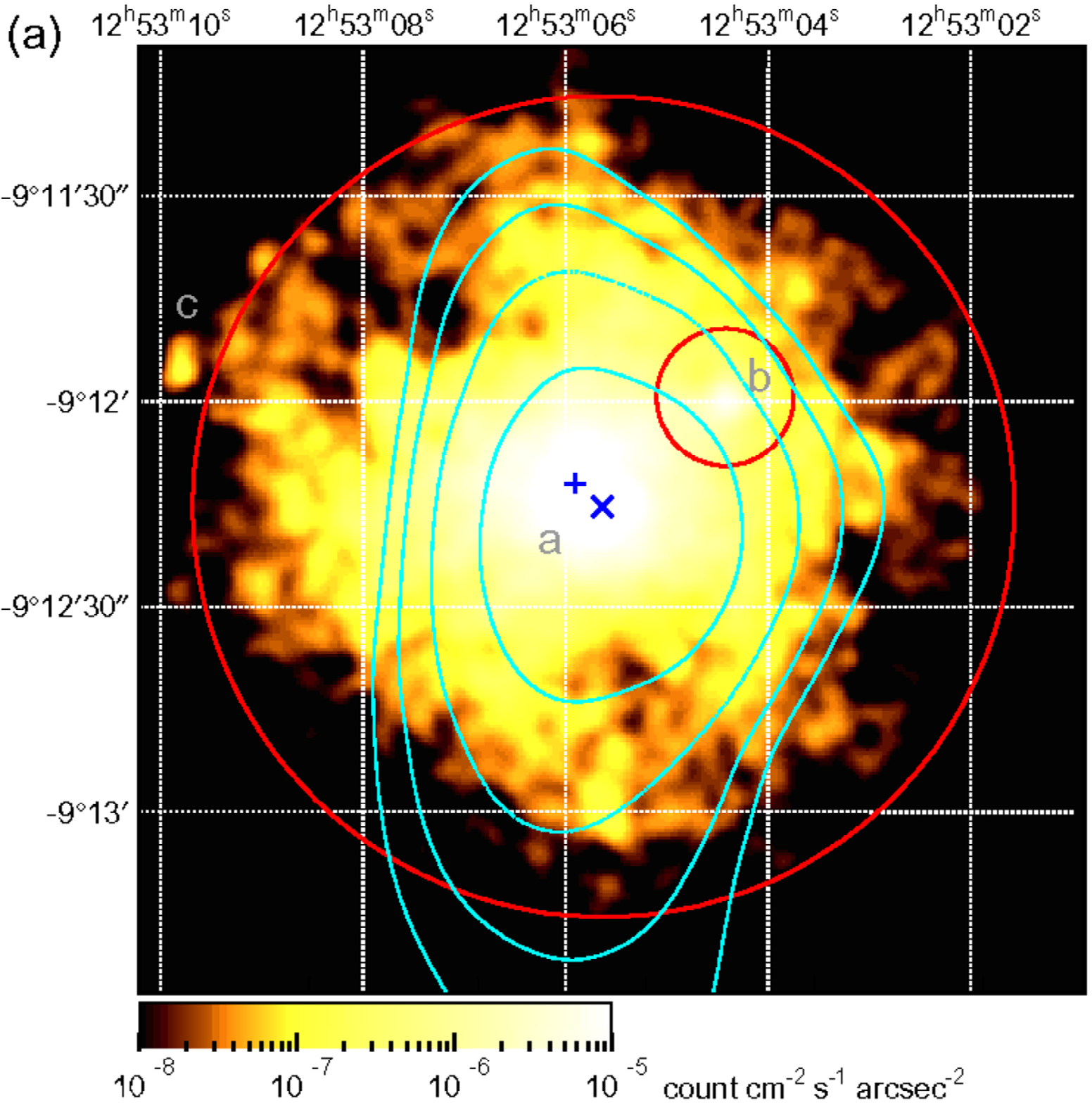}
\end{minipage}%
\begin{minipage}[t]{0.5\textwidth}
    \FigureFile(\textwidth,\textwidth){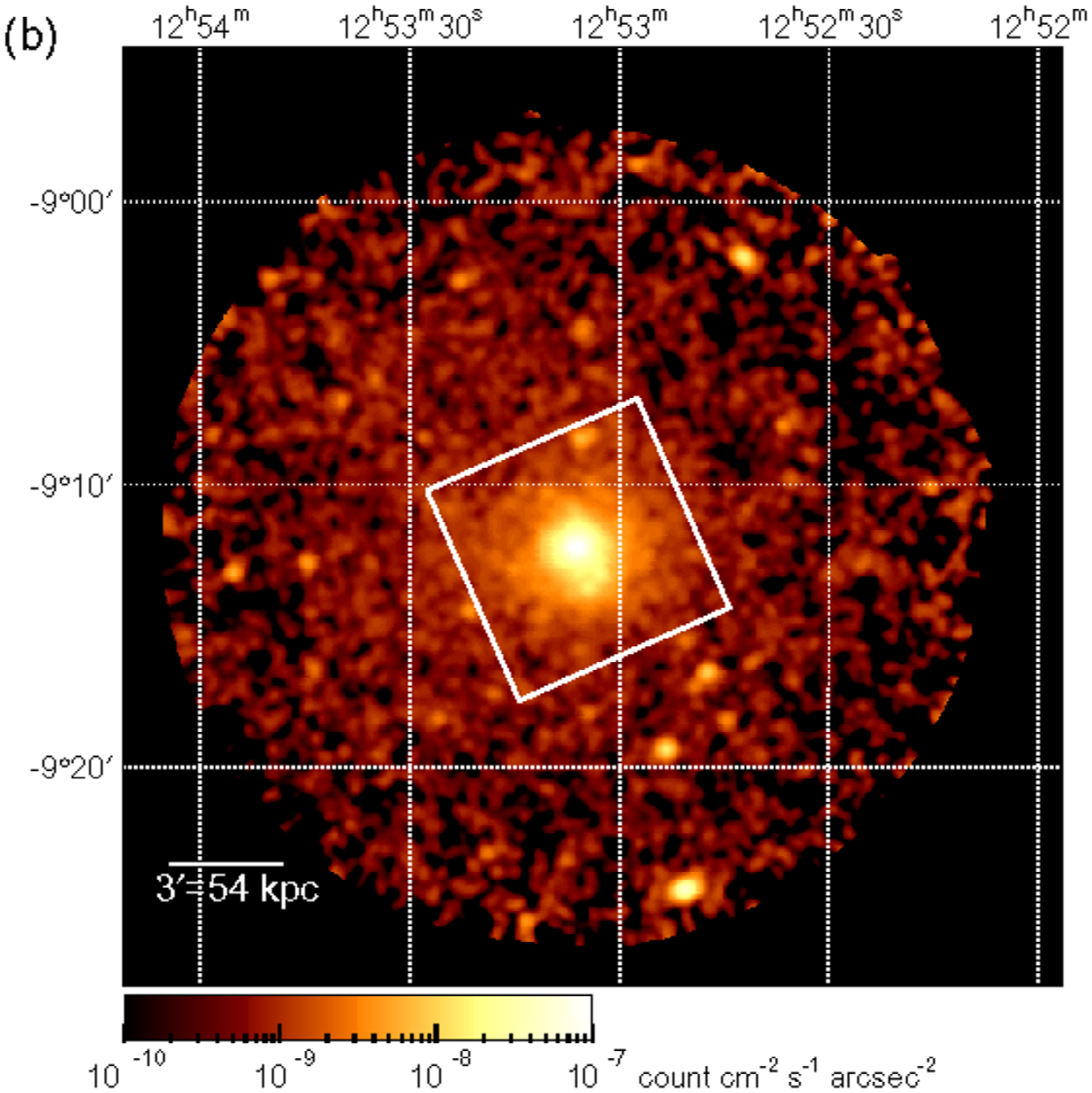}
\end{minipage}
\caption{
(a) Same as figure~\ref{fig:image-acis}, but the central region
of HCG~62 is expanded. The detected point sources other than
HCG~62a, 62b and 62c are excluded.
The central positions for the 2-dimensional 2-$\beta$
fit are denoted by cross and plus corresponding to
the narrower and the wider $\beta$ components.
The red circles indicate a radius of
$1'$ from HCG~62a and $10''$ from HCG~62b, respectively.
The overlaid contour represent a radio intensity map
at 1.4~GHz with 45$''$ FWHM resolution,
derived from the NRAO VLA Sky Survey (NVSS; \cite{Condon1998}).
(b) Combined image with MOS1 and MOS2 detectors in the 0.5--4~keV
band (J2000).  The image is Gaussian smoothed with $\sigma=10''$
and corrected for exposure and background.
A white square represents the ACIS-S3 field.
}\label{fig:image-acis-xmm}
\end{figure*}

A combined X-ray image taken with MOS1 and MOS2 detectors of XMM-Newton
in the 0.5--4~keV energy band is shown in figure~\ref{fig:image-acis-xmm}~(b).
We corrected for background and exposure,
and performed Gaussian smoothing to the image.
The IGM emission is very extended beyond the boundary of the
Chandra ACIS-S3 chip shown by a white square.
The cavities are not clearly seen because of the poorer spatial resolution.
Many point sources were detected in the outer
region by the SAS ``edetect'' task, but none of them outside of
$r = 100''$ were member galaxies of this compact group.
Point sources inside of $100''$ from HCG~62a were
excluded using the Chandra data (\S\,\ref{subsec:image-acis}),
and the XMM-Newton data were used to mask out the sources
outside of $100''$.

\subsection{Surface brightness profile}\label{subsec:radial}

\begin{table*}[tbg]
\caption{
Best fit parameters of the radial surface brightness profiles 
with ACIS-S3 ($r<4'$) and MOS1 ($1'<r<14'$) in 0.5--4~keV
by the 3-$\beta$ model.
First and second components were constrained to have common $\beta$.
The parameters of $\beta$ and $R_{\rm c}$ of the third component
were fixed at the ROSAT result.
}\label{table:radial}
\centerline{
\begin{tabular}{llll}\hline\hline
\hspace*{6.8cm} & 1 & 2 & 3 \\ \hline
$S_0$ ($10^{-6}$ photons~cm$^{-2}$~s$^{-1}$~arcsec$^{-2}$) $\dotfill$ &
0.83$^{+0.10}_{-0.08}$ & 0.29$\pm$0.03 & 0.0018$\pm$0.0001 \\
$\beta$ $\dotfill$ &
0.65$\pm$0.02 & $\leftarrow$ (fixed) & 0.63 (fixed) \\
$R_{\rm c}$ (arcmin/kpc) $\dotfill$ &
$0.10^{+0.02}_{-0.01}/1.7^{+0.3}_{-0.2}$ & $0.48^{+0.04}_{-0.03}/8.5\pm 0.6$ & 
$9.00/159.08$ (fixed) \\[0.5ex]
$\chi^2/$dof $\dotfill$ &
\multicolumn{3}{c}{755.39/570} \\[0.5ex]
\hline
\end{tabular}
}
\end{table*}

\begin{table*}[tbg]
\caption{
Best fit parameters by the 2-dimensional 2-$\beta$ model
for the central region ($r < 1'$) of HCG~62
with ACIS-S3 in 0.5--4 keV.
}\label{table:2Dradial}
\centerline{
\begin{tabular}{llll}
\hline\hline
\hspace*{9cm} & 1 (narrower) && 2 (wider) \\
\hline
$S_0$ ($10^{-6}$ photons~cm$^{-2}$~s$^{-1}$~arcsec$^{-2}$) $\dotfill$ &
0.79$^{+0.21}_{-0.16}$ && 0.63$\pm$0.06 \\
$\beta$ $\dotfill$ &
0.87$^{+0.86}_{-0.29}$ && 0.44$\pm$0.01 \\
$R_{\rm c}$ (arcmin/kpc) $\dotfill$ &
$0.07^{+0.05}_{-0.03}/1.3^{+0.9}_{-0.5}$ &&
$0.18^{+0.02}_{-0.01}/3.1\pm 0.3$ \\
R.A. (J2000) $\dotfill$ &
\timeform{12h53m05.63s} \hfill $\pm$0.4$^{\prime \prime}$ &&
\timeform{12h53m05.90s} \hfill $\pm$0.4$^{\prime \prime}$ \\
Dec. (J2000) $\dotfill$ &
\timeform{-09D12'15.5''} \hfill $\pm$0.4$^{\prime \prime}$ &&
\timeform{-09D12'11.6''} \hfill $\pm$0.3$^{\prime \prime}$ \\[0.5ex]
$\chi^2/$dof $\dotfill$ &
\multicolumn{3}{c}{132.07/122} \\[0.5ex]
\hline
\end{tabular}
}
\end{table*}

\begin{figure}[tbg]
\centerline{
\FigureFile(0.5\textwidth,0.5\textwidth){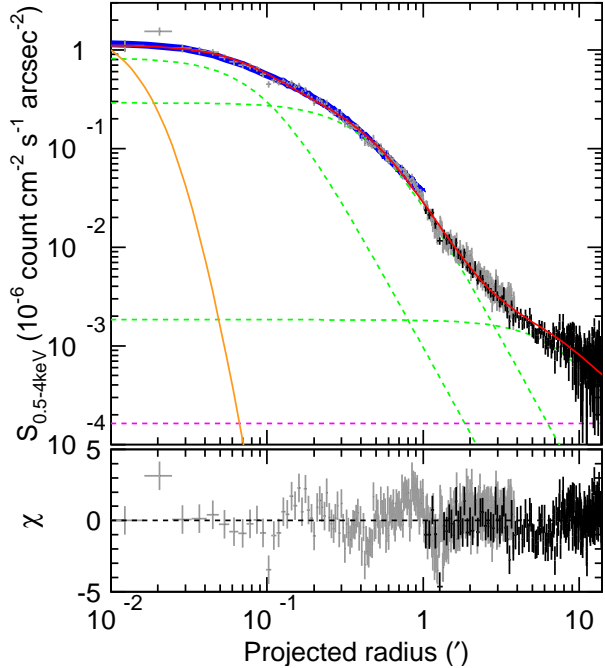}
}
\caption{
In the upper panel, radial profiles of the surface
brightness of HCG~62 in 0.5--4 keV around the central galaxy
HCG~62a are plotted for ACIS-S3 ($r<4'$) and MOS1 ($1'<r<14'$),
as indicated by grey and black lines, respectively.  
The orange solid line represents the point spread function of Chandra.
The best-fit 3-$\beta$ model is shown with red solid
line, and the three components are indicated with green dashed lines. 
For the outermost $\beta$ component,
parameters of $\beta$ and $R_{\rm c}$ are fixed to the ROSAT result
by \citet{Mulchaey_Zabludoff1998}.
Intensity for the sum of the soft Galactic and the hard emission is
indicated by horizontal magenta dashed line.  The blue thick line
represents the projected best-fit 2-dimensional double $\beta$ model
obtained with the Chandra image for the central region ($r<1'$).
In the bottom panel, the residuals of the fit are shown
in unit of $\sigma$.
}\label{fig:radial}
\end{figure}

We plot the radial surface brightness profile for the ACIS-S3 ($r<4'$)
and MOS1 ($1'<r<14'$) around HCG~62a in the energy range of
0.5--4~keV in figure~\ref{fig:radial}.
In this analysis, only the MOS1 data were used for the XMM-Newton
observation because of its low background and good spatial resolution.
HCG~62b was excluded with a circle of $10''$ radius.
Background subtraction was carried out separately for each instrument.
\citet{Mulchaey_Zabludoff1998} report an acceptable fit
with a double $\beta$ model for the ROSAT data.
As the first attempt, we fitted the radial profile with the
same ROSAT model, however, a large discrepancy was found in the central
region around $r\lesssim 1'$.
We then varied the parameters of the inner $\beta$ component.
The soft background and the hard component are included as fixed
constants in the fit.  The fit was still unacceptable because of a
large discrepancy in the central region.  Then, we fitted with
a 3-$\beta$ model by adding a narrow component with its $\beta$
parameter constrained to have the same value with the middle component,
and the fit was much improved.
The outermost component was fixed to the ROSAT value,
because the IGM emission of HCG~62 is much extended over the XMM-Newton 
field of view, and we could not constrain the parameters.

The fit results are summarized in table~\ref{table:radial}.
The derived core radius, $R_{\rm c}=0.\!\!'48^{+0.04}_{-0.03}$,
for the middle component are consistent with the value,
$R_{\rm c} = 0.\!\!'56^{+0.16}_{-0.16}$,
by \citet{Mulchaey_Zabludoff1998} with ROSAT PSPC,
while their $\beta=0.79^{+0.10}_{-0.09}$ is steeper than ours,
$\beta=0.65\pm 0.02$. 
This is due certainly to the narrow $\beta$ model component we have 
introduced, which is required because of the superior angular
resolution of Chandra ACIS-S3 than ROSAT PSPC.
The obtained $\chi^2$ value is still large,
however addition of yet another (4th) $\beta$ component does not
significantly improve the fit, and the derived $\beta$ values for the
inner two components are close to 3.0 which is obviously too steep.
Therefore, we conclude that it is difficult to improve the fit
better than the 3-$\beta$ model case described above, 
due mainly to the complicated spatial structure in the central region.

We next carried out a 2-dimensional fitting for the Chandra X-ray image.
The ACIS-S3 image shown in figure~\ref{fig:image-acis-xmm}~(a) suggests that
the brightest position may be slightly offset from the center of the 
extended group gas.  
To examine this, we fitted the 2-dimensional image in 0.5--4~keV 
within $r < 1'$ around HCG~62a with a 2-$\beta$ model (narrower and wider),
whose centers were varied as free parameters.
Due to the limited photon count in the image bin,
we chose the maximum likelihood method assuming
the Poisson statistics in the fitting.
The ``blank-sky'' background were included in the model as a constant 
surface brightness.
The data and the best fit model are projected around the common center 
(HCG~62a) as shown in figure~\ref{fig:radial}.  
The model (blue thick line) well describes the radial profile within $r\le 1'$,
and gives an acceptable fit with $\chi^2/\rm dof=132.07/122$.
The fit parameters are summarized in table~\ref{table:2Dradial}.  
We find that the centers of
the narrower and wider components are different by $5.44''$ (1.6~kpc).
The center of the narrower component is closer to the optical center
($2''$ offset) than the wider component (5$''$ offset).  

\section{X-ray Cavity}\label{sec:cavity}

In this section, we look into the X-ray structure and the spectral
characteristics of the cavities which are recognized in the Chandra image
in figure~\ref{fig:image-acis-xmm}~(a).  We broadly consider two
possibilities for the origin of the cavities as follows;
(1) depression of the X-ray flux is resulted by absorption
due to some intervening material, and
(2) X-ray emitted gas is deficient, possibly being
expelled by some process. We will examine the case (1) in
\S\,\ref{subsec:cavity spec} and (2) in \S\,\ref{subsec:hollow},
respectively.

\subsection{Flux depression \& temperature map}
\label{subsec:cavity image}

\begin{figure*}[bt]
\centerline{
\FigureFile(0.5\textwidth,0.5\textwidth){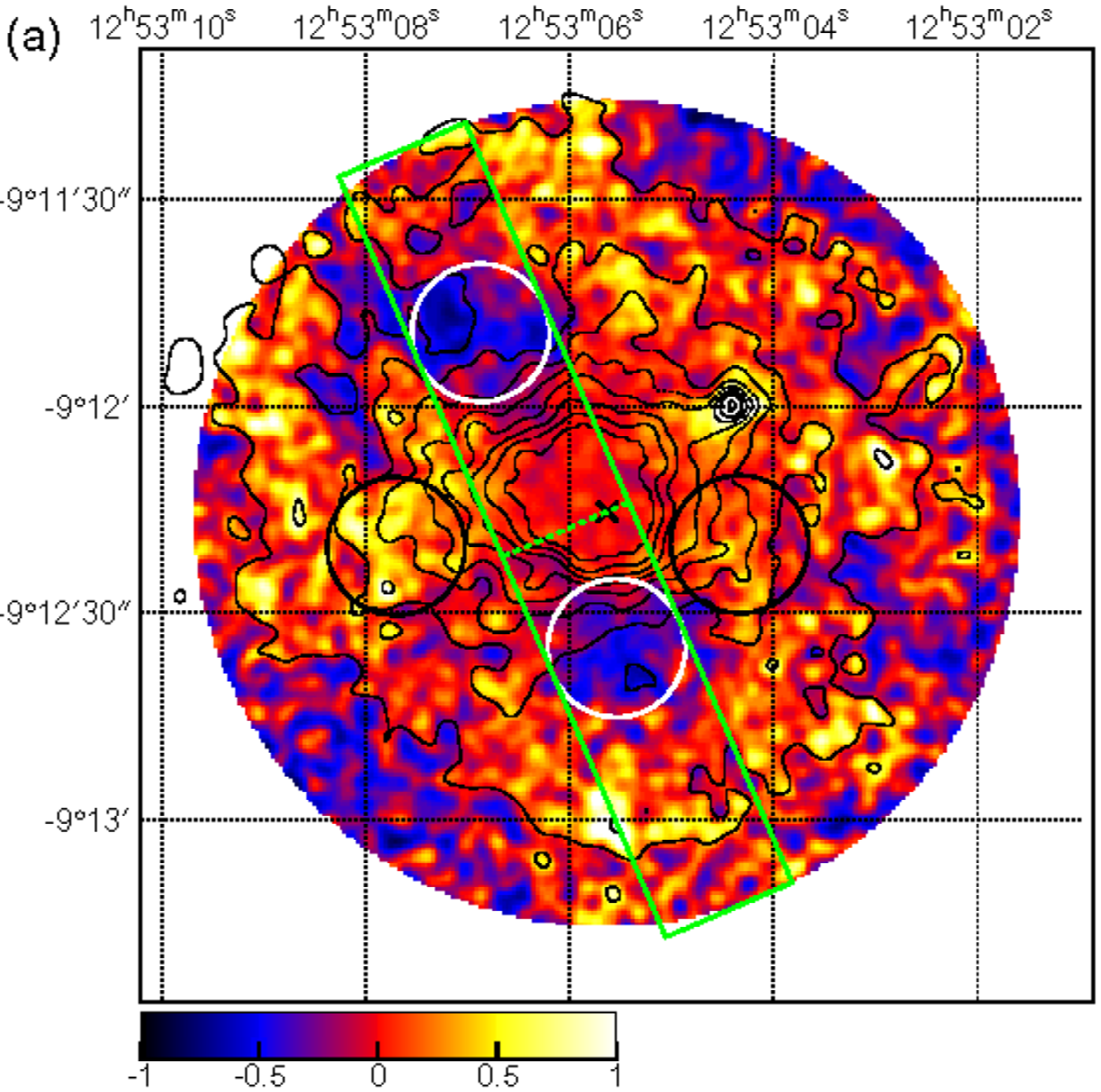}\hfill
\FigureFile(0.5\textwidth,0.5\textwidth){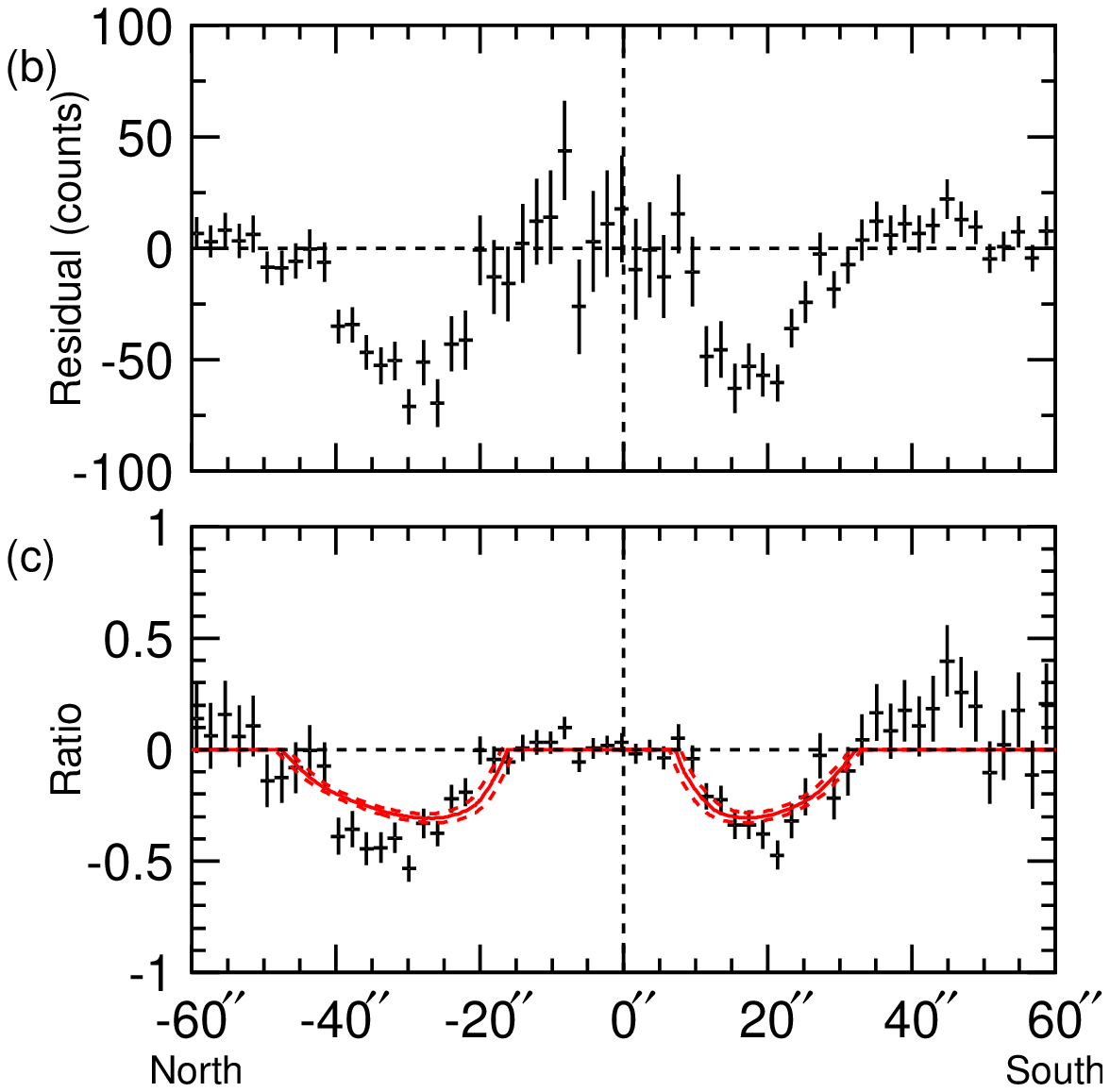}
}
\caption{
(a) Color-coded map of the relative deviation of the flux
from the best-fit 2-$\beta$ model based on
the 2-dimensional fit (table~\ref{table:2Dradial})
in 0.5--4~keV within a radius of $1'$ around HCG~62a with ACIS-S3 (J2000).  
X-ray contours shown by think black lines are the same in
figure~\ref{fig:image-acis-xmm}~(a).  White and black circles with a
radius of $10''$ represent regions where energy spectra are studied.
White ones are the cavities, and black ones are non-cavity regions
examined for comparison. The distances from HCG~62a to the centers of
the east and west light-blue circles are $20''$ and $30''$,
the same for the north and south cavities, respectively.
(b) Residual of the observed count to the 2-$\beta$ model
along the green rectangular region (20$''\times 120''$) in (a).
(c) Ratio of the residual counts divided by the 2-$\beta$ model.
See text \S\,\ref{subsec:cavity image} for red lines.
}\label{fig:cavity image}
\end{figure*}

\begin{figure}[tbg]
\centerline{
    \FigureFile(0.5\textwidth,0.5\textwidth){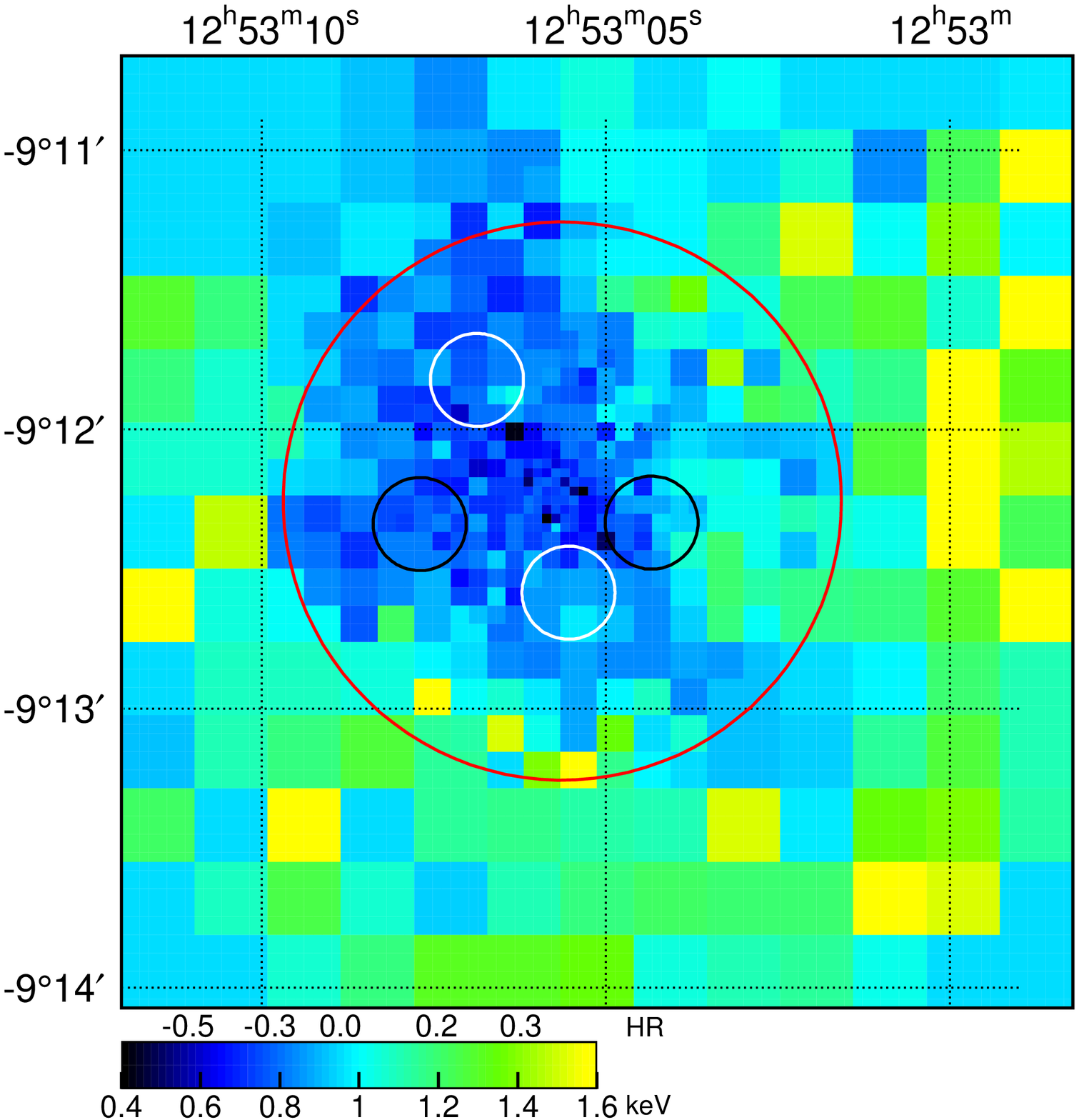}
}
\caption{
Color coded temperature map (J2000) based on the hardness ratio,
$\makebox{\it HR}\equiv (H-S)/(H+S)$, with ACIS-S3,
where $S$ and $H$ correspond to the 0.5--0.95~keV and 0.95--4~keV
counts, respectively.
The red circle has a radius of 1$'$
around HCG~62a, and the small circles (10$''$ in radius)
are the same as those in figure~\ref{fig:cavity image}.
Typical 90\% confidence error on temperature is
$\sim 0.2$~keV above 1~keV and $\sim 0.1$~keV below 1~keV\@.
}\label{fig:hr}
\end{figure}

To quantify surface brightness depression in the cavity regions,
we refer to the result of the 2-$\beta$ model fit of the 2-dimensional
image in \S\,\ref{subsec:radial}.  The relative deviation of the brightness
from the 2-$\beta$ model is color coded and plotted in
figure~\ref{fig:cavity image}.  Both of the cavity regions show a
brightness drop by as much as 50--70\% at the bottom as compared
with the level of surrounding regions.  If we approximate the cavity
shape by a sphere, their radii are $\sim 10''$ (3~kpc)
as indicated by white circles in figure~\ref{fig:cavity image},
with the north one slightly larger. This approximation is similar to
the one performed in B04.  The distances of the cavity
centers from HCG~62a are $\sim 30''$ (9~kpc) for the north and
$\sim 20''$ (6~kpc) for the south one, respectively.
If we take the distances from the central position of
the wider component of the 2-$\beta$ model,
the cavities are almost symmetrical in their distances $\sim 25''$ (7.4~kpc).
The positions and the distances of both cavities are
summarized in table~\ref{table:cavity non-cavity}.

Next, we look into temperature distribution using hardness ratios
({\it HR}\/).  After the point source and background subtractions,
we produced images in two energy bands, 0.5--0.95~keV ($S$) and
0.95--4~keV ($H$), 
which give nearly the same counts in both energy bands.
The {\it HR}\/ is defined as the ratio of the counts between these
bands as $\makebox{\it HR}\equiv (H-S)/(H+S)$.
We divided the region into small cells whose
sizes are determined to contain at least 50 counts in the 0.5--4.0~keV
($H+S$) band, and then the {\it HR}\/ values were calculated. 
The {\it HR}\/ values were then converted to temperature based on the response
matrices at the center of the field of view, assuming an absorption of
$N_{\rm H}=3.0\times 10^{20}$~cm$^{-2}$ and a metal abundance of 0.5
solar. Typical 90\% confidence error on {\it HR}\/ is $\sim 0.17$,
which corresponds to $\sim 0.2$~keV above 1~keV
and $\sim 0.1$~keV below 1~keV\@.
The resultant temperature map is shown in figure~\ref{fig:hr}.
A clear temperature drop down to $\sim 0.7$~keV is seen at the central region,
which is surrounded by a hotter gas with $kT \sim 1.4$~keV
at a radius greater than $1'$ (18~kpc).  
The central cool region shows an irregular shape, 
with an elongation to the directions of two cavities, 
although the temperature structure does not simply correlate with
the cavity regions.

\subsection{Spectral comparison between cavity \& non-cavity regions}
\label{subsec:cavity spec}

\begin{table*}[tbg]
\begin{center}
\caption{
Observed and calculated properties of the north and south cavities
and the east and west non-cavities.
}\label{table:cavity non-cavity}
\begin{tabular}{lllll}
\hline \hline
\hspace*{5.8cm} &
north cavity & east non-cavity &
south cavity & west non-cavity \\
\hline
Distance from HCG~62a $^\ast$ \dotfill &
31.9$''$ & 31.9$''$ &
19.7$''$ & 19.8$''$ \\
Distance from group core $^\dagger$ \dotfill &
26.5$''$ & 29.0$''$ &
24.1$''$ & 24.7$''$ \\
{\it HR} $^\ddagger$ \dotfill &
\makebox[0in][l]{$-0.13\pm 0.07$} & \makebox[0in][l]{$-0.21\pm 0.05$} &
\makebox[0in][l]{$-0.10\pm 0.06$} & \makebox[0in][l]{$-0.12\pm 0.05$} \\
Integration radius $^\S$ \dotfill &
15.7$''$ & $\leftarrow$ &
12.6$''$ & $\leftarrow$ \\
$N_{\rm c}$ or $N_{\rm nc}$ $^\|$ \dotfill &
$2225\pm 48$ & $2857\pm 58$ &
$2476\pm 63$ & $3316\pm 75$ \\
$F_{\rm proj} - F_{\rm sphere}$ or $F_{\rm proj}$ $^\sharp$ \dotfill &
50.63 & 64.93 & 48.31 & 63.46 \\
\hline\\[-1ex]
\multicolumn{5}{l}{\parbox{16cm}{\footnotesize
\par\noindent\footnotemark[$\ast$]
Position of HCG~62a is assumed to be coincide
with the center of the narrower component in table~\ref{table:2Dradial}.
\par\noindent\footnotemark[$\dagger$]
Position of the group core is assumed to be coincide with
the center of the wider component in table~\ref{table:2Dradial}.
\par\noindent\footnotemark[$\ddagger$]
Hardness ratio, $\makebox{\it HR}\equiv (H-S)/(H+S)$, within $r\le 10''$,
where $S$ (or $H$) corresponds to 0.5--0.95 (0.95--4)~keV counts.
\par\noindent\footnotemark[$\S$]
Integration radius is chosen to become
$N_{\rm c}/N_{\rm nc} = (F_{\rm proj} - F_{\rm sphere}) / F_{\rm proj}$
for each cavity and non-cavity pair.
\par\noindent\footnotemark[$\|$]
$N_{\rm c}$ for north and south cavities, and $N_{\rm nc}$ for non-cavities.
Each value is corrected for background and exposure.
\par\noindent\footnotemark[$\sharp$]
$F_{\rm proj} - F_{\rm sphere}$ for north and south cavities,
and $F_{\rm proj}$ for non-cavities  in arbitrary unit.
}}
\end{tabular}
\end{center}
\end{table*}

\begin{table*}[tbg]
\caption{
Results of spectral fits for the north and south cavity regions and
the east and west non-cavity regions within $r\le 10''$ in 0.5--3~keV
with a two temperature vMekal model.
See text \S\,\ref{subsec:cavity spec} for details.
}\label{table:spec cavity}
\centerline{
\begin{tabular}{lcllllc}
\hline \hline
\hspace*{5.5cm} &
$N_{\rm H, excess}$ &
$kT_1$ &
{\it Norm}$_1$\footnotemark[*] & {\it Norm}$_2$\footnotemark[*] &
$F_{0.5-4~\rm keV}$ & $\chi^2/$dof \\
 &
\hspace*{-2mm}($10^{20}$ cm$^{-2}$) &
(keV) & \multicolumn{2}{c}{($10^{-19}~\rm cm^{-5}$)} &
\multicolumn{2}{l}{\hspace*{-2mm}($10^{-14}~\rm ergs~cm^{-2}~s^{-1}$)} \\
\hline
\multicolumn{7}{l}{phabs ($N_{\rm H}=3.0\times 10^{20}~\rm cm^{-2}$ fixed) $\times$ 2-$T$ ($kT_2=1.4~\rm keV$ \& abundance fixed) vMekal model}   \\
\hline
north cavity \dotfill &
---  & $0.74^{+0.04}_{-0.05}$ & $1.3^{+0.3}_{-0.2}$ & $0.7^{+0.4}_{-0.4}$ & $4.0\pm 0.3$ & 18/14 \\
east non-cavity \dotfill &
---  & $0.75^{+0.04}_{-0.05}$ & $2.5^{+0.4}_{-0.4}$ & $0.6^{+0.5}_{-0.5}$ & $6.8\pm 0.4$ & 22/16 \\
south cavity \dotfill &
---  & $0.76^{+0.05}_{-0.06}$ & $1.6^{+0.3}_{-0.3}$ & $1.1^{+0.6}_{-0.4}$ & $5.2\pm 0.3$ & 18/20 \\
west non-cavity \dotfill &
---  & $0.76^{+0.04}_{-0.05}$ & $2.6^{+0.4}_{-0.4}$ & $1.3^{+0.7}_{-0.7}$ & $8.0\pm 0.4$ & 39/32 \\
\hline
\multicolumn{7}{l}{phabs ($N_{\rm H}=3.0\times 10^{20}~\rm cm^{-2}$ fixed) $\times$ zphabs $\times$ 2-$T$ ($kT_2=1.4~\rm keV$ \& abundance fixed) vMekal model}      \\
\hline
north cavity \dotfill &  
$\makebox[0in][r]{1}8.3^{+5.6}_{-4.6}$ & $0.70^{+0.06}_{-0.02}$ & $2.5^{+0.6}_{-0.4}$ & $0.7^{+0.3}_{-0.3}$ & $4.2\pm 0.3$ & \raisebox{-1.5ex}{\makebox[0cm][r]{$\big\}\;$} 42/31} \\[-1.5ex]
east non-cavity \dotfill &
$                   2.4^{+3.3}_{-2.4}$ & $\uparrow$             & $\uparrow$          & $\uparrow$          & $6.5^{+0.4}_{-0.3}$ &       \\
south cavity \dotfill &
$\makebox[0in][r]{1}1.7^{+3.6}_{-2.3}$ & $0.75^{+0.03}_{-0.04}$ & $2.5^{+0.4}_{-0.3}$ & $1.2^{+0.4}_{-0.4}$ & $5.5\pm 0.3$ & \raisebox{-1.5ex}{\makebox[0cm][r]{$\big\}\;$} 76/53} \\[-1.5ex]
west non-cavity \dotfill &
$                               < 2.8$ & $\uparrow$             & $\uparrow$          & $\uparrow$          & $7.6\pm 0.4$ &       \\
\hline\\[-1ex]
\multicolumn{7}{l}{\parbox{16cm}{\footnotesize
\footnotemark[*] Normalization for the Mekal model,
$\makebox{\it Norm}=\int n_{\rm e} n_{\rm H} dV/ (4\pi (1+z)^2 D_{\rm A}^{\,2})$, 
where $D_{\rm A}$ is the angular distance to the source.
}}
\end{tabular}
}
\end{table*}

\begin{figure}[tbg]
\centerline{
\FigureFile(0.48\textwidth,0.48\textwidth){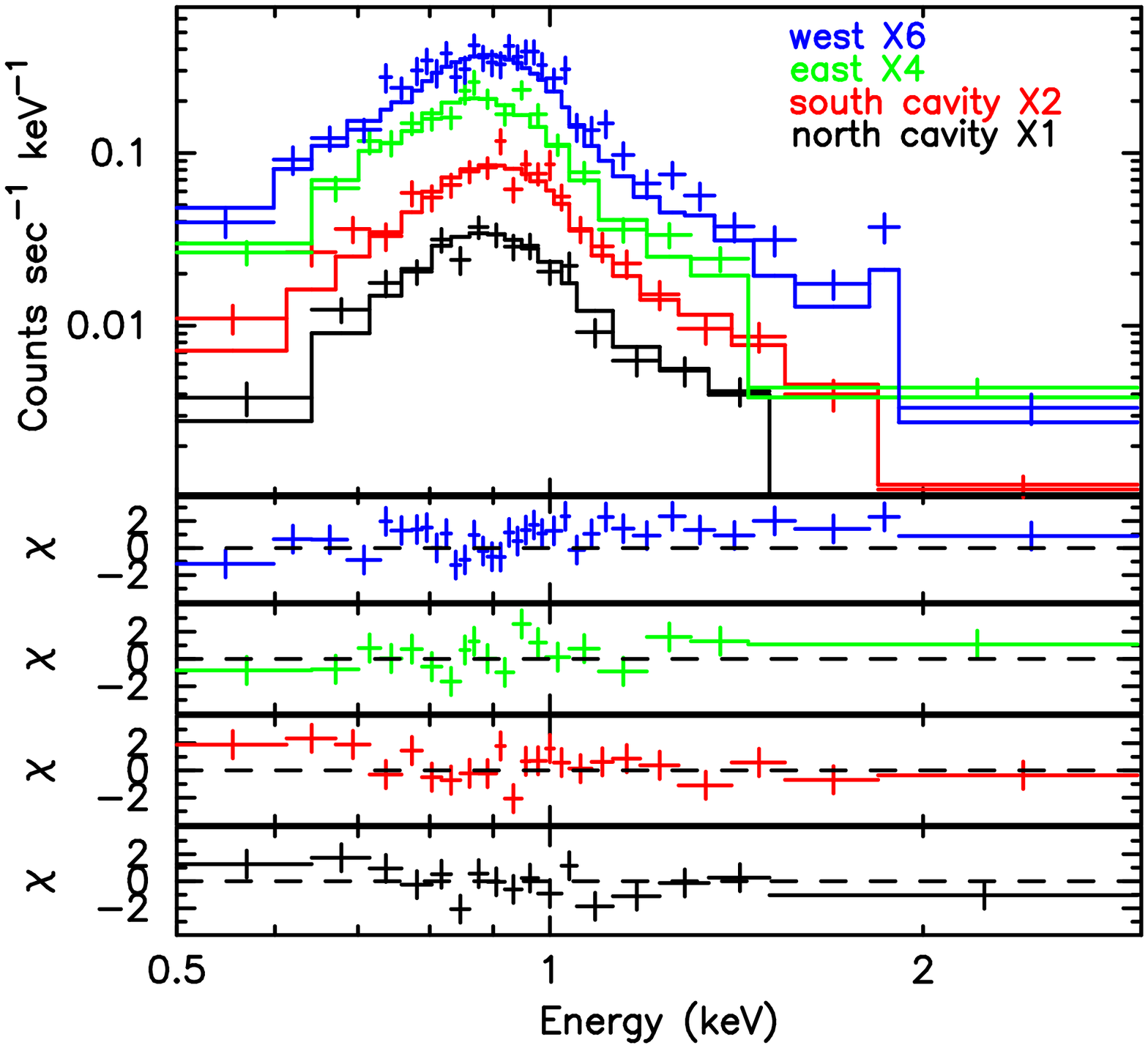}
}
\caption{
Chandra ACIS-S3 X-ray spectra for north (black) and south
(red) cavity regions, and east (green) and west (blue) non-cavity
regions, respectively.  Spectra of
the north cavity and east non-cavity pair and
the south cavity and east non-cavity pair are simultaneously fitted with 
common continuum parameters with individual excess absorption.
The spectra and the best-fit models are scaled by 2, 4, 6
for south, east, and west regions, respectively, for clarity.
The bottom four panels show the residuals of the fit. 
}\label{fig:spec cavity}
\end{figure}

In order to examine the possibility that absorption may be responsible for
the cavities, we look into the energy spectra in this subsection.
We extracted energy spectra for the two cavity regions from
the ACIS-S3 data, using circular regions with the same radius
of $10''$ (3~kpc) as shown in figure~\ref{fig:cavity image}.
To compare with them, we also extracted spectra for two non-cavity
regions with both $10''$ radii in the east and west at the
same distance from HCG~62a to the respective cavities.
Their precise locations are 
(\timeform{12h53m06.87s}, \timeform{-9D11'49.4''}),
(\timeform{12h53m05.54s}, \timeform{-9D12'35.1''}),
(\timeform{12h53m07.75s}, \timeform{-9D12'21.1''}) and 
(\timeform{12h53m04.33s}, \timeform{-9D12'20.1''}) (J2000)
for the north, south cavities and east, west non-cavity, respectively.
The non-cavity regions are indicated by black circles
in figure~\ref{fig:cavity image}, and their properties 
are summarized in table~\ref{table:cavity non-cavity}.
First of all, we calculated the hardness, {\it HR}, as defined
in the previous subsection, in these four regions.
The derived values with 90\% confidence errors were
$-0.130\pm 0.066$ vs.\ $-0.209\pm 0.052$
for the north cavity vs.\ east non-cavity pair,
and $-0.099\pm 0.058$ vs.\ $-0.117\pm 0.046$
for the south cavity vs.\ west non-cavity pair, respectively.
There is no significant difference in the hardness for both pairs.
However, both of the cavity regions suggest slightly larger hardness
than the respective non-cavity regions, which is in the same
sense as expected if absorption is responsible for the X-ray cavities.

To be more quantitative, we conducted a spectral fit for each spectrum.
All the spectra were binned to contain at least 30 counts in a bin.
The spectra were then fitted with a two temperature vMekal model
(2-$T$ model in \S\,\ref{sec:spec}) in the 0.5--3~keV band
with an absorption (phabs in XSPEC) fixed to the Galactic value,
$N_{\rm H} = 3.0\times 10^{20}$~cm$^{-2}$. 
Because of the limited statistics of those spectra, 
temperature of the hot component, $kT_2$, was fixed to 1.4~keV and 
the abundance of each element was fixed to the best-fit values for 
the 1-$T$ model in the annulus range of $0.4<r<0.6'$ 
in table~\ref{table:proj spec}.
See \S\,\ref{sec:spec} for details of the model.
The fit results are summarized in the first four rows of
table~\ref{table:spec cavity}.

All of the four spectra are well fitted by the 2-$T$ model
with acceptable $\chi^2$ values at the 90\% confidence limit.
The obtained temperatures exhibited quite similar values of
$kT_1 = 0.74$--0.76~keV, and we could not find any significant
difference between the cavity and non-cavity spectra except 
for the normalizations of the cool component, {\it Norm}$_1$.
It is interesting that the normalizations of the hot component,
{\it Norm}$_2$, are not different between them.
As described in the next section, the hot component is probably absent
from the central region of the group ($r \lesssim 0.6'$),
so that the hot component seen in these four regions are
mostly due to foreground or background emission.
It is therefore suggested that the cavities lie close to the
group core in our line of sight, and that only the cool component
is pushed away from the cavities.
The signature of hardening of {\it HR} above is
due probably to this effect.
As a remark, the fits are not acceptable 
when fitted with a one temperature vMekal model (1-$T$ model),
although the obtained temperatures are also very similar
among the four spectra.

Here, we evaluate how much $N_{\rm H, excess}$ is needed to reproduce
the observed cavity feature by a simple absorption at the source redshift.
For this purpose, we simultaneously fitted the cavity and
non-cavity spectra of each pair, by setting
all the parameters of the emission spectrum
to be the same including the normalization.
Thus, two new parameters, $N_{\rm H,excess}$ for each pair,
and three new constraint on {\it Norm}$_1$, {\it Norm}$_2$,
and $kT_1$ are added,
so that the degree of freedom (dof) increases by one.
The pulse-height spectra and fit residuals
for the four regions are shown in figure~\ref{fig:spec cavity}.
The $N_{\rm H, excess}$ for the
individual regions are summarized in the bottom four rows of
table~\ref{table:spec cavity}.
The obtained $\chi^2/\rm dof$ were
42/31 for the north cavity and east non-cavity pair, and
76/53 for the south cavity and west non-cavity pair.
Therefore both fits were not acceptable at the 90\% confidence limit,
although improvement of $\chi^2$ is marginal for the former pair.
The derived $N_{\rm H, excess}$ in cavities has to be $\sim 3$ times
larger than those in non-cavity regions for the necessary flux reduction.
The required mass of neutral hydrogen for north and south cavities
amount to $1.0\times 10^9 M_\odot$ and $0.4\times 10^9 M_\odot$,
respectively.  These values are comparable to the observed upper
limit of the H$_{\rm I}$ mass, $\sim 10^9 M_\odot$,
of the whole group including all member galaxies
by \citet{Verdes-Montenegro2001} and \citet{Stevens2004}.
It is therefore difficult to attribute the cavities to
the absorption by neutral gas.

\subsection{Hollow sphere model}\label{subsec:hollow}

Since the absorption model is found to be unlikely in the previous subsection,
we then consider an extreme case that a spherical cavity contains
no X-ray emitting gas at all, namely, the hollow sphere model.
It is assumed that the spherical cavities exist at the same distance
from us to the group core.
Even with this extreme assumption, there should be a certain flux
observed in the projected circular cavity region because of the foreground 
and background group emission.
This will give us a constraint on the physical size of the cavities. 
For an overall X-ray structure, we assume that the
3-dimensional X-ray emissivity obeys the 3-$\beta$ model shown in
table~\ref{table:radial}. 
Based on this model, we can estimate the emission from an arbitrary volume 
in the IGM and the emission from a column integrated along a certain 
line of sight.
Here a single cavity is considered. 
The observed projected counts corrected for background and exposure 
at the circular cavity region is denoted by $N_{\rm c}$, 
and the count in the corresponding non-cavity region by $N_{\rm nc}$, 
respectively. 
The non-cavity region has the same integration radius and distance from 
HCG~62a as the cavity region.
Then, based on the 3-$\beta$ model, we can calculate the flux,
$F_{\rm sphere}$, from the cavity volume which is assumed to be filled with 
hot gas.
The model also gives us the projected flux,
$F_{\rm proj}$, from the line-of-sight column passing
through the non-cavity region.  
These numbers allow us to estimate the counts purely emitted from 
the spherical cavity as,
$N_{\rm c}-N_{\rm nc}~(F_{\rm proj}-F_{\rm sphere}) /F_{\rm proj}$,
which should be zero on the assumption above.

Based on the observed cavity flux, we can derive the upper and lower
limits for the radius of the hollow spheres. Clearly, too small radius
will give too little depression in the X-ray flux at the cavity region,
and vise versa.  We applied the observed intensity and position of the
north and south cavities. The allowed ranges for the radii of the
hollow spheres were computed as $r=15.7\pm 0.9''$ and $r=12.6\pm 0.8''$
for the north and south cavities, respectively.
These sizes are almost consistent with the projected image as seen
in figure~\ref{fig:cavity image}~(a).
The observed and the calculated values of $N_{\rm c}$, $N_{\rm nc}$,
$F_{\rm proj}-F_{\rm sphere}$, and $F_{\rm proj}$ are summarized
in table~\ref{table:cavity non-cavity}.
The expected profile of the count rate deficiency with this assumption
is overlaid in figure~\ref{fig:cavity image}~(c) as a solid red line.
The dashed red lines correspond to the radii at the $\pm 90$\% 
confidence errors.
As seen in this plot, the spherical assumption of cavities gives
an asymmetric shape with deeper deficiency at the near side and
shallower at the far side, when projected to the 2-dimensional image.
Because the observed shapes of the cavities are almost symmetrical,
it is suggested that there must be an asymmetry in the shape and/or
the density inside the cavities. 
Namely, the far-side of the cavities should be larger in size and/or weaker 
in the X-ray emissivity.
It is also notable that even with this extreme assumption of hollow spheres 
at the same distance to the group core, the calculated deepness of 
the deficiency is nearly the same level or even slightly shallower than 
the observed ones.
This fact implies that the shape of cavities is probably elongated
in the direction of our line of sight. 
Another possibility is that the group gas itself has a flatter shape 
in the depth,
although it is implausible considering the redshift distribution
of the member galaxies \citep{Zabludoff2000} and that the projected 
X-ray image is quite symmetric.

\section{Radial Profiles of Temperature \& Abundance}\label{sec:spec}

\subsection{Consistency between Chandra \& XMM-Newton}
\label{subsec:consistency}

In order to obtain radial profiles of temperature and abundance of the IGM,
we examined energy spectra for each instrument (ACIS-S3, MOS1, MOS2,
and pn) from several circular annuls from the center of HCG~62a.
The center was taken from the narrower component in our 2-dimensional
fit carried out in \S\,\ref{subsec:radial}.
First of all, we checked the consistency between Chandra
and XMM-Newton using the spectra within $r\le 2'$ around HCG~62a.
Each spectrum was binned to contain at least 30 counts per bin 
to be tested with the $\chi^2$ fit.
The background data were taken from the same region in the
``blank-sky'' data for each instrument.
The hard emission and the soft background components are also
considered in the fitted model (\S\,\ref{subsec:soft bg}).
As for the XMM-Newton data, MOS1 and MOS2 spectra were summed up
and simultaneously fitted with the pn spectrum.
The energy range around the Al-K$_\alpha$ line (1.4867~keV), 
the instrumental background, was ignored for both MOS and pn.

We have fitted the spectra with a single temperature or
two temperature vMekal model including the Galactic photoelectric absorption.
Abundances of C, N, Na, and Al were fixed to be 1 solar.
We grouped several elements and constrained them to have a common abundance.
The first group contains O and Ne, the second group S, Ar and Ca,
and the third group is for Fe and Ni.  Among the other elements,
abundances of Mg and Si were determined separately.
In the case of the two temperature fit, we used the sum of two
vMekal models in which two components were
constrained to have common abundances.
The actual model formula are phabs$\times$(vMekal+powerlaw)+Mekal 
(1-$T$ model) and phabs$\times$(vMekal+vMekal+powerlaw) +Mekal (2-$T$ model).  
Here, phabs represents photoelectric absorption and $N_{\rm H}$ is fixed 
at the Galactic value of $3.0\times 10^{20}~\rm cm^{-2}$.  
The Mekal component is the soft background described 
in \S\,\ref{subsec:soft bg}, with $kT$ and $Z$
both fixed at 0.3~keV and 1~solar, respectively, and with the fixed
normalization.  The power-law represents the hard emission described in
\S\,\ref{subsec:soft bg} with its photon index, $\Gamma=1.5$,
and normalization both fixed.

\begin{figure*}[tbg]
\begin{minipage}[t]{0.48\textwidth}
\FigureFile(\textwidth,\textwidth){figure8a.eps}
\end{minipage}\hfill
\begin{minipage}[t]{0.48\textwidth}
\FigureFile(\textwidth,\textwidth){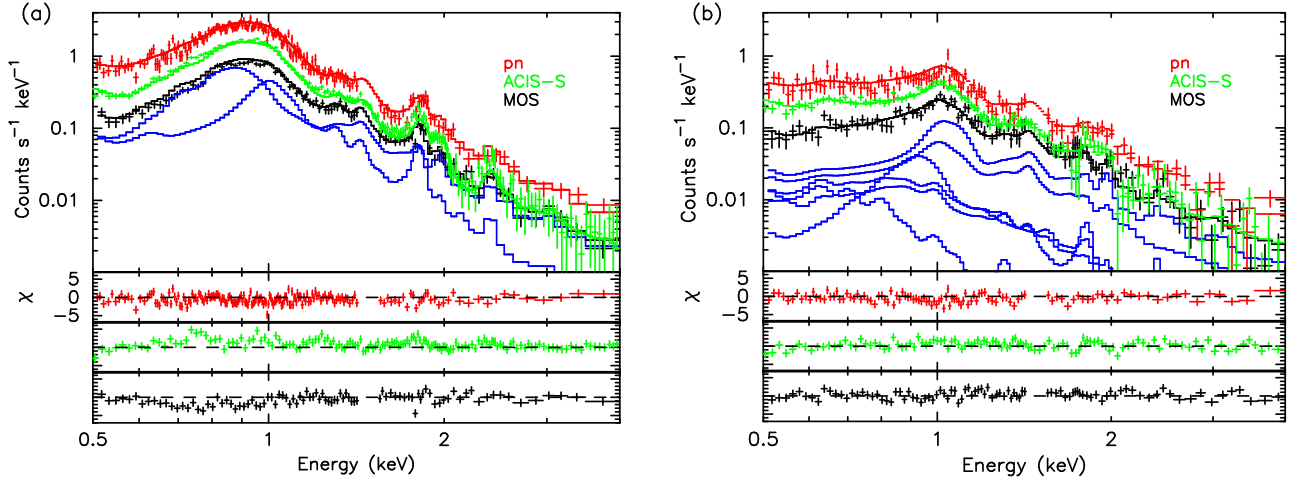}
\end{minipage}
\caption{
(a) Chandra ACIS-S3 (green), MOS1+2 (black), and pn (red) spectra
within $r\le 2'$ around HCG~62a in the 0.5--4~keV band.
The three spectra are simultaneously fitted with 2-$T$ model
without deprojection, and the best-fit models are indicated by solid lines.
Each component of the two temperature vMekal model
for MOS1+2 is indicated by blue lines.
The bottom three panels show the residuals of the fit.
(b) Same as (a), but in the 2--4$'$ annulus
fitted with the deprojected 2-$T$ model.
The outer contributions and fitting models for MOS1+2
are indicated by blue lines.
}\label{fig:spec}
\end{figure*}

The $r\le 2'$ spectra are shown in figure~\ref{fig:spec}~(a),
and the fit results with 2-$T$ model are summarized
in table~\ref{table:proj spec}.
The 1-$T$ model was rejected at high significance with 
$\chi^2/\rm dof = 2162/425$ (combined) for this region.
Three kinds of the fit results are listed in table~\ref{table:proj spec},
Chandra (ACIS-S3) only, XMM-Newton (MOS1+2, pn) only,
and the combined fit of all the instruments.
We have confirmed that most of the results from different
instruments were consistent at the 95\% confidence limit.
The abundances with ACIS-S3 were slightly larger than,
but mostly consistent with those with XMM-Newton,
while only Si gave inconsistent values at the 95\% confidence limit.
This is due probably to the systematic effect by the intrinsic Si line 
as seen in the ``blank-sky'' data for both Chandra and XMM-Newton. 
Since the Si line is most prominent in the IGM spectrum, 
the good statistics result in formally inconsistent abundance values.
In the annular range of 2--4$'$,
spectral fit for individual instruments (ACIS-S3, MOS1+2, and pn)
gave consistent results at the 90\% confidence limit,
due mainly to the poorer statistics.
Based on these results, data from all the instruments were
simultaneously fitted in the following analysis.

\subsection{Mekal vs APEC model \& notes on abundance}

We also summarize results when the spectra are fitted with
$\rm phabs\times (vAPEC+vAPEC+powerlaw)+Mekal$ (2-$T$ vAPEC model)
in table~\ref{table:proj spec}.
The vAPEC model (v1.3.1) gives smaller $\chi^2$ value especially for
the Chandra spectrum, and indicates slightly higher temperatures
by $\Delta T\lesssim 0.05$~keV
than the vMekal model. This is due probably to the better modeling
of the Fe-L line complex in the vAPEC model,
however it is confirmed that all the element abundances are
consistent within 90\% confidence errors between the two models.
We therefore adopt the vMekal model in the following analysis
considering comparison with previous literatures.

Table~\ref{table:abun free} represents a result
when all the element abundances in the vMekal or vAPEC model
are determined separately for the combined fit.
The errors of the Ne, Al, Ar, Ca, and Ni abundances are
significantly larger than those of other elements,
while other parameters agree with the values in 
table~\ref{table:proj spec} within 90\% confidence errors.
Thus abundances in table~\ref{table:proj spec}, which are 
linked together for (O, Ne), (Mg, Al), (S, Ar, Ca), and (Fe, Ni),
are essentially representing O, Mg, S, and Fe abundances, respectively.
There is an indication that the Ni abundance is higher
than the Fe abundance in both vMekal and vAPEC model fits.
This might be explained by the fact that the fitted spectra are
not deprojected so that they contain emission from outer region
in the group which exhibits higher temperature than the group core
as seen in \S\,\ref{subsec:temp-prof}.
On the other hand, it is known that Ni is synthesized
more abundantly than Fe in SN~Ia
when compared in the solar unit,
which will be discussed in \S\,\ref{sec:ICM-prop}.

\begin{table*}[tbg]
\caption{
Results of the projected spectral fit for $r\le 2'$ region
with two temperature vMekal and vAPEC models.
}\label{table:proj spec}
\centerline{
\begin{small}
\begin{tabular}{llllllllllc}
\hline\hline
& $kT_1$ & $kT_2$ & O, Ne & Mg, Al & Si & \makebox[0cm][l]{S, Ar, Ca} & Fe, Ni & 
\makebox[0cm][l]{{\it Norm}$_1$\footnotemark[*]} & \makebox[0cm][l]{{\it Norm}$_2$\footnotemark[*]} & $\chi^2/$dof \\
& (keV) & (keV) & (solar) & (solar) & (solar) & (solar) & (solar) & 
\multicolumn{2}{c}{($10^{-18}~\rm cm^{-5}$)} &           \\
\hline \hline
vMekal \\
\hline
Chandra & 
$0.72^{+0.02}_{-0.02}$ & $1.32^{+0.05}_{-0.06}$ & $0.36^{+0.12}_{-0.13}$ & $1.20^{+0.23}_{-0.19}$ & $1.23^{+0.19}_{-0.14}$ & 
$0.86^{+0.38}_{-0.29}$ & $0.89^{+0.09}_{-0.08}$ & $6.6^{+1.2}_{-1.0}$ & $9.3^{+1.3}_{-1.3}$ & 189/144 \\
XMM &
$0.77^{+0.02}_{-0.05}$ & $1.41^{+0.12}_{-0.11}$ & $0.23^{+0.17}_{-0.13}$ & $1.01^{+0.37}_{-0.30}$ & $0.76^{+0.19}_{-0.15}$ & 
$0.67^{+0.26}_{-0.23}$ & $0.81^{+0.16}_{-0.12}$ & $7.2^{+1.8}_{-1.7}$ & $8.5^{+2.0}_{-1.3}$ & 316/270 \\
combined &
$0.74^{+0.02}_{-0.03}$ & $1.36^{+0.06}_{-0.06}$ & $0.33^{+0.12}_{-0.10}$ & $1.16^{+0.18}_{-0.16}$ & $0.95^{+0.12}_{-0.10}$ & 
$0.75^{+0.20}_{-0.19}$ & $0.86^{+0.08}_{-0.07}$ & $6.8^{+0.8}_{-0.7}$ & $8.8^{+1.3}_{-1.3}$ & 821/423 \\
\hline
vAPEC \\
\hline
Chandra & 
$0.77^{+0.02}_{-0.02}$ & $1.33^{+0.15}_{-0.07}$ & $0.41^{+0.19}_{-0.13}$ & $1.20^{+0.22}_{-0.19}$ & $1.18^{+0.17}_{-0.15}$ & 
$1.21^{+0.37}_{-0.34}$ & $0.91^{+0.14}_{-0.08}$ & $7.7^{+0.9}_{-0.9}$ & $7.8^{+0.3}_{-1.1}$ & 158/144 \\
XMM &
$0.79^{+0.01}_{-0.01}$ & $1.47^{+0.11}_{-0.11}$ & $0.30^{+0.34}_{-0.22}$ & $0.97^{+0.44}_{-0.32}$ & $0.76^{+0.23}_{-0.17}$ & 
$0.71^{+0.29}_{-0.25}$ & $0.85^{+0.22}_{-0.16}$ & $7.4^{+2.1}_{-1.9}$ & $7.7^{+1.0}_{-1.0}$ & 311/270 \\
combined &
$0.78^{+0.01}_{-0.01}$ & $1.43^{+0.07}_{-0.07}$ & $0.40^{+0.15}_{-0.12}$ & $1.15^{+0.10}_{-0.17}$ & $0.93^{+0.12}_{-0.11}$ & 
$0.86^{+0.21}_{-0.20}$ & $0.90^{+0.09}_{-0.08}$ & $7.7^{+0.7}_{-0.7}$ & $7.5^{+0.5}_{-0.5}$ & 795/423 \\
\hline\\[-1ex]
\multicolumn{11}{l}{\parbox{16cm}{\footnotesize
\footnotemark[*] Normalization for the Mekal model,
$\makebox{\it Norm}=\int n_{\rm e} n_{\rm H} dV/ (4\pi (1+z)^2 D_{\rm A}^{\,2})$, 
where $D_{\rm A}$ is the angular distance to the source.
}}
\end{tabular}
\end{small}
}
\end{table*}

\begin{table*}[tbg]
\caption{
Same as table~\ref{table:proj spec},
besides the element abundances are determined separately
for the combined fit.
}\label{table:abun free}
\centerline{
\begin{small}
\begin{tabular}{llllllllllc}
\hline\hline
& $kT_1$ & $kT_2$ & O & Mg & Si & S & Fe & 
{\it Norm}$_1$ & {\it Norm}$_2$ & $\chi^2/$dof \\
& (keV) & (keV) & (solar) & (solar) & (solar) & (solar) & (solar) & 
\multicolumn{2}{c}{($10^{-18}~\rm cm^{-5}$)} &           \\
\hline \hline
vMekal &
$0.71^{+0.02}_{-0.01}$ & $1.31^{+0.14}_{-0.07}$ & $0.37^{+0.11}_{-0.10}$ & $1.17^{+0.21}_{-0.18}$ &
$1.06^{+0.16}_{-0.14}$ & $1.01^{+0.28}_{-0.25}$ & $0.86^{+0.08}_{-0.08}$ &
$6.2^{+0.8}_{-0.6}$ & $8.3^{+1.0}_{-1.3}$ & 801/418 \\
\hline
vAPEC &
$0.78^{+0.01}_{-0.01}$ & $1.44^{+0.08}_{-0.09}$ & $0.42^{+0.11}_{-0.09}$ & $1.13^{+0.23}_{-0.17}$ &
$0.95^{+0.16}_{-0.11}$ & $0.98^{+0.26}_{-0.22}$ & $0.83^{+0.11}_{-0.09}$ &
$8.1^{+1.1}_{-1.2}$ & $6.9^{+0.7}_{-0.8}$ & 783/418 \\
\hline\hline
& & & Ne & Al & Ar & Ca & Ni \\
\hline\hline
vMekal &
& & $0.49^{+0.50}_{-0.45}$ & $2.70^{+1.38}_{-1.32}$ & $0.67^{+1.02}_{-0.67}$ & $0.88^{+2.12}_{-0.88}$ & $2.33^{+0.77}_{-0.70}$ \\
\hline
vAPEC &
& & $0.72^{+0.34}_{-0.29}$ & $1.89^{+1.41}_{-1.21}$ & $0.72^{+1.01}_{-0.72}$ & $0.00^{+1.73}_{-0.00}$ & $1.80^{+0.84}_{-0.69}$ \\
\hline
\end{tabular}
\end{small}
}
\end{table*}

\subsection{Deprojection analysis}

\begin{table}
\caption{
Fractional contribution of the outer shells to the inner shells
in the deprojection analysis.
}\label{table:deproj-frac}
\centerline{
\begin{tabular}{rrrrrr}
\hline\hline
\hspace*{5em}   & 4--8$'$ & 2--4$'$ & 1--2$'$ & 0.6--1$'$ & 0--0.6$'$ \\
\hline
8--14$'$ $\dotfill$     & 0.373 & 0.075 & 0.018 & 0.004 & 0.002 \\
4--8$'$ $\dotfill$      & ----- & 0.273 & 0.056 & 0.012 & 0.007 \\
2--4$'$ $\dotfill$      & ----- & ----- & 0.273 & 0.048 & 0.026 \\
1--2$'$ $\dotfill$      & ----- & ----- & ----- & 0.238 & 0.109 \\
0.6--1$'$ $\dotfill$    & ----- & ----- & ----- & ----- & 0.531 \\
\hline
\end{tabular}
}
\end{table}

We conducted a deprojection analysis using energy spectra
from several circular annuls from the center of HCG~62a.
The inner and outer radii of the annuls were
$r= 0$--0.2$'$, 0.2--0.4$'$, 0.4--0.6$'$, 0.6--1$'$,
1--2$'$, 2--4$'$, 4--8$'$, and 8--14$'$,
in which $r$ represents the projected radius.
The Chandra ACIS-S3 data were used in the range of $r\le 4'$,
and the XMM-Newton data were used in $2'\le r\le 14'$.
Therefore, only the annulus of 2--4$'$ was simultaneously
examined by both satellites.
The spectra for $0.2'\lesssim r\lesssim 1'$ include two cavities.
Since the temperature and abundance of the two cavities are consistent
with that in the non-cavity region as shown previously
(\S\,\ref{subsec:cavity spec}),
we did not exclude the cavity region and simply analyzed the annular spectra.
Extraction of each spectrum and the treatment of the background
are conducted in the same way as described in \S\,\ref{subsec:consistency}.

Spherical symmetry was assumed in the deprojection procedure.
Starting from the outermost region, we fitted the projected
annular spectrum with 1-$T$ or 2-$T$ model.
Spectrum of the neighboring inner region was then fitted with
a model which contained contribution from the outer regions
with fixed model parameters at their best-fit values.
We repeated this procedure until the innermost region was reached.
Because the outer contributions are fixed in the inner model fitting, 
the propagation of errors is ignored in this method. 
However, the inner regions are always brighter than outer regions, 
hence this effect is negligible.
Table~\ref{table:deproj-frac} summarizes the fractional contribution
of the outer shells to the inner shells.
For example, the spectrum of the 4--8$'$ annulus was fitted
with adding 37.3\% flux of the best-fit model in the 8--14$'$ annulus.

\begin{figure*}[tbg]
\begin{center}
\FigureFile(0.5\textwidth,0.5\textwidth){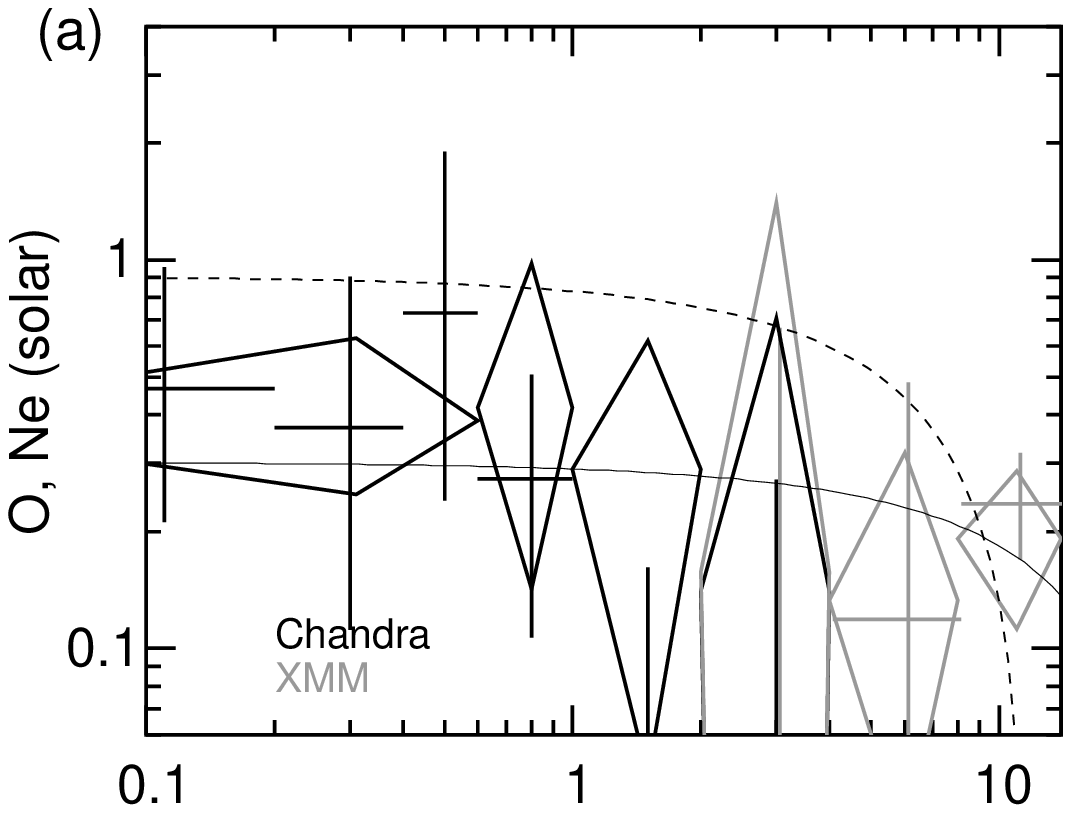}\hfill
\FigureFile(0.5\textwidth,0.5\textwidth){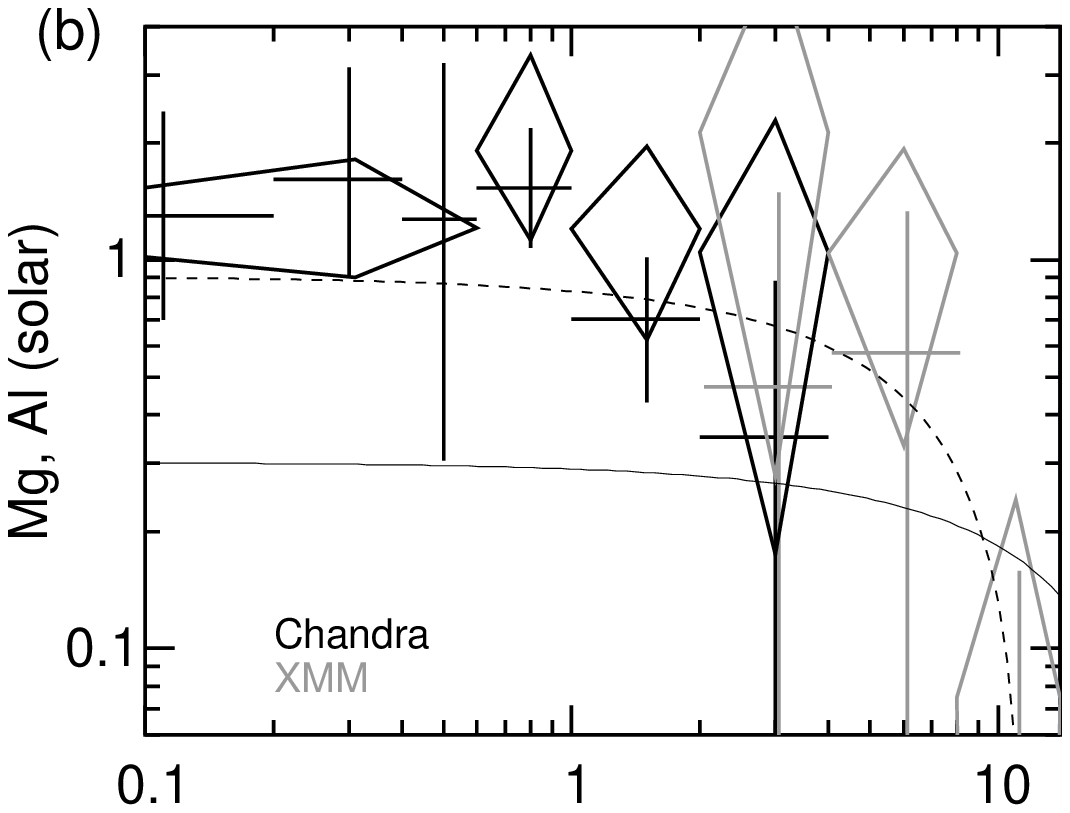}
\\[-4ex]
\FigureFile(0.5\textwidth,0.5\textwidth){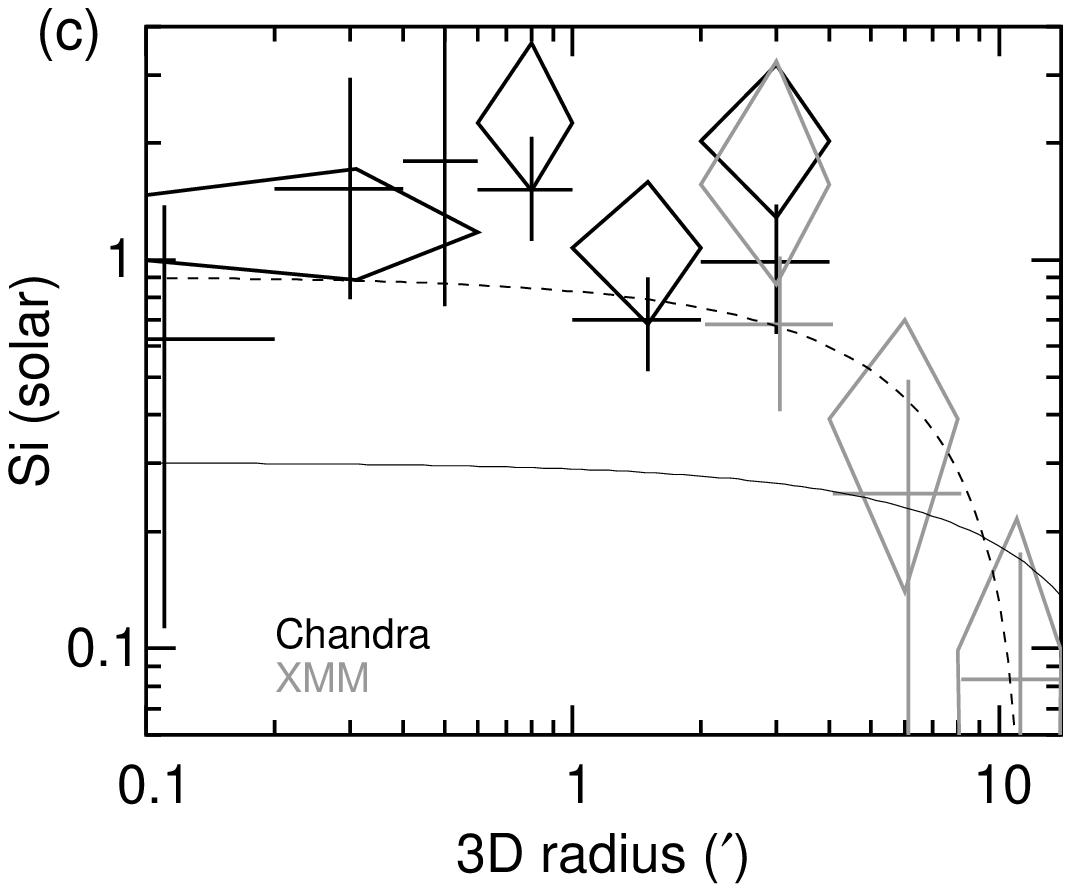}\hfill
\FigureFile(0.5\textwidth,0.5\textwidth){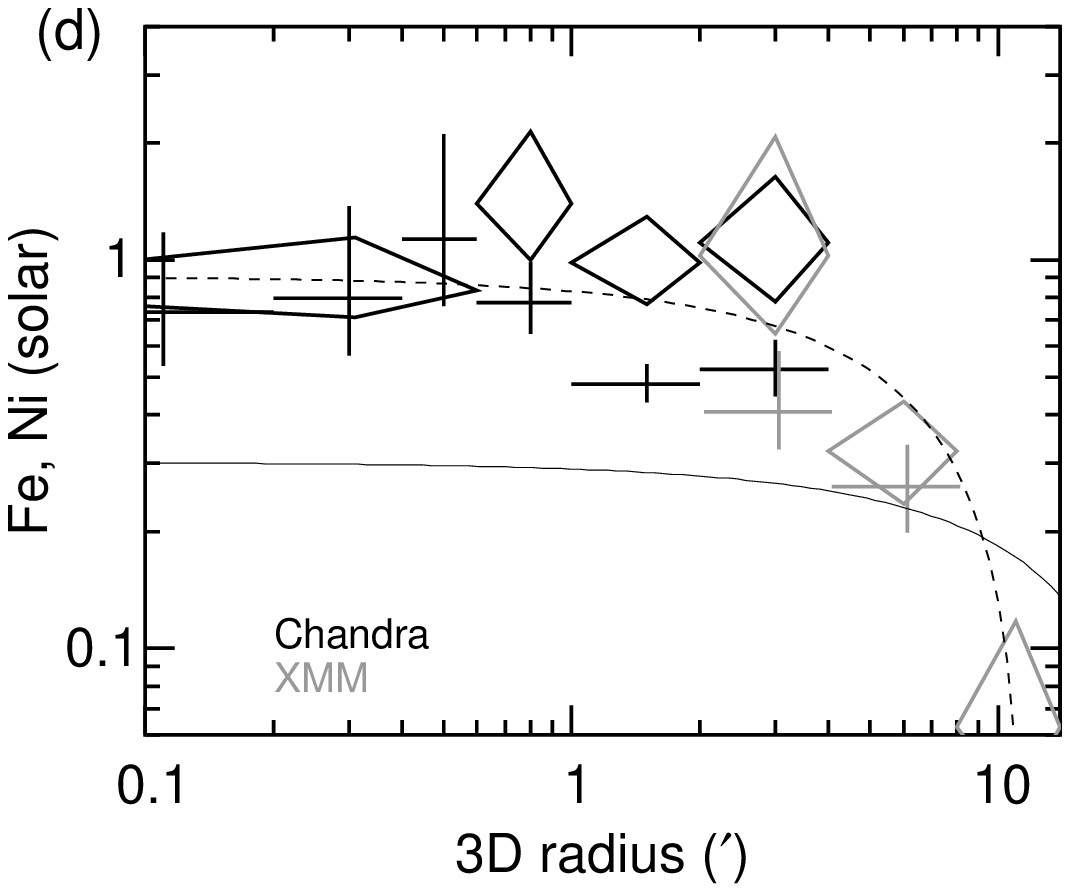}
\end{center}
\vspace*{-2ex}
\caption{
Abundance profiles of O (with Ne) (a), Mg (b), Si (c), and Fe (with Ni) (d)
using the 1-$T$ (crosses) and 2-$T$ (diamonds) vMekal models
with the deprojection analysis.
The ACIS-S3 (black) is used for $r<4'$, 
and MOS1, MOS2, and pn (gray) for $2'<r<14'$.
The solar photospheric value of $\rm [Fe/H]=4.7\times 10^{-5}$
\citep{Anders_Grevesse_1989} is adopted for the Fe abundance.
The solid and dashed lines correspond to the best fit relations 
of the deprojected O and Fe profile of the 2-$T$ results, respectively.
}\label{fig:plot-z}
\end{figure*}

Without the deprojection, the metal abundance of the group
center ($r\lesssim 1'$) is underestimated by about $30$\%,
while the temperature does not change significantly.
The sample spectra in the 2--4$'$ annulus fitted with
the deprojected 2-$T$ model are shown in
figure~\ref{fig:spec}~(b), and the fit results are
summarized in table~\ref{table:spec}.
Figures~\ref{fig:plot-kt}~\&~\ref{fig:plot-z} show the result of the
deprojection analysis, and the 1-$T$ results are quite similar to
the previous ASCA result \citep{Finoguenov_Ponman1999}.

\begin{table*}[tbg]
\caption{
Results of the deprojected spectral fit with 1-$T$ and 2-$T$ vMekal models.  
The Chandra data (ACIS-S3) are used for 0--4$'$, and the XMM-Newton
(MOS1+2, pn) data are used for 2--14$'$.
}\label{table:spec}
\centerline{
\begin{small}
\begin{tabular}{llllllllllc}
\hline\hline
$r$ & $kT_1$ & $kT_2$ & O, Ne & Mg, Al & Si & \makebox[0cm][l]{S, Ar, Ca} & Fe, Ni & \makebox[0cm][l]{{\it Norm}$_1$\footnotemark[*]} & \makebox[0cm][l]{{\it Norm}$_2$\footnotemark[*]} & $\chi^2/$dof \\
\makebox[0cm][l]{(arcmin)} & (keV) & (keV) &  (solar) & (solar) & (solar) & (solar) & (solar) & \multicolumn{2}{c}{($10^{-18}~\rm cm^{-5}$)} & \\
\hline\hline
\multicolumn{11}{c}{1-$T$ vMekal model} \\
\hline
\multicolumn{11}{c}{Chandra results} \\
0.0--0.6 &
$0.73\pm \makebox[0cm][l]{0.01}$         & \multicolumn{1}{c}{---} & $0.45^{+0.18}_{-0.14}$ & $1.20^{+0.39}_{-0.29}$ & $1.10^{+0.34}_{-0.26}$ & 
$<0.49$ & $0.76^{+0.16}_{-0.10}$ & $4.6^{+1.0}_{-1.2}$ & \multicolumn{1}{c}{---} & \makebox[0cm][r]{177}/\makebox[0cm][l]{96}   \\
0.0--0.2 &      
$0.67\pm \makebox[0cm][l]{0.02}$         & \multicolumn{1}{c}{---} & $0.47^{+0.49}_{-0.26}$ & $1.30^{+1.12}_{-0.60}$ & $0.63^{+0.75}_{-0.51}$ & 
$<1.03$ &              $0.74^{+0.44}_{-0.20}$ & $1.2^{+0.6}_{-0.7}$ & \multicolumn{1}{c}{---} & \makebox[0cm][r]{75}/\makebox[0cm][l]{57} \\
0.2--0.4 &      
$0.70^{+0.03}_{-0.02}$ & \multicolumn{1}{c}{---} & $0.37^{+0.53}_{-0.26}$ & $1.61^{+1.52}_{-0.70}$ & $1.53^{+1.43}_{-0.73}$ & 
$<1.47$                & $0.80^{+0.58}_{-0.23}$ & $1.2^{+0.6}_{-0.8}$ & \multicolumn{1}{c}{---} & \makebox[0cm][r]{96}/\makebox[0cm][l]{66} \\
0.4--0.6 &      
$0.77\pm \makebox[0cm][l]{0.03}$         & \multicolumn{1}{c}{---} & $0.73^{+1.17}_{-0.49}$ & $1.28^{+1.95}_{-0.97}$ & $1.79^{+2.14}_{-1.03}$ & 
$1.53$\makebox[0cm][l]{ $^\dagger$}         & $1.13^{+0.98}_{-0.37}$ & $0.7\pm \makebox[0cm][l]{0.3}$        & \multicolumn{1}{c}{---} & \makebox[0cm][r]{67}/\makebox[0cm][l]{67} \\
0.6--1.0 &      
$0.84\pm \makebox[0cm][l]{0.01}$         & \multicolumn{1}{c}{---} & $0.27^{+0.23}_{-0.17}$ & $1.53^{+0.65}_{-0.46}$ & $1.52^{+0.56}_{-0.40}$ & 
$<0.67$                & $0.78^{+0.21}_{-0.13}$ & $2.7^{+0.5}_{-0.5}$ & \multicolumn{1}{c}{---} & \makebox[0cm][r]{159}/\makebox[0cm][l]{92} \\
1.0--2.0 &      
$1.12\pm \makebox[0cm][l]{0.01}$         & \multicolumn{1}{c}{---} & $0.03^{+0.13}_{-0.03}$ & $0.71^{+0.31}_{-0.28}$ & $0.70^{+0.20}_{-0.18}$ & 
$0.49^{+0.51}_{-0.47}$ & $0.48^{+0.06}_{-0.05}$ & $6.2^{+0.6}_{-0.7}$ & \multicolumn{1}{c}{---} & \makebox[0cm][r]{126}/\makebox[0cm][l]{72} \\
2.0--4.0 &      
$1.37^{+0.03}_{-0.04}$ & \multicolumn{1}{c}{---} & $0.02^{+0.25}_{-0.02}$ & $0.35^{+0.53}_{-0.35}$ & $0.99^{+0.40}_{-0.34}$ & 
$0.55^{+0.76}_{-0.55}$ & $0.52^{+0.10}_{-0.08}$ & $5.6^{+0.8}_{-1.0}$ & \multicolumn{1}{c}{---} & \makebox[0cm][r]{128}/\makebox[0cm][l]{108} \\
\hline
\multicolumn{11}{c}{XMM-Newton results}      \\
2.0--4.0 &      
$1.43^{+0.17}_{-0.05}$ & \multicolumn{1}{c}{---} & $<0.67$                & $0.47^{+1.02}_{-0.47}$ & $0.68^{+0.34}_{-0.28}$ & 
$0.54^{+0.50}_{-0.47}$ & $0.41^{+0.18}_{-0.08}$ & $5.6^{+0.8}_{-1.2}$ & \multicolumn{1}{c}{---} & \makebox[0cm][r]{241}/\makebox[0cm][l]{199} \\
4.0--8.0 &      
$1.41^{+0.08}_{-0.07}$ & \multicolumn{1}{c}{---} & $0.12^{+0.37}_{-0.12}$ & $0.58^{+0.76}_{-0.58}$ & $0.25^{+0.24}_{-0.22}$ & 
$<0.15$                & $0.26^{+0.07}_{-0.06}$ & \makebox[0cm][r]{1}0.9$^{+1.3}_{-1.4}$ & \multicolumn{1}{c}{---} & \makebox[0cm][r]{164}/\makebox[0cm][l]{156} \\
8.0--14.0 &     
$0.64\pm \makebox[0cm][l]{0.03}$         & \multicolumn{1}{c}{---} & $0.24 ^{+0.08}_{-0.07}$ & $0.06 ^{+0.10}_{-0.06}$ & $0.08 ^{+0.09}_{-0.08}$ & 
$0.40 ^{+0.53}_{-0.40}$ & $0.04^{+0.01}_{-0.01}$         & \makebox[0cm][r]{4}7.4$^{+9.8}_{-8.9}$ & \multicolumn{1}{c}{---} & \makebox[0cm][r]{188}/\makebox[0cm][l]{180} \\
\hline
\multicolumn{11}{c}{combined results}      \\
2.0--4.0 &      
$1.39\pm \makebox[0cm][l]{0.03}$         & \multicolumn{1}{c}{---} & $<0.21$                 & $0.40^{+0.49}_{-0.40}$ & $0.79^{+0.27}_{-0.23}$ & 
$0.57 ^{+0.45}_{-0.42}$ & $0.47^{+0.08}_{-0.06}$ & $5.6^{+0.4}_{-0.6}$ & \multicolumn{1}{c}{---} & \makebox[0cm][r]{415}/\makebox[0cm][l]{314} \\
\hline\hline
\multicolumn{11}{c}{2-$T$ vMekal model}      \\
\hline
\multicolumn{11}{c}{Chandra results}      \\
0.0--0.6 &        
$0.66^{+0.03}_{-0.22}$ & 0.92$^{+0.18}_{-0.47}$ & 0.39$^{+0.24}_{-0.14}$& 1.21$^{+0.61}_{-0.31}$& 1.18$^{+0.54}_{-0.29}$& 
$<1.05$                & $0.84^{+0.31}_{-0.12}$ & 3.1$^{+0.7}_{-2.0}$    & 1.4$^{+2.8}_{-0.7}$   & \makebox[0cm][r]{151}/\makebox[0cm][l]{94}  \\
0.6--1.0 &      
$0.77^{+0.04}_{-0.07}$ & 1.36$^{+0.47}_{-0.20}$ & 0.42$^{+0.56}_{-0.27}$& 1.92$^{+1.45}_{-0.79}$& 2.25$^{+1.37}_{-0.74}$& 
$1.02^{+2.09}_{-1.02}$ & $1.40^{+0.75}_{-0.39}$ & 1.1$^{+0.5}_{-0.4}$    & 1.0$^{+0.5}_{-0.3}$   & \makebox[0cm][r]{78}/\makebox[0cm][l]{90}   \\
1.0--2.0 &      
$0.83^{+0.06}_{-0.15}$ & 1.36$^{+0.19}_{-0.11}$ & 0.29$^{+0.33}_{-0.24}$& 1.20$^{+0.76}_{-0.58}$& 1.08$^{+0.52}_{-0.39}$& 
$1.24^{+1.14}_{-0.89}$ & $0.99^{+0.31}_{-0.22}$ & 0.7$^{+0.5}_{-0.4}$    & 2.8$^{+0.8}_{-1.0}$   & \makebox[0cm][r]{69}/\makebox[0cm][l]{70}   \\
2.0--4.0 &      
$0.85^{+0.29}_{-0.23}$ & $1.59^{+0.69}_{-0.20}$ & $0.14^{+0.57}_{-0.14}$& $1.05^{+1.25}_{-0.87}$& $2.02^{+1.16}_{-0.74}$& 
$1.48^{+1.71}_{-1.25}$ & $1.11^{+0.54}_{-0.33}$ & $0.3^{+0.9}_{-0.2}$    & $3.2^{+0.8}_{-0.9}$   & \makebox[0cm][r]{121}/\makebox[0cm][l]{106} \\
\hline
\multicolumn{11}{c}{XMM-Newton results}      \\
2.0--4.0 &      
$0.90^{+0.27}_{-0.21}$ & $1.83^{+0.51}_{-0.22}$ & $0.16^{+1.24}_{-0.16}$& $2.13^{+3.95}_{-1.86}$& $1.57^{+1.68}_{-0.70}$& 
$1.07^{+1.32}_{-0.84}$ & $1.02^{+1.06}_{-0.38}$ & $0.3^{+0.6}_{-0.2}$    & $3.2 ^{+1.0}_{-1.1}$  & \makebox[0cm][r]{235}/\makebox[0cm][l]{197} \\
4.0--8.0 &      
$0.24^{+0.08}_{-0.04}$ & 1.40$^{+0.08}_{-0.06}$ & 0.13$^{+0.19}_{-0.10}$& 1.04$^{+0.89}_{-0.71}$& 0.39$^{+0.31}_{-0.25}$& 
$<0.22$                & $0.32^{+0.11}_{-0.09}$ & 3.2$^{+3.3}_{-1.7}$    & \makebox[0cm][r]{1}0.4$^{+1.7}_{-1.7}$& \makebox[0cm][r]{153}/\makebox[0cm][l]{154} \\
8.0--14.0 & 
$0.43^{+0.51}_{-0.09}$ & 0.89$^{+0.13}_{-0.56}$ & 0.19$^{+0.10}_{-0.08}$& 0.07$^{+0.17}_{-0.07}$& 0.10$^{+0.12}_{-0.10}$& 
$0.31^{+0.52}_{-0.31}$ & $0.06^{+0.05}_{-0.02}$ & \makebox[0cm][r]{2}3.9$^{+21.5}_{-11.5}$& \makebox[0cm][r]{2}2.9$^{+12.4}_{-11.5}$& \makebox[0cm][r]{185}/\makebox[0cm][l]{178} \\
\hline
\multicolumn{11}{c}{combined results}      \\
2.0--4.0 &      
$0.89^{+0.23}_{-0.12}$ & 1.71$^{+0.48}_{-0.15}$ & 0.13$^{+0.47}_{-0.13}$& 1.23$^{+1.10}_{-0.82}$& 1.68$^{+0.74}_{-0.52}$& 
$1.23^{+0.88}_{-0.70}$ & $1.05^{+0.39}_{-0.26}$ & 0.4$^{+0.6}_{-0.1}$    & 3.3$^{+0.6}_{-0.7}$   & \makebox[0cm][r]{402}/\makebox[0cm][l]{312} \\
\hline\\[-1ex]
\multicolumn{11}{l}{\parbox{17cm}{\footnotesize
\footnotemark[$\ast$]
Normalization for the vMekal model,
$\makebox{\it Norm}=\int n_{\rm e} n_{\rm H} dV/ (4\pi (1+z)^2 D_{\rm A}^{\,2})$, 
where $D_{\rm A}$ is the angular distance to the source.\\
\footnotemark[$\dagger$]
The S abundance was fixed to the best fit in calculating
errors due to an unresolved problem on XSPEC for this particular fit.
}}
\end{tabular}
\end{small}
}
\end{table*}

\subsection{Temperature profile}
\label{subsec:temp-prof}

\begin{figure*}[tbg]
\begin{center}
\vspace*{-2ex}
    \FigureFile(0.5\textwidth,0.5\textwidth){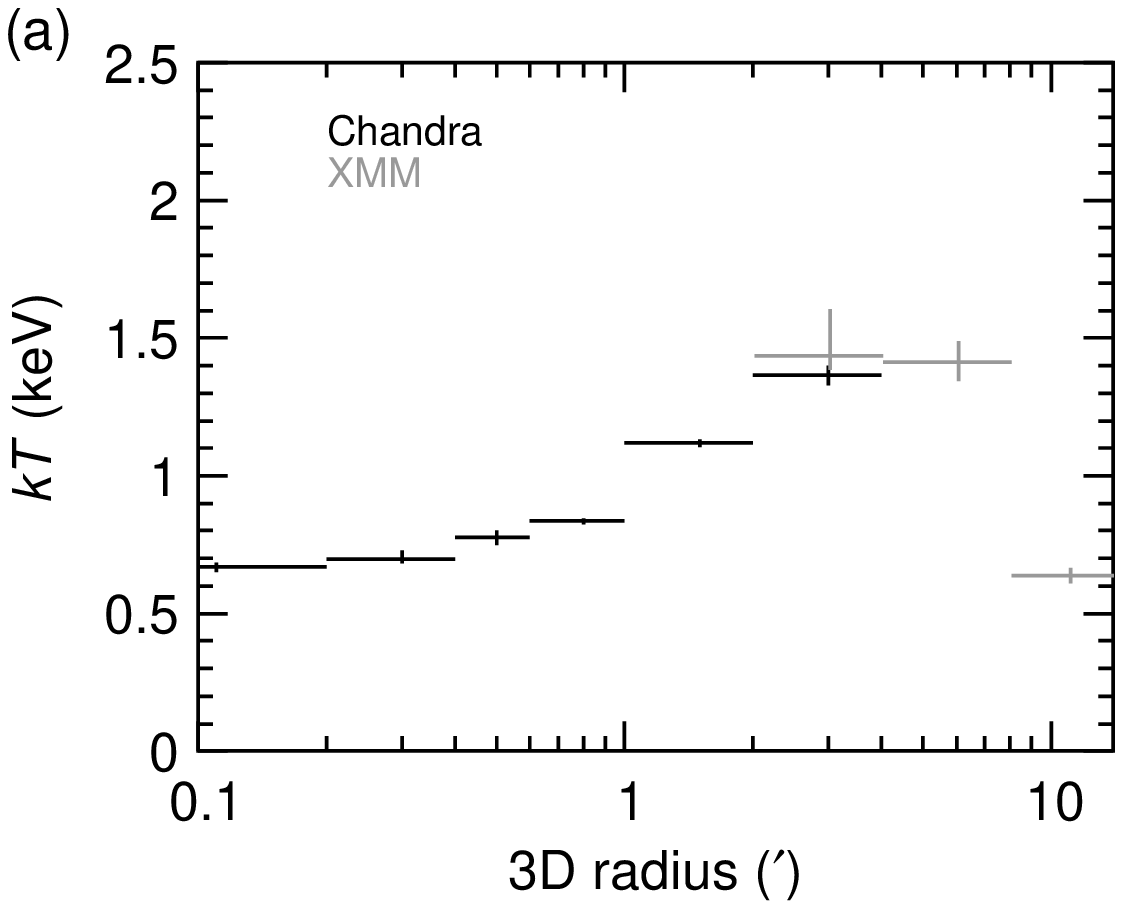}\hfill
    \FigureFile(0.5\textwidth,0.5\textwidth){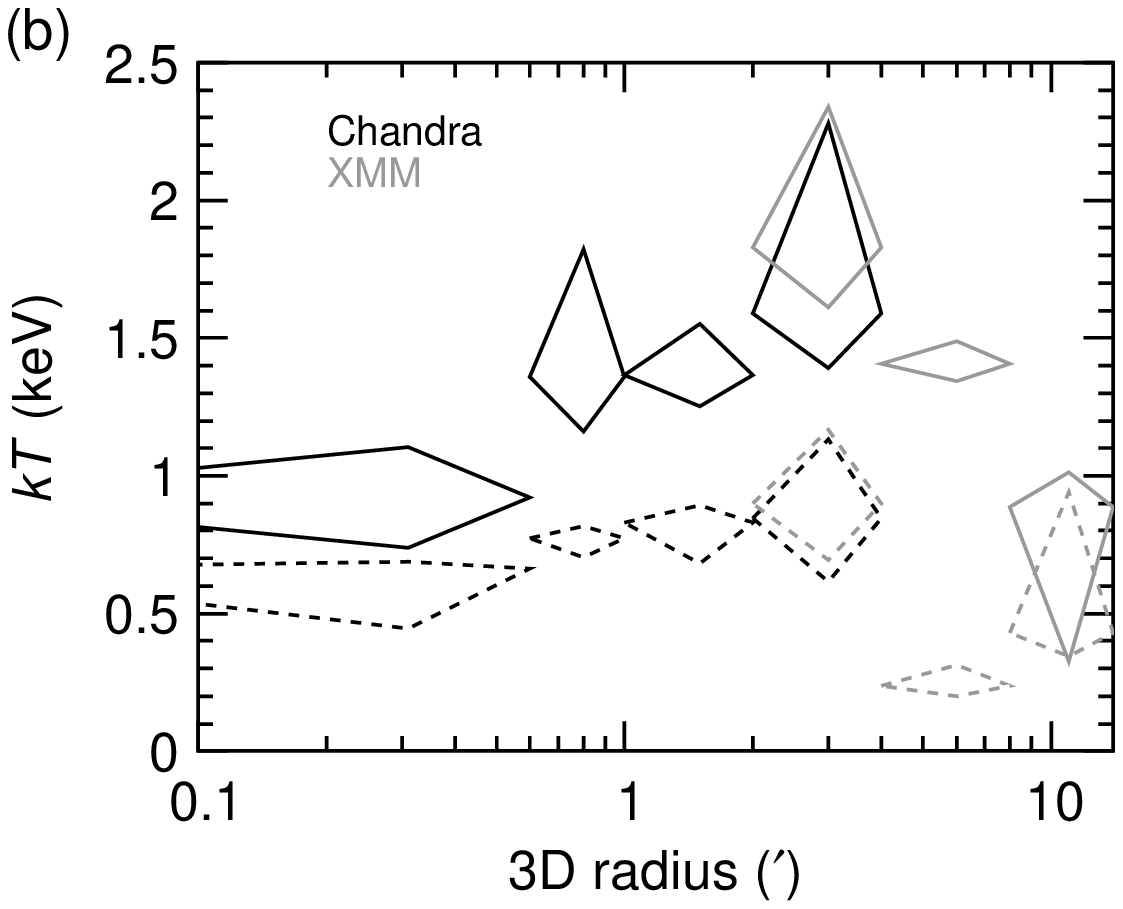}
\end{center}
\vspace*{-4ex}
\caption{
(a) Temperature profiles with the 1-$T$ vMekal models
based on the deprojected spectral fit.  
ACIS-S3 (black) is used for $r < 4'$, 
and MOS1, MOS2, and pn (gray) are used for $2'<r<14'$.
(b) Same as (a), but with the 2-$T$ vMekal models.
}\label{fig:plot-kt}
\end{figure*}

The temperature profiles obtained by both the 1-$T$ and 2-$T$ models
with the deprojection analysis are shown in figure~\ref{fig:plot-kt}.
For the radius range $0.6<r<2.0'$, the 2-$T$ model gives significantly
lower $\chi^2$ values ($\chi^2 = 78, 68$ for 0.6--1.0$'$, 1.0--2.0$'$,
respectively) than the 1-$T$ model ($\chi^2 = 159, 126$).
The significance was tested with the $F$-test,
and the 2-$T$ model was preferred with a significance higher than 5$\sigma$.
Within $r<0.6'$, the 2-$T$ model is preferred for the whole 0.0--0.6$'$ data
by $\Delta\chi^2 = 26$ (99.9\% confidence with $F$-test),
while the 1-$T$ model is also acceptable when we split the region
into smaller annuli, 0.0--0.2$'$, 0.2--0.4$'$, and 0.4--0.6$'$.
Furthermore, we have also tried the 2-$T$ model for the spectra
in the range 0--0.2$'$ and 0.2--0.4$'$, but the normalization of
the second component turned out to be nil.
This is due probably to the poorer statistics for the smaller annuli
and also to the fact that the cooler component is dominant
in the central region.
In the outer regions of $r>2'$, the $\chi^2$ values for the 2-$T$
model were also lower than the 1-$T$ case,
but both of the fit were acceptable.
For the outermost range 8--14$'$, the 2-$T$ model gave similar
temperatures for the hotter and cooler components within errors,
therefore the 1-$T$ fit turned out to be just enough. This is
because the hot component dominates the surface brightness.

For the 1-$T$ model, the temperature rises from $kT=0.7$~keV
at the center to $kT=1.4$~keV at larger radii, which is consistent
with the previous deprojection analysis for the ASCA data
\citep{Finoguenov_Ponman1999} and the ROSAT temperature
profile by \citet{Buote2000a}.
This profile in $r<2'$ is also seen in the temperature map
calculated from the {\it HR}\/ (figure~\ref{fig:hr}).
The temperature with the 1-$T$ model drops again at $r>8'$
from the intermediate level of $kT=1.4$~keV,
as shown in figure~\ref{fig:plot-kt}.
The temperature drop in our data is much steeper than those reported by
\citet{Finoguenov_Ponman1999} with ASCA and by \citet{Buote2000a}
with ROSAT\@. The ROSAT temperature at $r>8'$ is
$kT = 0.9$--1.0~keV, while ours is $kT=0.66\pm 0.03$~keV\@.
The influence of the background has been tested by changing
the background level within $\pm 5$\%,
and we found that the best fit temperature varied in 0.64--0.70~keV\@.
We have looked into the effects of point source contamination
and the influence of hard and soft components,
though none of them gives significant change in temperature.

For the 2-$T$ model, temperature of the cooler component
is nearly constant at $\sim 0.7$~keV, which is very close to
the central temperature obtained with the 1-$T$ fit.
Temperature of the hotter component is also nearly constant
at $\sim 1.4$~keV up to $r\sim 8'$.
These features are consistent with the ASCA results,
which indicate two temperature components at 0.7~keV and 1.4~keV
based on projected spectra within $r < 3'$ \citep{Buote2000b}.
The cool component is concentrated in the center and dominant within 1$'$,
while the hot component extends out to $\sim 4'$.
The extent of the cool component will be examined
in \S\,\ref{sec:mass-prof} more quantatively.

\subsection{Abundance profile}
\label{subsec:abun-prof}

The metal abundances were derived from the deprojected spectra based
on the 1-$T$ and 2-$T$ fits.
The Fe, Si, and Mg abundances are around one solar or more in $r<4'$.
On the other hand, the O abundance is always lower
than the solar value (table~\ref{table:spec} and figure~\ref{fig:plot-z}).
We could poorly constrain the S abundance due to the limited statistics.
The 2-$T$ fit gave significantly higher abundance than the 1-$T$ fit
in the intermediate range of $1.0'< r < 4.0'$,
while similar abundances between 1-$T$ and 2-$T$
are obtained for the inner ($r < 1.0'$) and outer ($4.0' < r$) regions.
Such discrepancy is often seen in the spectral fit, because the abundance
takes the lowest value at the temperature where the line emissivity
is the highest. This situation corresponds to the case of 1-$T$ fit,
and the 2-$T$ model tends to shift the temperature away from such a high
emissivity position.
The ASCA spectra of HGG~62 are fitted with the 1-$T$ Mekal model by
\citet{Finoguenov_Ponman1999}, and the present results for Fe, Si,
and Mg with 1-$T$ fits are in good agreement with
the ASCA deprojection results within the 90\% confidence limit at $r>1'$,

In figure~\ref{fig:plot-z}, the solid and dashed lines correspond to 
the best fit relations of the deprojected O (a) and Fe (d) profile of
the 2-$T$ results, respectively.
A simple linear model of $Z=-a\,r+b$ is assumed for each profile.
These best fit relations are overlaid to all the four plots for comparison.
O abundance of $\sim 0.3$~solar is significantly
lower than the other elements. It also indicates a flatter profile,
though significance is low due to the low abundance of oxygen.
The decline of Fe, Si, and Mg abundances with radius
previously reported by \citet{Finoguenov_Ponman1999} with ASCA,
is also confirmed with Chandra and XMM-Newton\@.
Our result shows that Fe, Si, and Mg abundances are
1--2 solar at the center and drop to about 0.1 solar at $r\sim 10'$.
These abundance levels are about twice larger than the values derived
with the ASCA spectra ($\sim 0.6$ solar by \cite{Finoguenov_Ponman1999}).
This is due primarily to the much better angular resolution
of the Chandra X-ray telescope (0.5$''$) than ASCA (3$'$)
at the very central region.
In addition, our 2-$T$ model gives higher abundances
in the intermediate range of $1.0'< r < 4.0'$.
As described in \S\,\ref{subsec:temp-prof},
the 2-$T$ model is significantly needed in this region,
so that the ASCA abundances are likely to be underestimated.
We also note that many authors recently take the solar abundance to be
the value given by \citet{Grevesse_Sauval1998},
i.e.\ $\rm [Fe/H]=3.2\times 10^{-5}$,
which is obtained from the measurements of solar system meteorites.  
The solar photospheric value of $\rm [Fe/H]=4.7\times 10^{-5}$ 
\citep{Anders_Grevesse_1989}, which is adopted in our analysis,
gives the Fe abundance approximately by a factor of 1.47 smaller.
Considering this effect, all the Fe, Si, and Mg abundances have
similar value around 1.2 solar at the center ($r < 0.6'$).

\section{Mass Profiles}
\label{sec:mass-prof}

\subsection{Formulation of two-phase gas}
\label{subsec:formulation-cool-hot}

Based on the radial distributions of the cool and hot components, 
we will derive mass distributions of gas and dark matter.
Since the estimated gas mass contained
in a single cavity volume, if it is filled up, is $\sim 6$\%
of the mass in the shell of $7''<r<32''$,
the assumption of the spherical symmetry gives relatively small errors.
However, because two temperatures are required for each shell
within the deprojected radius of 0.6$'$--4.0$'$
as seen in the previous section, we need an additional assumption
between the hot and cool components.
Here we assume a pressure balance between the two phase,
i.e.\  $P_{\rm gas} = n_1\, kT_1 = n_2\, kT_2$, 
where $n_1$, $n_2$, $T_1$ and $T_2$ are
the cool and hot gas densities and temperatures, respectively,
at each 3-dimensional radius of $R$\@.
Such a pressure balance is previously adopted by
\citet{Ikebe1999} for the Centaurus cluster,
and by \citet{Xue2004} for the RGH~80 galaxy group.

We also introduce a volume filling factor, $f(R)$, of the cool component,
namely a fractional volume, $V_1\equiv f\, V$,
in the total volume, $V$, is filled with the cool gas,
and $V_2\equiv (1-f)\, V$ is filled with the hot gas.
This means that the cool and hot gas are not completely mixed,
instead the cool component is somewhat patchy or localized
with rather an irregular shape. Such a hypothesis may be supported
by the 2-dimensional temperature map (figure~\ref{fig:hr})
and the existence of the cavities.

\begin{figure*}[tbgp]
\begin{center}
\FigureFile(0.45\textwidth,0.45\textwidth){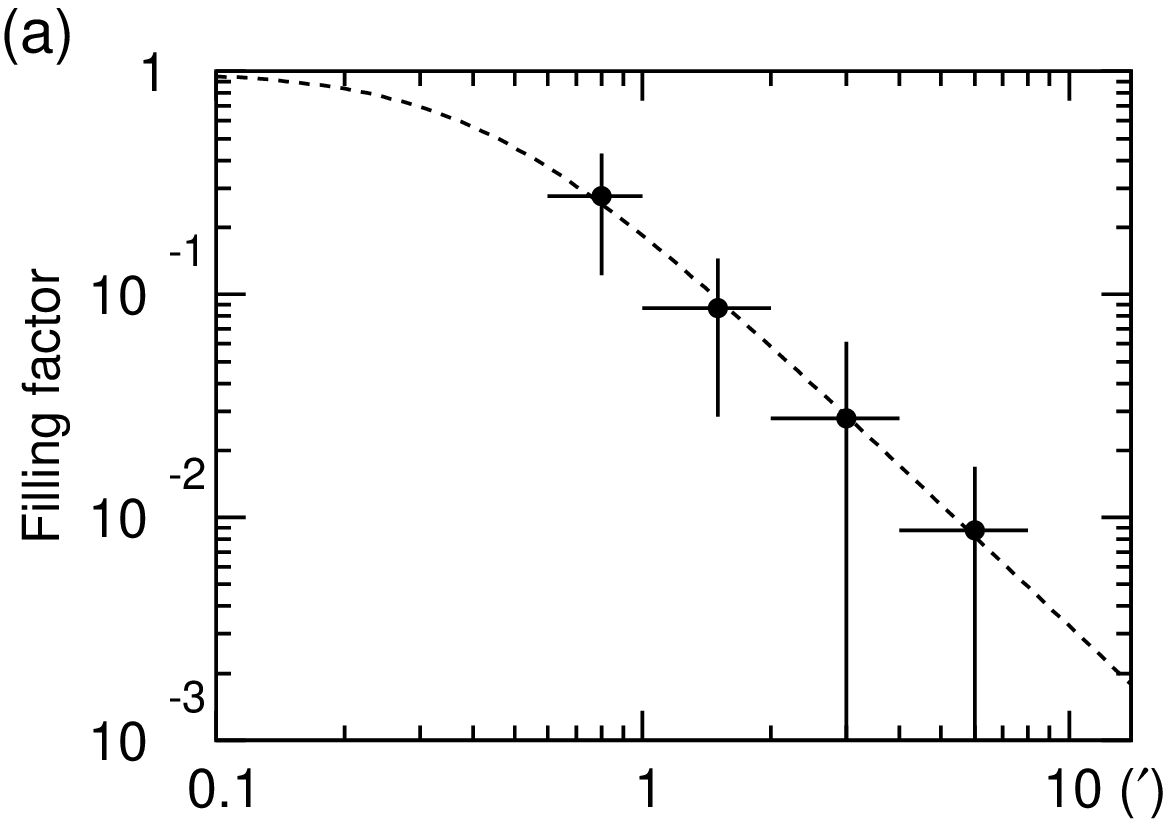}
\hspace{1em}
\FigureFile(0.45\textwidth,0.45\textwidth){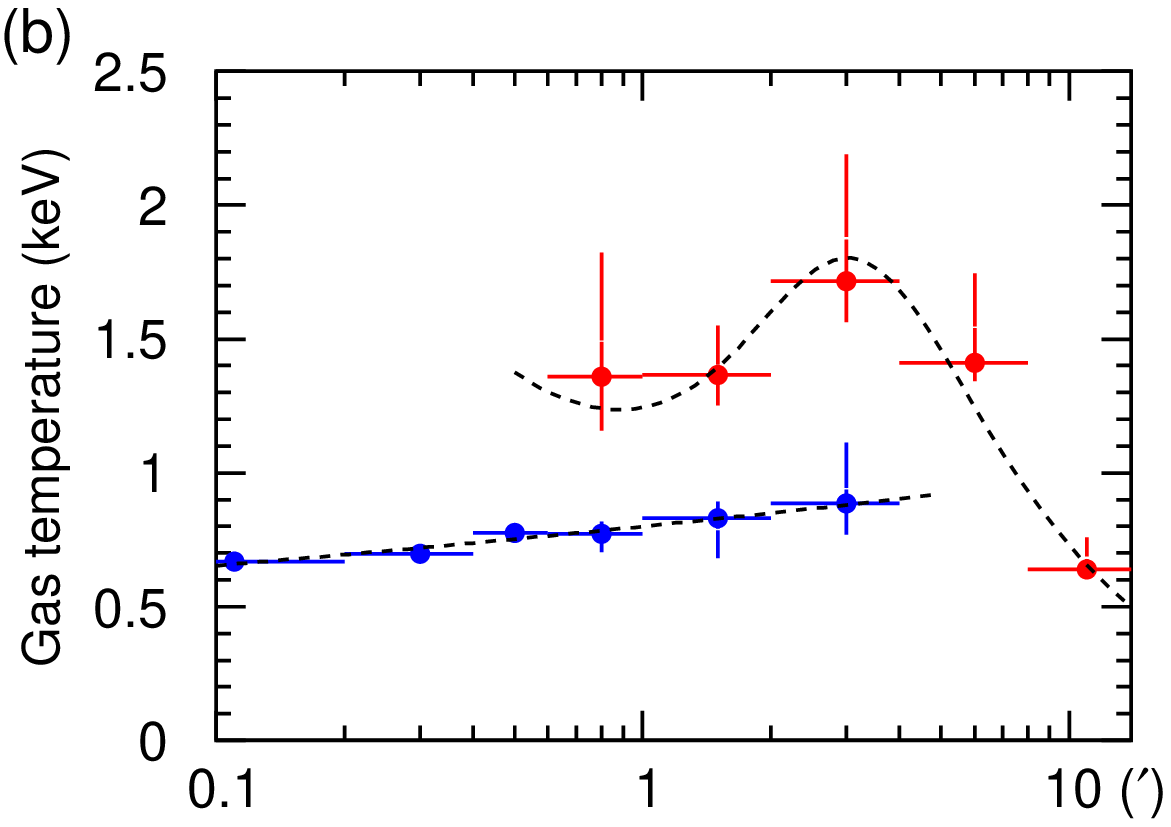}
\\[-4.0ex]
\FigureFile(0.45\textwidth,0.45\textwidth){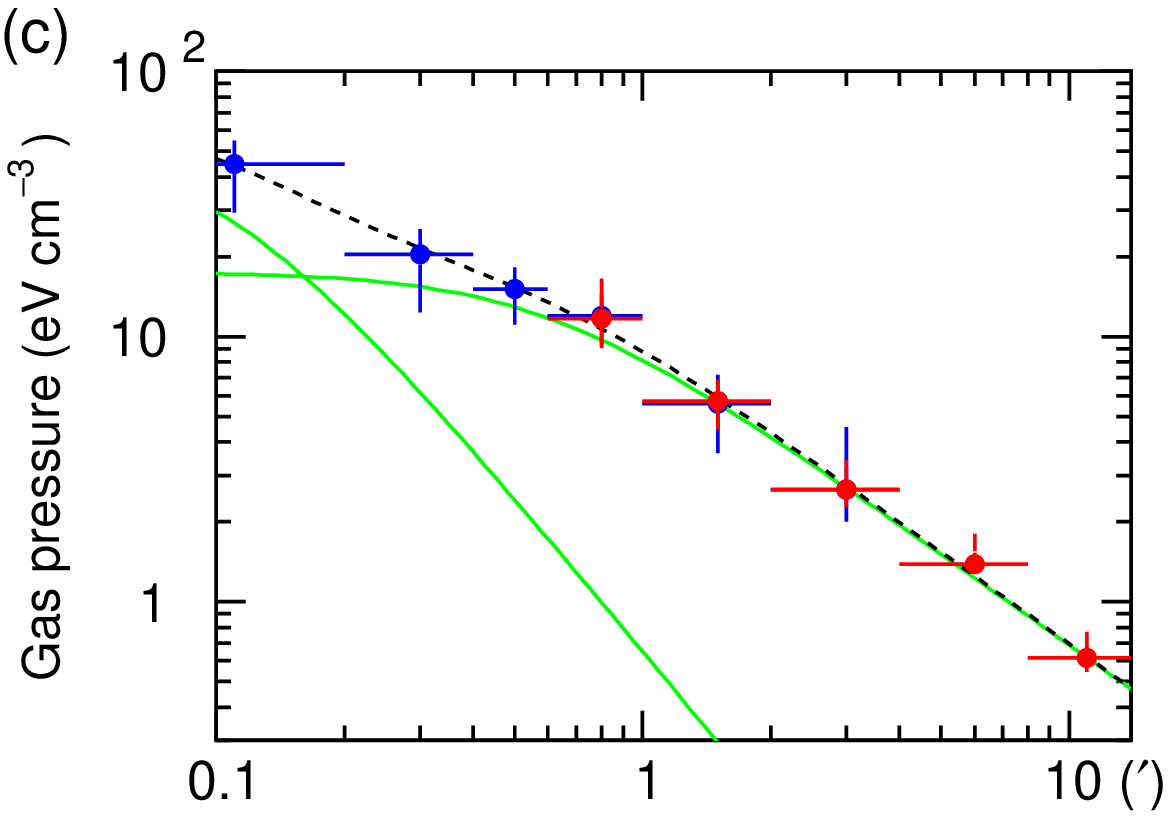}
\hspace{1em}
\FigureFile(0.45\textwidth,0.45\textwidth){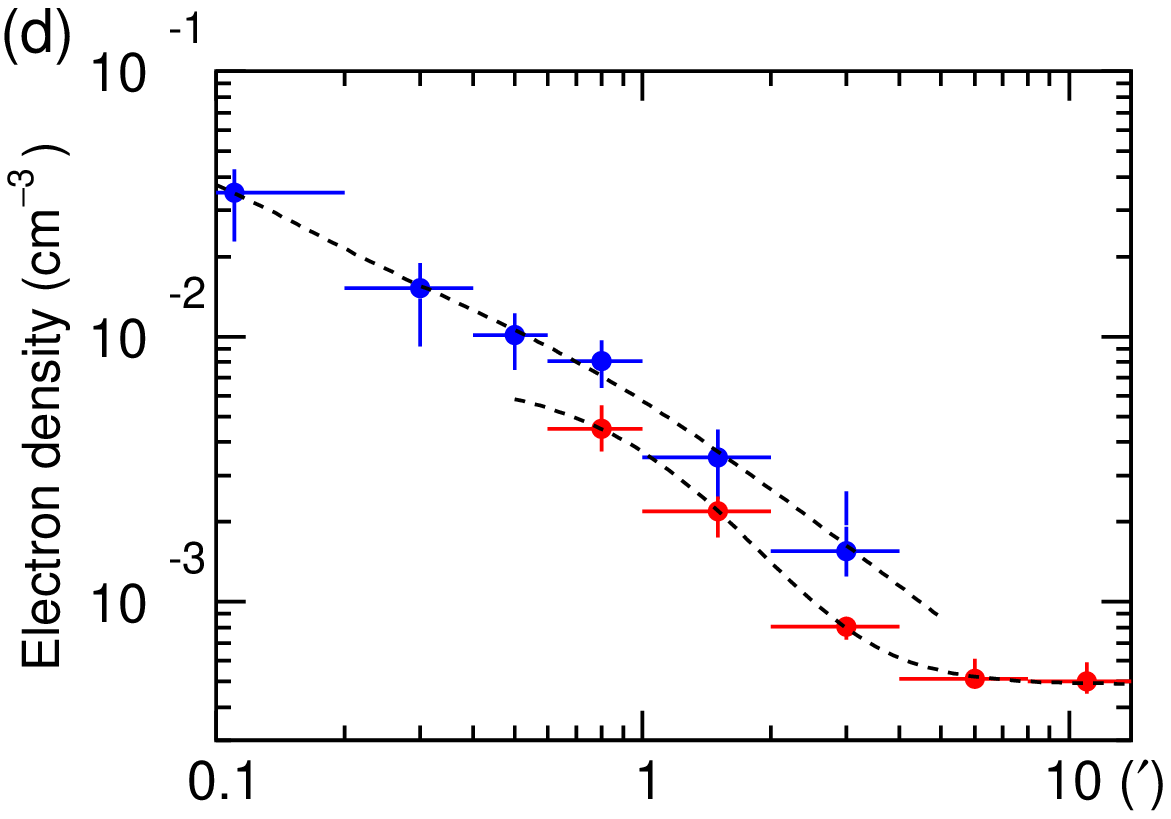}
\\[-4.5ex]
\FigureFile(0.45\textwidth,0.45\textwidth){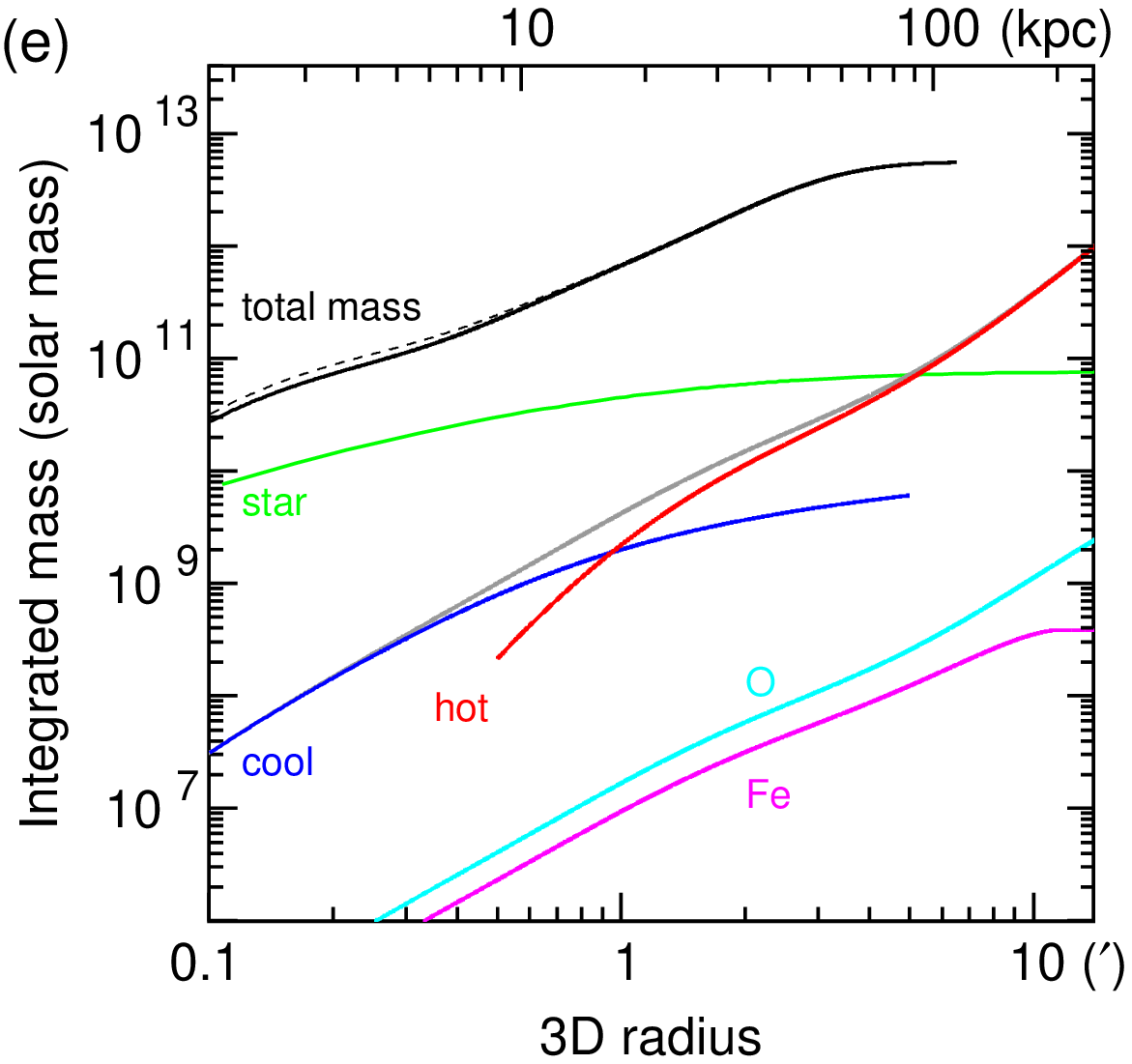}
\hspace{1em}
\FigureFile(0.45\textwidth,0.45\textwidth){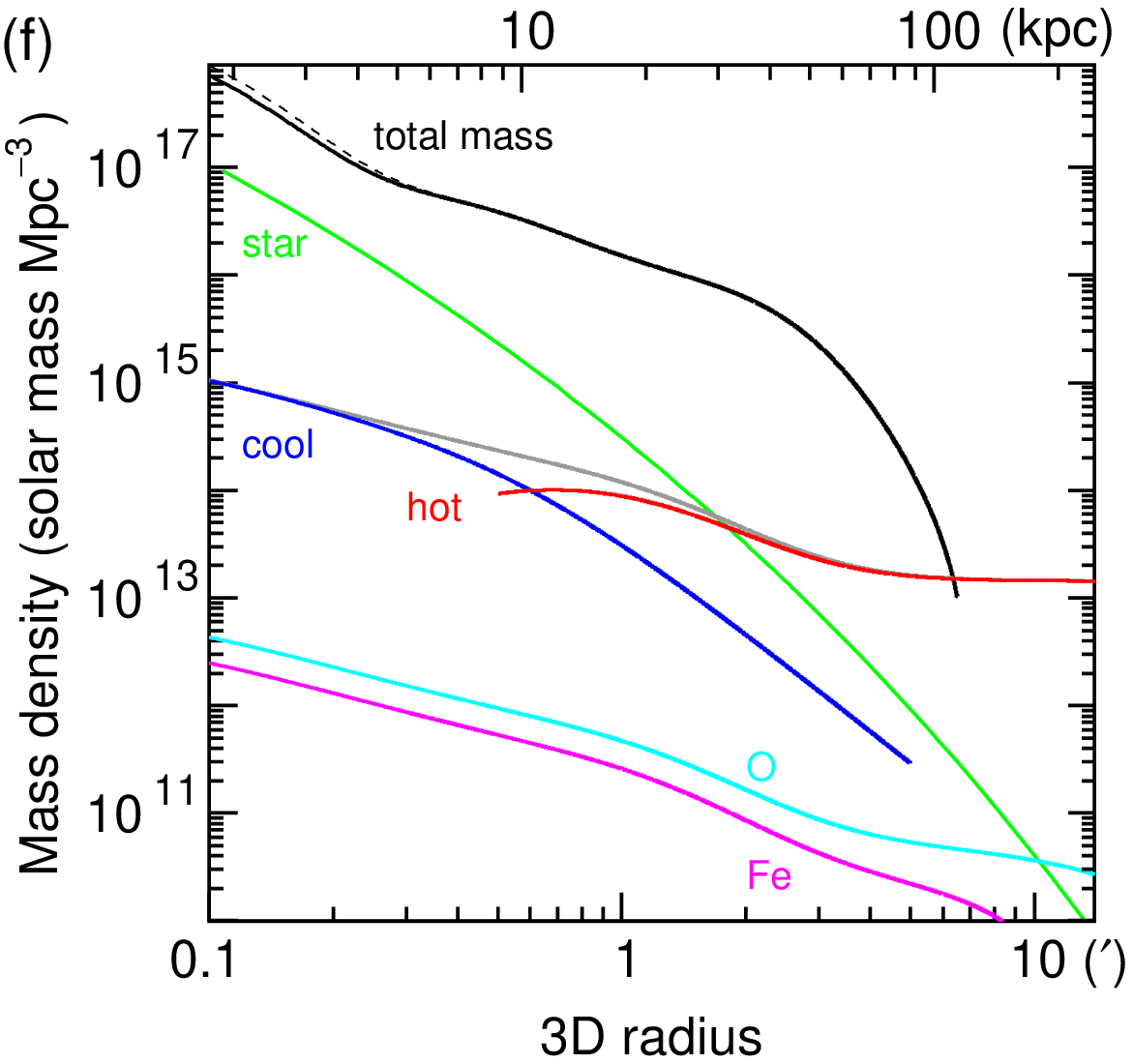}
\end{center}
\vspace*{-2ex}
\caption{
(a) Volume filling factor $f(R)$ of the cool component,
(b) cool or hot gas temperature $kT_1$ or $kT_2$, and
(c) gas pressure $P_{\rm gas}$,
(d) cool or hot electron density $n_{\rm e1}$ or $n_{\rm e2}$,
plotted against the 3-dimensional radius $R$\@.
Blue points represent the cool component and red points the hot.
In panels (b)--(d), 2-$T$ fit results are adopted
in the range of 0.6--4$'$ and 1-$T$ fit results for others.
The systematic errors when the background level is increased or
decreased by $\pm 5$\% are considered in the error bars for
the outermost two points.
The dashed lines in panels (a)--(d) and green curves in panel (b)
represent the best fit models. See text in details.
(e) A black line indicates the total integrated gravitational mass,
$M_{<R}$, estimated from the best fit models.
The solid line is calculated by eq.~(\ref{eq:Mtot}),
and dashed line is by eq.~(\ref{eq:Mtot'}).
Red and blue lines represent the gas mass 
of the hot and cool components, respectively.
The O and Fe mass contained in either of the gas
is plotted by cyan and pink lines.
The stellar mass estimated from the $R$ band photometry
of the HCG~62a galaxy by \citet{Tran2001} is indicated by a green line.
See text for details.
(f) Same as (e), but for the differential mass density profiles. 
}\label{fig:mass-prof}
\end{figure*}

Because the normalizations of the 2-$T$ vMekal models are
expressed as ${\it Norm}_1 = C_{\rm 12}\, n_1^{\,2}\, V_1$ and
${\it Norm}_2 = C_{\rm 12}\, n_2^{\,2}\, V_2$
using a certain common constant, $C_{\rm 12}$,
the volume filling factor,
$f(R)$, can be solved under the pressure balance as,
\begin{eqnarray}
f(R) = \left[\; 1 + 
\left( \frac{T_2}{T_1} \right)^2 \frac{{\it Norm}_2}{{\it Norm}_1} 
\;\right]^{-1}.
\end{eqnarray}
We have calculated this formula for each shell
in 0.6--1.0$'$, 1.0--2.0$'$, 2.0--4.0$'$, and 4.0--8.0$'$
using the best-fit values in table~\ref{table:spec} to plot
$f(R)$ against the 3-dimensional radius $R$ in
figure~\ref{fig:mass-prof}~(a).
This plot clearly indicates that the cool component is
dominant at the central region, while it occupies only
$\lesssim$~1\% at $R>4'$.
We have fitted these four points with a 3-dimensional $\beta$-model function,
$f(R) = \left[ 1 + (R/R_{{\rm c,}f})^2 \right]^{-3\,\beta_f/2}$,
and obtained $R_{{\rm c,}f} = 0.43'\,^{+0.19}_{-0.16}$ and
$\beta_f = 0.60^{+0.19}_{-0.13}$.

Considering the projection effect,
we can calculate the volume occupied by the cool gas
at each 2-dimensional ring of
$\theta_{\rm in}$--$\theta_{\rm out}$ in unit of radian, as
\begin{eqnarray}
V_1 = f\; \frac{4}{3}\;\pi\; D_{\rm A}^{\,3}\;
(\theta_{\rm out}^3 - \theta_{\rm in}^3)\;
(1-\theta_{\rm r}^2)^{3/2}/\;(1-\theta_{\rm r}^3),
\end{eqnarray}
where $\theta_{\rm r}\equiv \theta_{\rm in} / \theta_{\rm out}$,
and $D_{\rm A} = 61$~Mpc is the angular diameter distance to HCG~62.
Because the vMekal normalization is defined as
${\it Norm}=\int n_{\rm e} n_{\rm H} dV/ (4\pi (1+z)^2 D_{\rm A}^{\,2})$,
the electron density of the cool gas, $n_{\rm e1}$, is computed as
\begin{eqnarray}
n_{\rm e1} = \sqrt{1.2\times 4\pi\; (1+z)^2\; D_{\rm A}^{\,2}\, /\; V_1},
\end{eqnarray}
assuming $n_{\rm e}=1.2\; n_{\rm H}$ for a fully ionized gas with
hydrogen and helium mass fraction of $X=0.7$ and $Y=0.28$.
The electron density of the hot gas, $n_{\rm e2}$,
is similarly derived by replacing $f$ into $(1-f)$.
The ion density including helium is $n_{\rm i} = 0.92\; n_{\rm e}$,
therefore the gas pressure is calculated as
\begin{eqnarray}\label{eq:Pgas}
P_{\rm gas} = 1.92\; n_{\rm e1}\; kT_1 = 1.92\; n_{\rm e2}\; kT_2.
\end{eqnarray}

The derived gas pressure $P_{\rm gas}$,
cool or hot gas temperature $kT_1$ or $kT_2$, and
cool or hot electron density $n_{\rm e1}$ or $n_{\rm e2}$
are plotted in figures~\ref{fig:mass-prof}~(b)--(d).
The 2-$T$ fit results are adopted for points
in the range of 0.6--4$'$ and the 1-$T$ fit results for others.
The systematic errors when the background level is increased or
decreased by $\pm 5$\% are considered in the error bars for
the outermost two points.
It is supposed that blue points belongs to the cool component
and red points to the hot one.

We must be careful in dealing with these plots because data points are
not independent with each other.
We first fitted the gas pressure plot by combining both the cool and hot
data points, with a 3-dimensional double $\beta$-model,
$P_{\rm gas} =
P_1 \left[ 1 + (R/R_{{\rm c,}P_1})^2 \right]^{-3\,\beta_{P_1}/2}+
P_2 \left[ 1 + (R/R_{{\rm c,}P_2})^2 \right]^{-3\,\beta_{P_2}/2}$.
Due to the small number of data points, we have fixed
$R_{{\rm c,}P_1}=0.1'$ and $\beta_{P_1}=0.65$ using the best-fit
values in table~\ref{table:radial} obtained with the radial
surface brightness profile fit.
The fitted parameters are $P_1 = 58\pm 130$ eV~cm$^{-3}$,
$P_2 = 18\pm 32$ eV~cm$^{-3}$,
$R_{{\rm c,}P_2} = 0.60\pm 1.44'$, and
$\beta_{P_2} = 0.38\pm 0.23$.
The best-fit model is indicated by a dashed line in
figure~\ref{fig:mass-prof}~(c).
The green curves represent each component of the double $\beta$-model.

We then fitted the cool gas temperature with a power-law model,
$kT_1(R) = a\, R^{\,b}$, and the hot gas electron density with
a 3-dimensional $\beta$-model and a constant,
$n_{\rm e2}(R) = n_{\rm e2,0}\;
\left[ 1 + (R/R_{{\rm c,}n_{\rm e2}})^2 \right]^{-3\,\beta_{n_{\rm e2}}/2}
+ C_{n_{\rm e2}}$.
The best-fit values are
$a=0.80\pm 0.17$~keV,
$b=0.09\pm 0.20$,
$n_{\rm e2,0} = 0.007\pm 0.020$~cm$^{-3}$,
$R_{{\rm c,}n_{\rm e2}} = 1.4\pm 7.2'$,
$\beta_{n_{\rm e2}} = 1.2\pm 5.8$, and
$C_{n_{\rm e2}} = (0.49\pm 0.48)\times 10^{-3}$~cm$^{-3}$.
The best-fit models for $kT_1$ and $n_{\rm e2}$ are
drawn by dashed lines.
The model functions for $kT_2$ and $n_{\rm e1}$ are derived using the
relation $kT_2 = P_{\rm gas} / n_{\rm e2}$ and
$n_{\rm e1} = P_{\rm gas} / (kT_2)$, which are
indicated by another dashed line in each panel.
Although errors of these parameters are quite large,
these functions can reproduce the observed properties
of the cool and hot gases fairly well.

\subsection{Dark matter and gas mass}
\label{subsec:gas-dm-mass-prof}

The gas mass density $\rho_{\rm gas}$ is expressed as 
\begin{eqnarray}
\rho_{\rm gas} = 1.92\;
\mu\, m_{\rm p}\; [\, f\, n_{\rm e1} + (1-f)\, n_{\rm e2} \,],
\end{eqnarray}
where $\mu=0.62$ is the mean molecular weight,
and $m_{\rm p}$ is the proton mass.
Assuming the hydrostatic equilibrium,
the total integrated gravitational mass, $M_{< R}$,
within the 3-dimensional radius of $R$ is given by
\begin{eqnarray}
M_{<R} = - \frac{R^2}{\,\rho_{\rm gas}\,G\,}\frac{dP_{\rm gas}}{dR},
\label{eq:Mtot}
\end{eqnarray}
in which $G$ is the gravitational constant.
The differential mass density, $M(R)$, is given by
\begin{eqnarray}
M(R) = \frac{1}{4\pi R^2} \frac{dM_{<R}}{dR}.
\end{eqnarray}
Figures~\ref{fig:mass-prof}~(e) and (f) show the integrated and
differential mass profiles (black lines) as a function of
the 3-dimensional radius in unit of arcmin or kpc.
We also overlaid the gas mass (gray) for the hot (red) and cool (blue)
components in the same panels, by integrating the gas density,
$\rho_{\rm gas}$.

However, in the two-phase model under the pressure balance,
the hot gas is lighter in mass density than the cool gas,
and becomes buoyant. 
The hot gas would escape from the group core if it is not enclosed 
by other mechanisms (see \S\,\ref{sec:supporting-cavities} in detail).
In such a case, hydrostatic equilibrium may have been broken,
hence we have also calculated the integrated gravitational mass
by treating the hot and cool components separately. 
Namely,
\begin{eqnarray}
\nonumber
M_{1,{<R}} &=& \makebox{$-\frac{R^2}{\,\rho_1\,G\,}\frac{dP_{\rm gas}}{dR}$},
\quad\rho_1 = 1.92\;\mu\, m_{\rm p}\, n_{\rm e1},\\
\nonumber
M_{2,{<R}} &=& \makebox{$-\frac{R^2}{\,\rho_2\,G\,}\frac{dP_{\rm gas}}{dR}$},
\quad\rho_2 = 1.92\;\mu\, m_{\rm p}\, n_{\rm e2},\\
M_{<R} &=& f\; M_{1,{<R}} + (1-f)\; M_{2,{<R}}.
\label{eq:Mtot'}
\end{eqnarray}
The gravitational mass indicated by dashed lines
in figures~\ref{fig:mass-prof}~(e) and (f)
are calculated using eq.~(\ref{eq:Mtot'}).
There are only small differences between the two at the group center;
the latter is larger by about 15\% at $R=0.1'$ (1.8~kpc).

We have encountered one severe problem in these plots.
The integrated total mass saturates at $R\sim 6'$ (100~kpc),
and the differential mass becomes even smaller than the gas mass.
This is physically unrealistic, suggesting that some presumption(s)
we have supposed might not be realized around this radius.
The direct source of the problem is caused by the fact
that the hot gas temperature, $kT_2$, drops very steeply
from $\sim 1.5$~keV to 0.64~keV in this radial range.
Therefore, the electron density, $n_{\rm e}$, almost saturates,
while the gas pressure, $P_{\rm gas}$, decreases monotonously.
This might in part be affected by the smaller field of view
in our observations than the extended group emission.
We cannot deproject the outermost shell in 8--14$'$,
so that X-ray flux in this shell is slightly overestimated.
Nevertheless, the major origin of the problem is due to the
steep temperature drop. It should be examined more precisely
by the Suzaku satellite, which has lower and more stable background
than XMM-Newton, and much superior low-energy sensitivity than ASCA\@.

Another possibility, which is astrophysically more interesting,
is that the hydrostatic equilibrium is not reached there.
The flattening of the electron density suggests an outflow
of the hot gas. The steep temperature drop can be reversely recognized
as the temperature rise in the boundary region between the cool and hot gas.
This suggests a shock heating of accreting ICM,
and/or remnants of past lifted cavities
(see \S\,\ref{sec:supporting-cavities}).
The apparent inconsistency in the total mass may therefore indicate
traces of the dynamical evolution.

Aside from the above problem, the derived gravitational mass is about
$M_{<R}=5\times 10^{12}M_\odot$ within 100 kpc,
close to the value given by \citet{Pildis1995},
$M_{\rm tot} = 2.9\times 10^{13}M_\odot$ within $r<15'$ (270~kpc)
with the ROSAT PSPC observation assuming a single-phase gas model.
We also plot the O (cyan) and Fe (pink) mass contained in the whole gas.
The simple linear fit plotted in figure~\ref{fig:plot-z} is adopted
to calculate the O and Fe mass.
The derived gas and Fe masses of
$M_{{\rm gas}<R}=1.6\times 10^{12} M_\odot$ and
$M_{{\rm Fe}<R}=4\times 10^{8} M_\odot$ within $r<300$~kpc
are consistent with the result from ASCA and ROSAT
by \citet{Finoguenov_Ponman1999} of
$M_{{\rm gas}<R}=(1.7\pm 0.1)\times 10^{12} M_\odot$ and
$M_{{\rm Fe}<R}=(3.8\pm 2.0)\times 10^{8} M_\odot$.
However, our values are based on the extrapolation out to
the XMM-Newton field of view, so that they have larger errors
by a factor of $\sim 2$.

\subsection{Stellar mass}
\label{subsec:stellar-mass}

The stellar mass in figures~\ref{fig:mass-prof}~(e) and (f)
indicated by green lines is estimated 
from the $R$ band photometry of the HCG~62a galaxy by \citet{Tran2001},
assuming the spherical symmetry and
the surface brightness profile obeying the de Vaucouleurs
$r^{1/4}$ law,
\begin{eqnarray}
\Sigma(r)\propto 10^{-3.33071\,\left[(r/r_{\rm eff})^{1/4}-1\right]},
\end{eqnarray}
where $r_{\rm eff}=32''$ is the effective radius, or the half light radius.
Because the $r^{1/4}$ law is for the surface brightness,
deprojection for the 3-dimensional radius must be performed.
We utilized a numerical table computed by \citet{Young1976}.
The mass-to-light ratio, $M_{\rm star}/L_B=8\; (M_\odot /L_{B,\odot})$,
is assumed, and we use $B-R=2.0$~mag by \citet{Hickson1989},
and $A_B=0.224$ and $A_R=0.139$ after NASA/IPAC Extragalactic Database
(NED) to calculate the $B$ band absolute magnitude,
$L_B = L_R + (B-R) - A_B + A_R = -19.46$~mag.
Using the value $L_{B,\odot}=5.48$~mag, the stellar mass of the HCG~62a
galaxy is calculated as $M_{\rm star}=7.6\times 10^{10} M_\odot$.

The $B$ magnitude within the effective diameter of 59.90$''$ isophotes
(table~\ref{tab1}) is also given by \citet{Hickson1989},
however, we found that $M_{\rm star}$ calculated from this value
was by about a factor of four larger than the value calculated above,
if we assume that $r_{\rm eff}$ is the same for both bands.
Then, the stellar mass would significantly exceed the total gravitational 
mass at the central region.
We therefore have adopted the stellar mass based on the $R$ band photometry.
\citet{Tran2001} also report that there is $19\%$ of disk component
which can be fitted by $\Sigma(r)\propto\exp(-r/r_{\rm d})$ for HCG~62a.
This gives asymmetric image residual flux $R_A=19$\%,
and the total residual fraction of light $R_T=18$\%.
Therefore, uncertainty on the stellar mass estimation is
at least $\sim 20$\%.

We notice several features in the mass profiles. The gas mass of the
cool component is taken over by the hot component at $\sim 1'$ (18~kpc).
The stellar mass is overcome by the gas
mass at $\sim 2.5'$ (45~kpc), indicating that the cool
component is very concentrated in the group center.  The region of
high metal abundance almost coincides with the volume dominated by the
stellar mass, which supports the natural view that stars are
responsible for the production of excess iron and silicon around the
group center.

\section{ICM properties}\label{sec:ICM-prop}

The temperature structure in HCG~62 can be characterized by a mixture
of hot ($\sim 1.4$ keV) and cool ($\sim 0.7$ keV) components.
As shown in figure \ref{fig:mass-prof},
the cool component is dominant within $r < 0.8'$ and then
a cool and hot mixture appears in $r = 0.8$--4$'$.
The hot component is dominant in the outer region.
The two-phase nature seems to be preferred than the
single-phase structure such as that seen in M~87 \citep{Matsushita2002}.
However, this apparent two-phase does not necessarily mean that
the cool and hot gas co-exist at the intermediate region,
instead it probably represents that they are patchy and/or
have irregular shapes.
As seen in figure~\ref{fig:hr}, the cool region is elongated
from northeast to southwest, roughly corresponding to
the direction of the two cavities.
In this sense, HCG~62 group is different from
NGC~1399, NGC~5044 and RGH~80 galaxy groups
\citep{Buote2002,Buote2003a,Xue2004},
in which strong evidence for a multi-phase gas is suggested
at the central region.
On the other hand, $T_{\rm cool}$ is similar to these groups,
and $T_{\rm cool}\simeq T_{\rm hot}/2$ is recognized among them.
It is claimed that the temperatures of the cool component
seen in these three groups are close to the kinetic temperature of the stars, 
although the stellar velocity dispersion is not known for the HCG~62a galaxy.

Mass density of the cool component exceeds the hot-component density
within about 10~kpc, in which the stellar mass density is orders of
magnitude higher.  This again implies that the cool component is
probably connected with the stars concentrated around the central
galaxy HCG~62a.  We also note that the cool component mass density
is always smaller than the stellar mass density.
On the other hand, the hot component exceeds the
stellar mass density around $r\sim 2'$ (30~kpc),
indicating that the gravitational potential of the galaxy group is
traced by the hot component. Its temperature (1.4~keV) is also typical
for groups of galaxies.

In the outer region ($r> 8'$), the temperature of the hot component
drops fairly sharply. The radius corresponds to about $0.15\, r_{\rm
vir}$, with $r_{\rm vir} \approx 1.1\; (T/1.5~{\rm keV})^{1/2}$ Mpc.
This causes a peak in the temperature profile at $r \approx 5'$ (90
kpc).  Similar temperature profiles have been observed in other galaxy
groups such as RGH 80, NGC 2563, and NGC 5044 (\cite{Xue2004},
\cite{Mushotzky2003}, \cite{Buote2003a}). The drop of temperature
suggests either that the dark matter is confined within this small
radius or that the gas is yet to be heated in this region. In either
case, there is a certain boundary of the group around $0.2\, r_{\rm
vir}$, and it is suggested that these groups are young and forming
systems.

The heavy elements are enriched by SN~Ia and SN~II
\citep{Tsujimoto1995}.
The former process dominantly yields Fe group
\citep{Iwamoto1999},
and the latter produces lighter elements like O, Ne and Mg
\citep{Thielemann1996,Nomoto2006}.
Both supernova contribute to Si and S\@.
Since O is produced mainly by SN~II,
the flatter O abundance profile compared with those of
Si and Fe implies enhanced contribution of SN~Ia in the central region
and/or that the shallow potential of HCG~62 is unable to confine
SN~II products which should have been supplied to the intracluster
space in the form of galactic winds.
The products of SN~Ia, on the other hand, are considered to be 
brought in to ICM by gas stripping.

The ratio, Mg/O, is $3.3\pm 2.2$ times solar within 1$'$ (18~kpc)
based on the 2-$T$ fit. 
Though the error is large, this value is similar or somewhat larger than 
those in other groups:
$2.5\pm 0.4$ solar for NGC~5044 \citep{Xu2002},
$1.3\pm 0.2$ for NGC~4636 \citep{Tamura2003}, both measured with RGS,
and $\sim 2$ solar in RGH~80 \citep{Xue2004}.
In M~87, this ratio is $1.3\pm 0.1$ solar within 20~kpc
\citep{Matsushita2003}.
Since both O and Mg are mainly produced by SN~II, 
the marginally high Mg/O value may reflect the difference
in the stellar initial mass function (IMF)\@.
Theoretical calculations \citep{Nomoto2006,Thielemann1996}
predict that a lighter-mass progenitor of SN~II synthesizes
enhanced Mg compared with O, therefore HCG~62 might have had the stellar IMF
with smaller number of massive stars ($M\gtrsim 20 M_\odot$)
than our galaxy and other groups above.
However, the dependence of the Mg/O ratio on the progenitor mass
is typically $\sim 20$\% at most, so that we have to think about
other possibilities for the Mg/O ratio greater than $\sim 1.2$.

If excess Fe in the center is caused by SN~Ia activity of HCG~62a,
it should be more extended at least than the extent of the central 
cool component which is the gas directly bound by the central galaxy.
The possible enhancement of the Ni/Fe ratio at $r<2'$
as seen in \S\,\ref{subsec:consistency} supports this scenario.
For example, the CDD2 model in \citet{Iwamoto1999}
predicts the Ni/Fe ratio of 1.8 solar.
The region of high metal abundance almost covers the volume
dominated by the stellar mass, which supports the natural view
that stars are responsible for the production of excess
Fe and Si around the group center.
It seems to range even to the 3D radius of $\sim 6'$ (100~kpc)
in figure~\ref{fig:plot-z}, where the hot gas mass exceeds the
stellar mass as seen in figure~\ref{fig:mass-prof}~(f).
This suggests either that the Fe production has
occurred in a wider region than the present location of the galaxy
or that there have been an outflow of Fe.
The presence of cavities in HCG~62 suggests that part of
the central metal-rich gas may have been lifted from the galaxy.

The observed abundance gradients indicate that there have been no
strong mixing occurred in the core region. We also note that there is
no significant change of temperature or abundance across the cavity
regions. These features imply that the process of cavity creation
causes mild, subsonic gas motion occurring in a fairly limited volume.
This implication is consistent with the result by \citet{Brueggen2002}
who showed that mixing by buoyant bubbles gave relatively weak impact
on the metallicity gradients based on numerical simulations.

\section{Supporting mechanisms of X-ray cavities}
\label{sec:supporting-cavities}

B04 have systematically studied 16 clusters,
1 group (HCG~62), and 1 galaxy (M84), in which prominent
X-ray surface brightness depressions (cavities or bubbles) are observed.
They find that a mechanical (kinetic) luminosity seems to correlate
with the 1.4~GHz synchrotron luminosity. However, its ratio ranges
widely between a few and several thousand, and they have concluded
that the radio luminosity is an unreliable gauge of
the mechanical power of the AGN jets.
\citet{Dunn2004} and D05
have studied 21 clusters, 3 galaxies, in which HCG~62 is not included,
and find that the ratio of an energy factor, $\mathcal{K}$,
to a volume filling factor, $\mathcal{F}$, shows a large scatter,
$1\lesssim \mathcal{K}/\mathcal{F} \lesssim 1000$, for active cavities
associated with radio lobes, and that it becomes
even larger for ghost cavities.
The factor, $\mathcal{K}$, accounts for the additional energy
from relativistic particles accompanying the electrons (e.g., protons).
The parameter of $\mathcal{F}$ represents the volume filling factor
of the relativistic particles, and is supposed not to vary far from unity.
Typical value of $\mathcal{K}$ in literature is $\mathcal{K}=100$,
according to the measurements of cosmic-rays around the solar system.
However, there is no direct evidence indicating
that such a high energy density is really carried by protons.
The reason why $\mathcal{K}/\mathcal{F}$ varies so largely
from cluster to cluster is a mystery, which might imply
that there are several ways in supporting mechanism and/or
formation of the X-ray cavities.

In this section, we examine whether this non-thermal pressure support
scenario is realistic or not for HCG~62.
In \S\,\ref{subsec:cavity:characteristics},
we summarize our X-ray results, as well as optical and radio observations.
In \S\,\ref{subsec:cavity:gas}--\ref{subsec:cavity:energetics},
standard indices of cavities are calculated for HCG~62,
and difficulties in the non-thermal support scenario is
considered in \S\,\ref{subsec:cavity:difficulties}.
In \S\,\ref{subsec:cavity:clump} and \S\,\ref{subsec:cavity:gal},
other possibilities of the supporting mechanism and
formation of cavities are investigated.

\subsection{Characteristics in X-ray, optical, and radio}
\label{subsec:cavity:characteristics}

We have confirmed the X-ray cavities reported by
\authorcite{Vrtilek2001}~(\yearcite{Vrtilek2001},\yearcite{Vrtilek2002})
in the northeast and southwest regions of HCG~62 
in the Chandra image as shown in
figures~\ref{fig:image-acis-xmm}~(a) and  \ref{fig:cavity image}.
As described in \S\,\ref{subsec:cavity spec},
absorption is unlikely to be the origin of cavities.
There are no significant spectral differences
between cavity and non-cavity regions,
nor any traces of the shock-heated gas around the cavities,
as seen in the temperature map (figure~\ref{fig:hr}).
According to our hollow sphere model analysis (\S\,\ref{subsec:hollow}),
both the cavities should be aligned side by side with the group core
to a fairly good degree, namely, the projected distances to the group core
are close to their real ones. It is also suggested that
the shape of cavities is probably elongated in the direction of
our line of sight, and that the side of the cavities 
which is farther from the core should be 
larger in size and/or weaker in the X-ray emissivity.

With regard to the central HCG~62a galaxy,
we found no evidence for the AGN activity in the X-ray data.
The best-fit spectrum of HCG~62a is a thermal emission rather than
a power-law one.  No point sources were recognized at the center of
HCG~62a in both soft and hard bands.
The upper limit of the AGN emission can be placed around
$L_{\rm X}\lesssim 10^{39}$~erg~s$^{-1}$ (0.5--4~keV)\@.
We also found that the location of HCG~62a was slightly shifted
from the group core by about 5$''$ (1.5~kpc) on the projected sky image
(table~\ref{table:2Dradial}).
This positional shift may imply the effect of gas stripping
and/or the dynamical motion of the galaxy.

The ASCA has detected a spatially extended ($\sim 10'$)
excess hard X-ray emission above $\sim 4$~keV
in the HCG~62 group \citep{Fukazawa2001},
which is supposed to be due to the
relativistic electrons with Lorentz factor
$\gamma\sim 10^3$--$10^4$ and/or sub-relativistic particles.
This may also have some relation to the X-ray cavity,
however, we could not confirm nor reject this result
due to the higher non X-ray background of
Chandra and XMM-Newton than ASCA\@.

In the optical band, \citet{Coziol1998} and \citet{Shimada2000}
have detected weak [NII] and [OI] emission lines in
the spectrum of HCG~62a, and classified it as
a low luminosity AGN (LLAGN)\@.
On the other hand, \citet{Coziol2004} have assigned
the lowest activity index of $-5$ (quiescent, intermediate
and old stellar populations) to HCG~62a, according to
the equivalent width measurement of the H$\alpha$ absorption lines.
They have also distinguished HCG~62 group among 27 compact groups
of galaxies as type C, which comprises groups with high velocity
dispersions and are dominated by elliptical galaxies with no activity,
presumably corresponding to the later stage of the evolution.

In the radio band, as shown in figure~\ref{fig:image-acis-xmm}~(a),
weak radio emission at 1.4~GHz is detected around HCG~62a,
however its luminosity is as small as
$L_{\rm radio}=1.8\times 10^{38}$~erg~s$^{-1}$ (10 MHz--5 GHz),
smallest in table~1 of B04.
The radio emission does not show a clear association with
the cavities, though the angular resolution of 45$''$
FWHM is not sufficient to see detailed structures.

\subsection{Gas pressure \& non-thermal pressure}
\label{subsec:cavity:gas}

We take the distances to both cavities from the group core
to be $R_{\rm cav}=25''$ (7.4~kpc) and their radii to be 
$r_{\rm cav}=13.5''$ (4.0~kpc)
for simplicity. This approximation can be justified by the similarity
between north and south cavities measured from the group core
(table~\ref{table:cavity non-cavity}) with the center position
determined from the center of the wider component in 2-dimensional
2-$\beta$ fit (table~\ref{table:2Dradial}).
Assuming those parameters, the ICM gas pressure
at $R_{\rm cav}$ is calculated to be
$P_{\rm gas} = 17$ eV~cm$^{-3}$, from eq.~(\ref{eq:Pgas}).

The non-thermal pressure of relativistic particles can be
estimated from the radio intensity.
Here, we assume that half of the radio flux
comes from a single cavity, namely,
$L_{\rm radio}=9\times 10^{37}$~erg~s$^{-1}$ (10 MHz--5 GHz)
according to B04.
Assuming that the radio emission is due to synchrotron radiation
by relativistic electrons,
the total non-thermal pressure can be calculated as
$P_{\rm tot} = \mathcal{K} P_{\rm e} + \mathcal{F} P_{\rm B}$,
where $P_{\rm e}$ is a pressure of the relativistic electrons,
and $P_{\rm B}$ is a pressure of the magnetic field,
following the convention adopted by D05. 

The relativistic electron pressure is calculated to be
$P_{\rm e} = C_{\rm e} L_{\rm radio} B^{-3/2}/V$
(e.g., \cite{Govoni_Fertti2005}),
where $B$ is a magnetic flux density,
$V\equiv\frac{4}{3}\pi r^3_{\rm cav}$ is the volume of the cavity,
and $C_{\rm e}$ is a constant depending on the spectral index $\alpha$
of the radio emission, as
$C_{\rm e}=\sqrt{27m_{\rm e}^5 c^9/(2\pi e^7)}\;(\nu_1^{-1/2}-\nu_2^{-1/2})/
(\log\nu_2-\log\nu_1)=1.4\times 10^{8}$ [cgs] for $\alpha=1$,
$\nu_1=10$~MHz, and $\nu_2=5$~GHz. 
Therefore,
\begin{eqnarray}\nonumber &&
\mathcal{K} P_{\rm e} = 3.3\times
\\\nonumber && \hspace*{-1em}
\left(\frac{\mathcal{K}}{100}\right)
\left(\frac{L_{\rm radio}}{9{\scriptstyle\times} 10^{37}\,\makebox{cgs}}\right)
\left(\frac{r_{\rm cav}}{4\,\makebox{kpc}}\right)^{-3}
\left(\frac{B}{10\,\makebox{$\mu$G}}\right)^{-\frac{3}{2}}
~\makebox{eV~cm$^{-3}$}.
\end{eqnarray}
On the other hand, the magnetic pressure is
\begin{eqnarray}\nonumber
\mathcal{F} P_{\rm B} = \mathcal{F}\, \frac{B^2}{8\pi} = 2.5\;
\left(\frac{\mathcal{F}}{1}\right)
\left(\frac{B}{10\,\makebox{$\mu$G}}\right)^2
~\makebox{eV~cm$^{-3}$}.
\end{eqnarray}
Therefore, the total non-thermal pressure,
$P_{\rm T}\equiv \mathcal{K} P_{\rm e} + \mathcal{F} P_{\rm B}$,
takes the minimum at a certain value of $B=B_{\rm eq}$,
the so-called equipartition condition, which is calculated as
\begin{eqnarray}\nonumber &&
B_{\rm eq} = 11\;
\left(\frac{\mathcal{K}/\mathcal{F}}{100}\right)^\frac{2}{7}
\left(\frac{L_{\rm radio}}{9{\scriptstyle\times} 10^{37}\,\makebox{cgs}}\right)^\frac{2}{7}
\left(\frac{r_{\rm cav}}{4\,\makebox{kpc}}\right)^{-\frac{6}{7}}
~\makebox{$\mu$G},
\\\nonumber &&
P_{\rm T,eq}=\frac{7}{4}P_{\rm e,eq}=\frac{7}{3}P_{\rm B,eq}=5.8\times
\\\nonumber && \hspace*{1em}
\left(\frac{\mathcal{K}/\mathcal{F}}{100}\right)^\frac{4}{7}
\left(\frac{L_{\rm radio}}{9{\scriptstyle\times} 10^{37}\,\makebox{cgs}}\right)^\frac{4}{7}
\left(\frac{r_{\rm cav}}{4\,\makebox{kpc}}\right)^{-\frac{12}{7}}
~\makebox{eV~cm$^{-3}$}.
\end{eqnarray}

Thus the derived pressure at the equipartition condition is
less than half of the ICM gas pressure,
while the required equipartition magnetic flux density is typical
for radio lobes of radio-loud AGNs \citep{Kataoka2005}.
They have also found that the equipartition condition is
achieved at least for the radio lobes of 40 radio galaxies.
The discrepancy between $P_{\rm T,eq}$ and $P_{\rm gas}$
may be explained by the underestimation of $\mathcal{K}/\mathcal{F}\sim 100$
or $L_{\rm radio}\sim 9\times 10^{37}$~erg~s$^{-1}$,
or by non-equilibrium situation.
To make a balance between them, $\mathcal{K}/\mathcal{F}=690$ is needed.
This value is consistent with those obtained by D05
for active bubbles in other clusters.
It is also plausible that the radio intensity is getting dimmer in time 
due to the synchrotron cooling of relativistic electrons.
We also note that the $L_{\rm radio}$ value used for the
calculation is probably overestimated,
because the observed radio intensity is likely to include
emission from the core regions.
Radio observation with higher angular resolution is desired.

\subsection{Time scales}
\label{subsec:cavity:time}

For discussion of cooling and non-equilibrium effects,
comparison of several time scales is important.
First of all, synchrotron electron ($\gamma_{\rm e}\gtrsim 10^4$)
must be long-lived since the cavity is not fueled now from the central AGN.
The synchrotron cooling time is calculated as
\begin{eqnarray}\nonumber
t_{\rm sync} = \frac{9 m_{\rm e}^3 c^5}{4 e^4 B^2 \gamma_{\rm e}} = 25\;
\left(\frac{B}{10\,\makebox{$\mu$G}}\right)^{-2}
\left(\frac{\gamma_{\rm e}}{10^4\rule[-0.5ex]{0mm}{0mm}}\right)^{-1}
~\makebox{Myr}.
\end{eqnarray}
Since rims of the cavities have not been shock heated,
the cavities are supposed to have expanded at a velocity less than
the sound speed,
$v_{\rm s}=\sqrt{\gamma kT/(\mu m_{\rm p})}=425$~km~s$^{-1}$,
where we have taken $kT=0.7$~keV, $\gamma=5/3$ and $\mu=0.62$.
Therefore, age of the cavities must be longer than
the expansion time of
\begin{eqnarray}\nonumber
t_{\rm s} = \frac{r_{\rm cav}}{v_{\rm s}} = 9.2\;
\left(\frac{r_{\rm cav}}{4\,\makebox{kpc}}\right)
~\makebox{Myr},
\end{eqnarray}
which is shorter than $t_{\rm sync}$.
If we assume $\mathcal{K}/\mathcal{F}=690$,
$t_{\rm sync}$ can be shortened by a factor of
$(\mathcal{K}/\mathcal{F}/100)^{-4/7}=0.33$,
and becomes comparable to $t_{\rm s}$.
The time scale for the cavities to collapse when internal pressure
has disappeared is also considered to be about $t_{\rm s}$.
Such hollow cavities are buoyant and rise up outwards,
even when pressure balance between inside and outside of the cavities
is conserved. \citet{Churazov2001} have given the terminal velocity
as $v_{\rm t} = \sqrt{2 g V / SC}$, where $S\equiv \pi r^2_{\rm cav}$
is the cross section of the bubble, $g\equiv GM_{<R_{\rm cav}}/R_{\rm cav}^2$ 
is the gravity at the bubble, and $C=0.75$ is the drag coefficient.
The travel time of the cavities to the current position is estimated to be
\begin{eqnarray}\nonumber &&
t_{\rm buoy} = R_{\rm cav}/v_{\rm t} = 15\times
\\\nonumber && \hspace*{1em}
\left(\frac{R_{\rm cav}}{7.4\,\makebox{kpc}}\right)^2
\left(\frac{r_{\rm cav}}{4\,\makebox{kpc}}\right)^{-1/2}
\left(\frac{M_{<R_{\rm cav}}}{2{\scriptstyle\times} 10^{11} 
M_\odot}\right)^{-1/2}
~\makebox{Myr},
\end{eqnarray}
which is comparable to or somewhat longer than
$t_{\rm sync} = 8$~Myr with $\mathcal{K}/\mathcal{F}=690$.
The refill time of the cavity is also given by \citet{McNamara2000} as,
\begin{eqnarray}\nonumber &&
t_{\rm refill} = 2 R_{\rm cav}\, \sqrt{r_{\rm cav}/(GM_{<R_{\rm cav}})} 
= 31\times
\\\nonumber && \hspace*{1em}
\left(\frac{R_{\rm cav}}{7.4\,\makebox{kpc}}\right)
\left(\frac{r_{\rm cav}}{4\,\makebox{kpc}}\right)^{1/2}
\left(\frac{M_{<R_{\rm cav}}}{2{\scriptstyle\times} 10^{11} M_\odot}
\right)^{-1/2}
~\makebox{Myr}.
\end{eqnarray}
These considerations indicate that
the cavity age of $t_{\rm s}$ or $t_{\rm buoy}\sim 10$~Myr
is much shorter than the group age of $\sim$~Gyr.
Our time scales are consistent with B04, but in which they use 
$R_{\rm cav}$ instead of $r_{\rm cav}$.

\subsection{Energetics}
\label{subsec:cavity:energetics}

On the assumption that past AGN activity has produced the two cavities
in about $(t_{\rm buoy}/2)$, we can estimate the required
mechanical (kinetic) power of the AGN jets.
A work to generate two cavities, $W_{\rm mech}\equiv 2 P_{\rm gas} V$,
divided by $(t_{\rm buoy}/2)$ is called the mechanical luminosity (B04),
and calculated to be
\begin{eqnarray}\nonumber &&
L_{\rm mech} = 4\, P_{\rm gas}V / t_{\rm buoy} = 1.8\times 10^{42}\times
\\\nonumber && \hspace*{1em}
\left(\frac{R_{\rm cav}}{7.4\,\makebox{kpc}}\right)^{-2}
\left(\frac{r_{\rm cav}}{4\,\makebox{kpc}}\right)^\frac{7}{2}
\left(\frac{M_{<R_{\rm cav}}}{2{\scriptstyle\times} 10^{11} M_\odot}
\right)^\frac{1}{2}
~\makebox{erg~s$^{-1}$}.
\end{eqnarray}
It is supposed that this level of the AGN activity must continue for about
$-t_{\rm buoy}< t < -0.5\; t_{\rm buoy}$,
and that it became inactive since $-0.5\; t_{\rm buoy} < t$.
This value is comparable to the X-ray emission typically observed for LLAGN,
although our upper limit on the X-ray luminosity is much lower,
$L_{\rm X}\lesssim 10^{39}$~erg~s$^{-1}$ (0.5--4~keV)\@.
The observed radio emission around HCG~62a,
$L_{\rm radio}=1.8\times 10^{38}$~erg~cm$^{-1}$,
is also much lower than $L_{\rm mech}$, as pointed out by B04.

Because there are no evidence for the strong AGN activity
at present time, the central AGN, if exists, should have made
a final outburst of total energy $\sim 10^{57}$~erg
within $\sim 20$~Myr ago. Clearly, a single supernovae cannot
account for this size of energy.
Such an absence of strong X-ray or radio emission
at the core is also noticed in NGC~4636
($L_{\rm radio}=1.4\times 10^{38}$~erg~s$^{-1}$;
$L_{\rm X}<2.7\times 10^{38}$~erg~s$^{-1}$;
\cite{Ohto2003,O'Sullivan2005}),
which shows a disturbed X-ray halo containing cavities
associated with small-size jets.
This might suggest that the AGN activity continues only for
a short time scale, which is difficult to understand within
the popular paradigm of AGN with a steady accretion disk.
Formation of ghost cavities with an impulsively episodic activity
of $\ll 10^5$~yr is discussed by \citet{Wang_Hu2005}.
Considering captures of red giant stars by a super massive black hole
($M_{\rm BH} > 2\times 10^8 M_\odot$),
its feedback energy can amount to $2.4\times 10^{52}$~erg
with a frequency of a few $10^{-5}$~yr$^{-1}$,
which is too small to supply sufficient energy for the formation of cavities.
It is probably true that HCG~62a also contains a massive black hole
of $M_{\rm BH}\sim 10^8 M_\odot$,
because observations of the centers of nearby early-type galaxies
show almost all have massive black hole \citep{Tremaine2002}.
However, we have to consider switching (on\,$\rightarrow$\,off)
of the AGN activity.

\subsection{Difficulties in non-thermal pressure support by jets}
\label{subsec:cavity:difficulties}

In \S\,\ref{subsec:cavity:gas}--\,\ref{subsec:cavity:energetics},
we have examined the properties of the cavities
and the central AGN from the point of view that 
the non-thermal pressure of relativistic particles were
provided through symmetrical jets of the past AGN activity.
However there are several difficulties in this scenario.

As seen in \S\,\ref{subsec:cavity:gas},
$\mathcal{K}/\mathcal{F}=690$ is needed to balance
the ambient gas pressure with the internal non-thermal pressure
at the equipartition condition.
This value is in proportion to the inverse of $L_{\rm radio}$,
which is probably overestimated because significant fraction
of $L_{\rm radio}$ should be attributed to the core region
instead of the radio robes (figure~\ref{fig:image-acis-xmm}~(a)).
Then $\mathcal{K}/\mathcal{F}=690$ becomes still larger,
which is probably unrealistic.
One possibility is that the relativistic electrons are
fading due to the synchrotron loss, which may be
justified by the fact that $t_{\rm sync}$
under the equipartition magnetic field of $B_{\rm eq}=17$~$\mu$G
($\mathcal{K}/\mathcal{F}=690$) is comparable to the estimated
cavity age of $t_{\rm buoy}$.
The relativistic electrons have much shorter synchrotron cooling time
than that of protons, hence apparently large $\mathcal{K}/\mathcal{F}$
might have been attained.
In this case, the cavities are starting to collapse
and losing the internal pressure support.

It is notable that both the two cavities observed in HCG~62 are
pretty circular when projected in the sky
(figure~\ref{fig:image-acis-xmm}~(a)).
The relative deviation image of figure~\ref{fig:cavity image}
is remarkably smooth around the group core,
and there are no obvious trails toward the cavities.
This fact is quite difficult to understand considering the scenario that
a pair of radio robes produced by symmetrical jets
from the AGN has pushed away the IGM\@.
In practice, clusters or galaxies hosting cavities usually show
irregular or filamentary structures around the cluster core and cavities.
The most prominent example is those of M~87
\citep{Churazov2001,Young2002}, in which the ``trails'' of
the rising two radio bubbles are clearly seen in the X-ray image.
Furthermore, our hollow sphere analysis suggests that the shape of
cavities is probably elongated in the direction of our line of sight.
It appears to be difficult for the symmetrical jets to make
an elongation in such direction.
We also note that shock heating is unlikely to be 
the origin of the cavities, because there is no significant
evidence of heating at the edge of the cavities.

These considerations indicate that there might exist
another supporting mechanism and/or formation scenario
of the X-ray cavities, at least for HCG~62.
In the following subsections,
we consider these possibilities.

\subsection{Another supporting mechanism --- hot gas clump ---}
\label{subsec:cavity:clump}

One possible supporting mechanism is a clump of hotter gas
than the surrounding ICM ($kT=0.7$~keV)\@.
For example, if we assume that the temperatures
in the cavities are higher than that of ICM by three times,
namely $T_{\rm hc}=3\, T$, the required density, $n_{\rm e,hc}$,
is three times smaller.
The emitted X-ray is roughly in proportion to
$n_{\rm e,hc}^2 \sqrt{T_{\rm hc}} = 0.2\, n_{\rm e}^2 \sqrt{T}$,
hence the observed X-ray intensity can be as low as 20\%.
Since we could have detected such hot emission if its flux were $\sim 20$\%
of the $F_{\rm sphere}$ in table~\ref{table:cavity non-cavity},
we can derive the lower limit of the hot clump temperature,
as $T_{\rm hc}\gtrsim 3\, T$.

\citet{Schmidt2002} have done a similar discussion
and ruled out volume-filling X-ray gas with temperature
below 11~keV for a cavity in the Perseus cluster.
With regard to the origin of the hot gas clump,
it is suggested that the jets may intrinsically contain
protons, or that they may have captured ambient thermal
protons possibly shock-heated in the very initial phase
of the cavity formation (D05).
If we observe in higher energy band, such hot emission can
be detected as a hard tail of the spectrum.
The hard $\Gamma=1.5$ power-law component observed with
ASCA \citep{Fukazawa2001} might originate from this kind
of hot thermal emission, although the detected sky area of
the power-law component is much more extended than the cavities.

If such a hot clump really exists, it undergoes a cooling by
thermal conduction. 
Assuming $kT_{\rm hc}=3\; kT \simeq 2$~keV
and $n_{\rm e,hc}=n_{\rm e}/3 \simeq 10^{-3}$~cm$^{-3}$,
the thermal conductivity for a non-magnetized plasma
is given by \citet{Spitzer1962} as,
\begin{eqnarray}\nonumber &&
\kappa_{\rm S}=10^{30}\;
\left(\frac{kT_{\rm hc}}{2\,\makebox{keV}}\right)^\frac{5}{2}
\left(\frac{n_{\rm e,hc}}{10^{-3}\,\makebox{cm$^{-3}$}\rule[-0.5ex]{0mm}{0mm}}\right)^{-1}
\left(\frac{\ln\Lambda}{36}\right)^{-1}
~\makebox{cm$^2$~s$^{-1}$},
\end{eqnarray}
where $\ln\Lambda$ is the Coulomb logarithm.
Therefore, the cooling time of the cavities is roughly calculated to be
\begin{eqnarray}\nonumber &&
t_{\rm cond}\simeq \frac{r_{\rm cav}^2}{\kappa_{\rm S}}=0.5\;\times
\\\nonumber && \hspace*{1em}
\left(\frac{r_{\rm cav}}{4\,\makebox{kpc}}\right)^2
\left(\frac{kT_{\rm hc}}{2\,\makebox{keV}}\right)^{-\frac{5}{2}}
\left(\frac{n_{\rm e,hc}}{10^{-3}\,\makebox{cm$^{-3}$}\rule[-0.5ex]{0mm}{0mm}}\right)
~\makebox{Myr},
\end{eqnarray}
which is much shorter than the estimated cavity life span of $t_{\rm buoy}$.

It is claimed that the thermal conductivity may become
5--10 times smaller than the Spitzer value
under the turbulent magnetic fields (e.g., \cite{Chandran2004}),
although $t_{\rm cond}$ seems to be still smaller considering this effect.
On the other hand, from the observational point of view,
we do see this kind of temperature variations in clusters.
The most prominent example is the ``cold front'',
first reported by \citet{Markevitch2000} for A2142.
\citet{Ettori2000} have pointed out that it requires the classical
Spitzer thermal conductivity to be reduced at least by a factor
of 250--2500. \citet{Markevitch2003} also find that
$\sim 40$ times reduction is needed for A754.
For the magnetic field expected in the radio robe ($B\sim 10$~$\mu$G),
the electron and proton gyro radii are by 11--12 orders of magnitude
smaller than their Coulomb mean free paths, therefore the
effective conductivity strongly depends on the topology
of the field. 
If the magnetic field encloses the hot clump like a cage,
the hot clump may survive for more than $\sim 20$~Myr.

\subsection{Another formation scenario --- galaxy motion ---}
\label{subsec:cavity:gal}

As we found in \S\,\ref{subsec:radial} with the 2-dimensional
2-$\beta$ model fit, the location of HCG~62a (narrower component)
is slightly offset from the group core (wider component)
by about 5$''$ (1.5~kpc) on the projected sky image.
This fact inevitably leads to the idea that the HCG~62a galaxy
is moving around the group core. It is natural to consider that HCG~62a
is performing a pseudo-Kepler motion, gradually decreasing its
distance to the group core by a dragging force.
From this point of view, the two cavities might be
a piece of the trail of the HCG~62a orbit.
Here, we assume that the orbital plane of HCG~62a
is nearly in parallel to our line of sight, with a circular orbit.
and that its orbital radius is close to
the distance, $R_{\rm cav}$, of the two cavities to the group core.

It is interesting that the two cavities are located almost
at the same distance from the group core,
whereas the south cavity is by 1.6 times closer to HCG~62a
than the north cavity. As indicated by a cross in 
figure~\ref{fig:image-acis-xmm}~(a), HCG~62a is shifted
from the group core indicated by a plus mark
(center of the wider component) roughly toward the direction 
of the south cavity.
These two facts support the assumption above.
The measured redshift of HCG~62a is consistent with
that of the group (table~\ref{tab1}).
On this assumption, the rotation speed is calculated
to be $v_{\rm rot}=\sqrt{G M_{\rm <R_{\rm cav}} / R_{\rm cav}} 
= 340$~km~s$^{-1}$,
which is comparable to the sound speed, $v_{\rm s}=425$~km~s$^{-1}$.
The period of rotation is
\begin{eqnarray}\nonumber
t_{\rm rot} = \frac{2\pi R_{\rm cav}}{v_{\rm rot}} = 133\;
\left(\frac{R_{\rm cav}}{7.4\,\makebox{kpc}}\right)^\frac{3}{2}
\left(\frac{M_{<R_{\rm cav}}}{2{\scriptstyle\times} 10^{11} M_\odot}\right)^{-\frac{1}{2}}
~\makebox{Myr}.
\end{eqnarray}

Even though it is at least shorter than the group age, 
is much longer than other time scales,
and the ``tunnel'' seems to be filled relatively quickly.
In this scenario, the ICM gas would have experienced the encounter
with the HCG~62a galaxy several times periodically.
If some processes, e.g., freezing the plasma with magnetic field
or the pressure support with hotter gas,
have slowed the collapse of the cavities,
this effect may be worthwhile consideration.
In terms of the energetics, the kinetic energy of
the galaxy motion amounts to $\sim 10^{59}$~erg,
therefore it can supply sufficient
energy to produce the cavities by depositing
7\% of the kinetic energy per orbit.
Such a motion of the galaxy also would have played an important
role on the mixing and the metal enrichment of the ICM,
which has been discussed in the previous section.

\section{Conclusion}\label{sec:conc}
\begin{itemize}
\item
We have carried out a detailed study on the hot-gas properties of the
group of galaxies HCG~62, based on the data from Chandra and
XMM-Newton. We confirmed the two cavities located almost symmetrically
around the central galaxy.
\item
The size of spherical hollow cavities are constrained from the surface
brightness structure to be 12$''$--17$''$. The agreement with the
observed angular size suggests that the gas density in the cavity is
very low, less than 1/3 and consistent with zero.
\item
The spectral fit indicated that the cavities were not caused by X-ray
absorption. The observed temperature in the cavity region is
consistent with that in the surrounding region.
\item
The spectrum within $4'$ from the center requires two temperatures:
0.7~keV and 1.4~keV\@. The cool component is centrally concentrated,
narrower than the hot component, suggesting its association with
the central galaxy HCG~62a.
\item
The mass profiles were obtained for the gas and stars. The hot
component is much more extended than the stars, and thought to trace
the gravitational potential of the galaxy group.
\item
The gravitational mass density drops steeply at about 5$'$ from the
center. This is caused by the observed sharp drop of the temperature.
There is a possibility that these regions are not in the hydrostatic
equilibrium.
\item
Abundance of O is $\sim 0.3$~solar, $\sim 3$ times less abundant
than Fe and Si, and shows a flatter profile.
The shallow potential of HCG~62 is unable to confine the SN~II
products which should have been escaped in the form of galactic winds.
The marginally higher Mg/O ratio of $3.3\pm 2.2$ implies steeper IMF\@.
\item
Abundances of Fe and Si show concentration in the central region,
and a high Ni/Fe ratio is suggested.
These results are consistent with that they are synthesized
by SN~Ia in the central galaxy.
\item
The non-thermal energy density necessary to support the cavity implies
$\mathcal{K}/\mathcal{F}=690$, namely almost 700 times larger energy
than that of electrons needs to be contained the cavity. The lack of
the central AGN or the trailing radio feature seems to suggests that
the origin of the cavity in HCG~62 may not be directly related to
AGN activities.
\item
We looked into alternative scenarios for the cavity creation. A clump
of very hot gas and fast motion of the central galaxy were considered,
but more observational evidences are necessary to perform a
quantitative evaluation.
\end{itemize}

\bigskip
Thanks are given to an anonymous referee for useful comments which improved 
the original manuscript.
Part of this work was financially supported by a Research Fellowship for 
Young Scientists from JSPS and Grant-in-Aid for Scientific Research 
(No.\ 16340077) from the Japan Society for the Promotion of Science,
and also by a Grant-in-Aid of the Ministry of Education, Culture, Sports,
Science and Technology (14079103; 16340077).
N.~O.\ acknowledges support from the Special Postdoctoral
Researchers Program of RIKEN.

\end{document}